\documentclass[12pt,a4paper,oneside]{article}

\usepackage[a4paper, total={6.5in,9in}]{geometry}
\usepackage{amsmath,amssymb,amsfonts}
\usepackage{graphicx}
\usepackage{mathtools}
\usepackage{subcaption}
\usepackage[font={normalsize}]{caption}
\usepackage{float}
\usepackage{blindtext}
\usepackage{dblfloatfix}
\usepackage{balance}
\usepackage{flushend}
\usepackage{commath}
\usepackage{bbold}
\usepackage{hyperref}
\usepackage{mathbbol}
\usepackage{microtype}

\begin{document}

\begin{center}
\Large{{\noindent Accurate 2D Reconstruction for PET Scanners based on the Analytical White Image Model}}
\end{center}
\begin{center}
    Tomislav Matuli\'{c}*, Damir Ser\v{s}i\'{c}
\end{center}
\begin{center}
    \small{{\noindent * Corresponding author}}
\end{center}
\begin{center}
Department of Electronic Systems and Information Processing, University of Zagreb Faculty of Electrical Engineering and Computing, Unska 3, 10000 Zagreb, Croatia
\end{center}
\noindent\rule{\textwidth}{1pt}
\begin{abstract}
In this paper, we provide a precise mathematical model of crystal-to-crystal response which is used to generate the white image - a necessary compensation model needed to overcome the physical limitations of the PET scanner. We present a closed-form solution, as well as several accurate approximations, due to the complexity of the exact mathematical expressions. We prove, experimentally and analytically, that the difference between the best approximations and real crystal-to-crystal response is insignificant. The obtained responses are used to generate the white image compensation model. It can be written as a single closed-form expression making it easy to implement in known reconstruction methods. The maximum likelihood expectation maximization (MLEM) algorithm is modified and our white image model is integrated into it. The modified MLEM algorithm is not based on the system matrix, rather it is based on ray-driven projections and back-projections. The compensation model provides all necessary information about the system. Finally, we check our approach on synthetic and real data. For the real-world acquisition, we use the Raytest ClearPET camera for small animals and the NEMA NU 4-2008 phantom. The proposed approach overperforms competitive, non-compensated reconstruction methods.
\end{abstract}
\noindent\rule{\textwidth}{1pt}

\section{Introduction}
Image reconstruction in Positron Emission Tomography (PET) has been well investigated over the last 50 years. Textbook analytical methods for image reconstruction, like Filtered Back-Projection (FBP) and Back-Projection Filtering (BPF), are well known, but rarely used in practice. They usually solely rely on Radon transformation and do not take into account a precise physical model of the PET scanner. Thus, reconstructed real-world images tend to be distorted. Still, analytical methods can consider the physical model of the PET scanner, as described in \cite{Matulic2021}, which results in less image degradation. 

Besides analytical methods, there are several iterative reconstruction algorithms, which prevail in practical applications. Important representatives are Maximum
Likelihood Expectation Maximization (MLEM) \cite{Shepp1982}, and its more efficient version Ordered Subset Expectation Maximization (OSEM) \cite{Hudson1994}. Their essential part is the projection model. Advanced applications take into the account detailed physical model of the scanner. Even when using a detailed physical model of the scanner, reconstructed images are noisy. Some techniques regarding the noise cancellation can be seen in \cite{Seri2016}, \cite{Seri2014}, \cite{Tomic2013}, and \cite{Tomic2012}. Moreover, state-of-the-art algorithms use a machine learning approach for the enhancement of the reconstructed image. An iterative ML algorithm approach using a Convolutional Neural Network (CNN) can be seen in \cite{Gong2019a}. The goal is to produce high-quality images using a low dosage of radioactive tracer. This idea was expanded in \cite{Xie2020a}, where a Generative Adversarial Network (GAN) was used. In \cite{Gong2019b}, the approach based on the deep image prior was investigated. The idea behind the deep image prior is to enhance the reconstructed image without pre-training. Another example of using CNN in image reconstruction can be seen in \cite{Ida2019a}. A CNN performs the mapping between sinogram and image space. On the same note, an interesting approach that is based on GAN is in \cite{Yang2020GAN}. Moreover, the application of neural networks in PET imaging can be seen through attenuation correction (\cite{Han2017a}, \cite{Liu2018a}, \cite{Nie2017a}, \cite{Wolterink2017a}), scatter correction (\cite{Berker2018a}, \cite{Qian2017a}) and denoising (\cite{Zhang2017a}, \cite{Wang2018a}). Another interesting approach in image reconstruction is compressive sensing. Compressive sensing paradigm is actively investigated in many areas (\cite{Zhu2013a, Lee2018ab, Fati2019, Vlasic2022, Ralai2020, Ralasic2018a}), but only a handful of papers deal with PET imaging and compressive sensing. For example, in \cite{Hanif2013} authors study the image reconstruction from a reduced set of measurements thus exporting a compressive sensing paradigm. Last, but not least, a parametric model shows a promising area of research for image reconstruction. The iterative approach combined with the Gaussian mixture model in PET imaging is described in \cite{Tafro2019a}.

The common property of all encountered iterative methods is that they depend on the model of the system: better the model - better the reconstruction. The model is typically represented by the system matrix: it maps the image space to the projection (or the measurement data) space. The system matrix can be decomposed into simpler matrices such that each sub-matrix models a different aspect of the actual measurement. One of the sub-matrices is geometric: it models mapping between pixels or voxels of the reconstructed image and each pair of scintillating detectors (e.g. two crystals).

Hence, an important challenge is how to obtain an accurate system model. Once it is available, it can be used for the high-quality PET image reconstruction. In the sequel, we describe some well-known approaches for the system matrix estimation, as well as the proposed method. There are roughly three different ways how to calculate and estimate the system matrix.

Point source measurements are the most accurate way to derive the system matrix. Unfortunately, this approach is impractical. In \cite{Panin2006}, the authors estimated that around 2.6 years are needed for acquisition to fully generate the system matrix. 

Monte Carlo (MC) simulations provide a fairly accurate model of the PET system. In contrast to the physical point source approach, MC simulations are based on virtual measurements conducted on a model of the system. There are several publicly available software solutions for PET simulation and reconstruction that are based on MC simulation such as PETSIM \cite{Thompson1992}, SIMSET: Simulation System for Emission Tomography \cite{Harrison}, GATE, the GEANT4 Application for Tomographic Emission \cite{Jan2004},\cite{Sarrut2021}, and many other. Still, there are several major drawbacks of the MC simulation. It is slow, and a large number of pseudo-events are required to reduce the statistical noise. Due to the long computation time, the system matrix is often computed in advance and pre-stored. Pay attention that in 3D PET scanners the system matrix can be very large, containing over than $10^{12}$ elements. This makes a problem either with storing or accessing the pre-computed model. To reduce the storage space, sparsity and symmetries of the system matrix can be exploited, as described in \cite{Herraiz2006},\cite{Yamaya2008},\cite{Iriarte2016}. 

Finally, the analytical approach allows the computation of the system matrix on-the-fly, since it is usually very fast. The downside of the analytical approach is that it often models only the geometric component of the system matrix, thus overlooking some major physical effects that happen during PET measurement (scatter, etc.).

PET is a non-invasive functional imaging modality based on electron-positron annihilation, which happens after injection and dispersion of the radioactive tracer. A measurement event in PET is a simultaneous detection (in a given time window) of a pair of crystals hit by high-energy photons (511\,keV) going in opposite directions. A virtual line that connects the two excited crystals is called the Line Of Response (LOR), since it represents all possible positions of the annihilation source point for the given event.
A simple geometry based on the intersection of the LOR-s is used in \cite{Siddon1985}, while tubes of response with voxels and blob basis functions are used in \cite{Lougovski2014} and \cite{Lougovski2015}. Bilinear and trilinear interpolation is exploited in \cite{Rahmim2005} and \cite{Hu2007}. A multi-ray method was investigated in \cite{Moehrs2008}. The highly investigated approach is based on a calculation of solid angles - an angle between a point source and a pair of crystals. In \cite{Qi1998}, the authors calculated the geometric projection matrix based only upon solid angles between the two crystals. Similarly, an algorithm for the derivation of the geometric projection matrix can be seen in \cite{Terstegge}. This approach was further extended in \cite{delaPrieta2006}, considering the effective detector polygon.

The goal of this paper is to provide a simple and fast, but very accurate 2D reconstruction for the PET scanners. Instead of modeling the entire system matrix, we generate the so-called white image. It describes the probability of detecting a point source by the entire PET scanner including the rotation of all crystals around the object that is being measured. In this work, we consider that the ring of crystals makes a $2\pi$ radians rotation, or its multiple. Notice that if the observed object was a radioactive source uniformly distributed over the measurement area, the result of the measurement would be such a white image. Our white image is generated using mathematical modeling and additionally checked by the MC approach, as well as by the real measurement.

Furthermore, the white image model is used for compensation in the MLEM algorithm. To obtain it, the scanner geometry must be known, but only a few parameters are needed. We denote the center of rotation as the origin of the 2D model. The parameters are radii from origin to each crystal, length of crystals, and distances from all possible lines that connect centers of two crystals to the origin. Our crystal-to-crystal response is not based on solid angles, as in \cite{Qi1998}. 

In this paper, we provide a detailed crystal-to-crystal response with an explicit mathematical expression written in the Cartesian coordinate system in its most general form. Some parts of the derived model are exact, while the others are given in the approximate, but very accurate form. 

We start from a crystal-to-crystal response, given as probability density function (PDF) of its annihilation point source. The support of the 2D PDF is of the rectangular shape. 
To calculate a white image, we need to rotate each possible response around the origin and integrate the result. We give a closed-form solution for the special case when the origin matches the center of symmetry of the PDF. Then, we introduce a very simple approximate model and prove its accuracy using the exact one. Finally, it is used to generate the final white image model. Rotating and integration of the approximate responses result in a closed form of the overall solution. To prove the validity of the result, an additional MC analysis was conducted. To demonstrate this technique, we incorporated a white image in the MLEM algorithm, originally proposed in \cite{Shepp1982}. The use of the proposed model outperforms the competitive methods in many aspects.

The paper is organized in the following way. A mathematical model of crystal-to-crystal response are given in Section 2. In Section 3 we discuss approximations of the crystal-to-crystal response. In Section 4 we introduce the white image and its role in the MLEM algorithm. Finally, we demonstrate and validate our approach in Section 5. We present results on simulated and real PET data. Section 6 concludes the paper.

\section{Crystal-to-crystal response and its rotation}

\begin{figure}[t]
\centering
\begin{subfigure}{0.485\textwidth}
  \centering
  \includegraphics[width=1\linewidth]{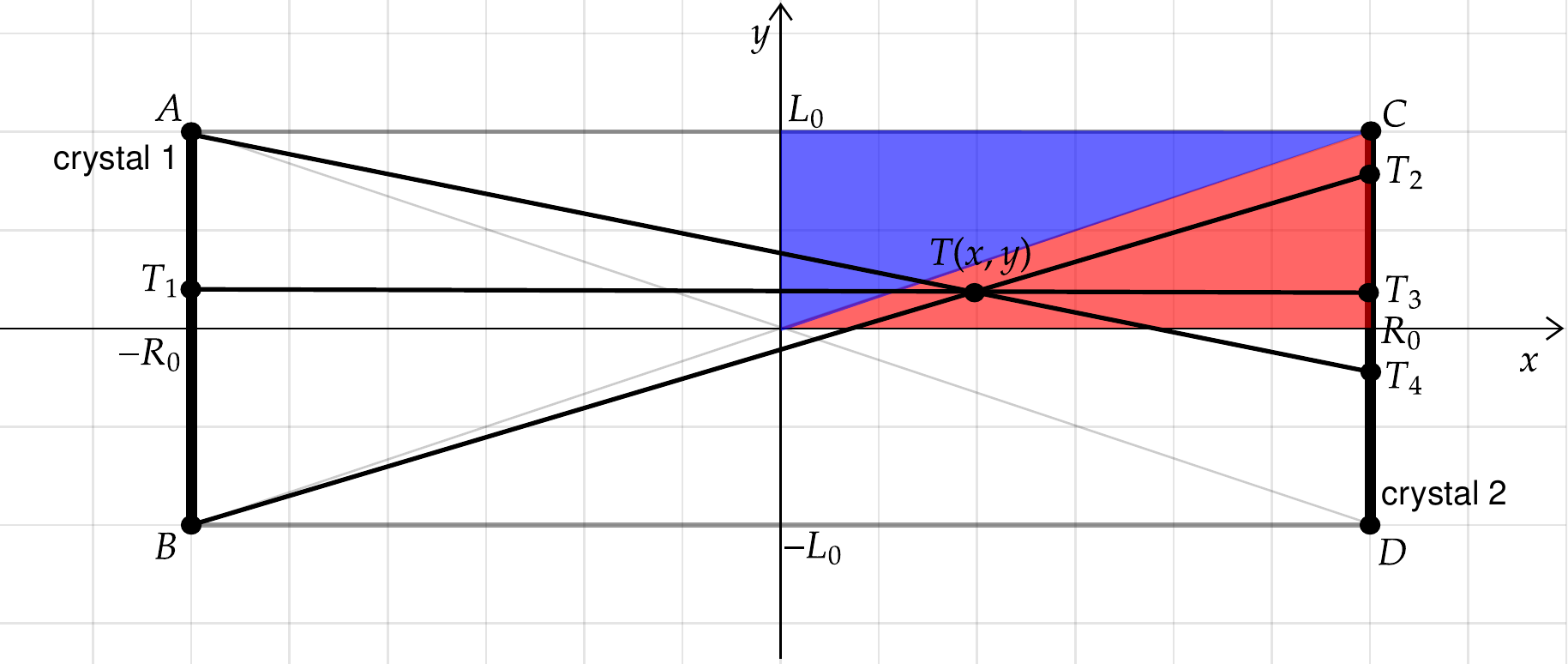}
  \caption{The first region}
  \label{fig:derive_1}
\end{subfigure}\hspace{1mm}
\begin{subfigure}{0.485\textwidth}
  \centering
  \includegraphics[width=1\linewidth]{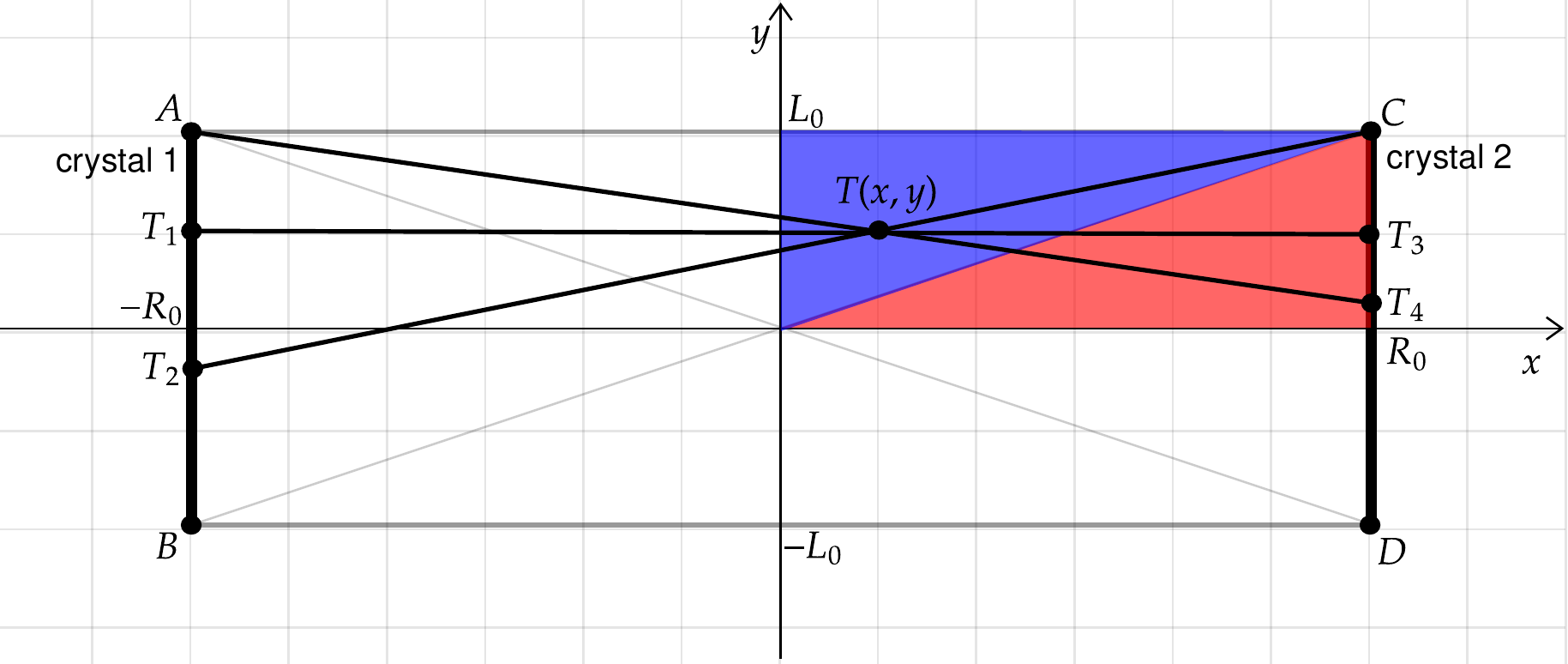}
  \caption{The second region}
  \label{fig:derive_2}
\end{subfigure}
\centering
\label{fig:derive_first}
\centering
\caption{Response between two crystals for two regions.}
\end{figure}

In this section, we calculate the probability of annihilation at some point $T$ if the event was detected by a given pair of crystals. The crystals are depicted in Fig. \ref{fig:derive_1} and Fig. \ref{fig:derive_2} as $AB$ and $CD$ lines. The 2D PDF support is considered rectangular. In practice, the crystals are usually not parallel, but the size of crystals ($AB$ and $CD$ length $= 2L_0$) is much smaller than its mutual distance ($AC$ and $BD$ length $= 2R_0$), and a simple projection is good enough to obtain an accurate model (details given in Section 4, Figure 9).

We calculate the probability for the first quadrant only, which is depicted in two colors. Due to geometric symmetries, we can easily derive explicit expressions for all four quadrants. 

Calculation is conducted for each region separately. Both triangular regions, as well as their geometric relations, are shown in Fig. \ref{fig:derive_1} and Fig. \ref{fig:derive_2}. Non-normalized probability density function $\hat{P}$ is: 

\begin{equation}
    \hat{P}(T)=\hat{P}(x,y)=\frac{L_{hit}(x,y)}{L_{tot}},
    \label{eq:prob_define}
\end{equation}
where $L_{tot}=4\cdot L_0$ is a cumulative length of both crystals and $L_{hit}$ is a portion of crystals' length that can be hit with a ray that originates at point $T(x,y)$.

For the first region, we have $L_{hit}(x,y) = 2\cdot L_0 + |T_2T_3| + |T_3T_4|$. $|T_mT_n|$ denotes length of $T_mT_n$ line. Notice two pairs of similar triangles: $ \Delta ATT_1 \sim \Delta T_4TT_3$ and $\Delta BTT_1\sim \Delta T_2TT_3$. Thus:
\begin{equation}
\begin{gathered}
    \frac{R_0+x}{L_0-y} = \frac{R_0-x}{|T_3T_4|},\;\;\;\;\;\frac{R_0+x}{L_0+y} = \frac{R_0-x}{|T_2T_3|}. 
\end{gathered}
\label{eq:derive_red_region_ratio}
\end{equation}
After expressing $|T_2T_3|$ and $|T_3T_4|$ from (\ref{eq:derive_red_region_ratio}) and inserting in  $L_{hit}(x,y) = 2\cdot L_0 + |T_2T_3| + |T_3T_4|$ and (\ref{eq:prob_define}) we get:
\begin{equation}
    \hat{P}(x,y) = \frac{R_0}{R_0+x}.
\label{eq:red_region_prob}
\end{equation}

Similarly, for the second region we can write $L_{hit}(x,y) = |AT_1| + |CT_3| + |T_1T_2| + |T_3T_4| = 2\cdot(L_0-y) + |T_1T_2| + |T_3T_4|$. After spotting that $\Delta ATT_1 \sim \Delta T_4TT_3$ and $\Delta T_2TT_1\sim \Delta CTT_3$, we can write:
\begin{equation}
\begin{gathered}
    \frac{R_0+x}{L_0-y} = \frac{R_0-x}{|T_3T_4|},\;\;\;\;\;\frac{R_0+x}{|T_1T_2|} = \frac{R_0-x}{L_0-y} 
\end{gathered}
.
\label{eq:derive_blue_region_ratio}
\end{equation}

Finally, non-normalized probability density function for the second region is:
\begin{equation}
    \hat{P}(x,y) = \frac{R_0^2}{R_0^2-x^2}\frac{L_0-y}{L_0}.
\label{eq:blue_region_prob}
\end{equation}
Outside of rectangle $ABCD$, the probability is equal to zero.

The expression:
\begin{equation}
    P(x,y) = \frac{1}{2R_0L_0}
    \begin{cases} 
        \frac{R_0}{R_0+|x|}, & (x,y) \in A\\
        \frac{R_0^2}{R_0^2-x^2}\frac{L_0-|y|}{L_0}, & (x,y) \in B\\
        0, & elsewhere
   \end{cases}
\label{eq:prob_two_crystals}
\end{equation}

is the the result for all quadrants (notice the absolute values). We denoted regions as $A = \big\{ (x,y) \in \mathbb{R}^2 \;\; \big| \;\;  |y| < \frac{L_0}{R_0}|x|, |y| \leq L_0, |x| \leq R_0 \big\}$ and $B = \big\{(x,y) \in \mathbb{R}^2 \;\; \big| \;\;  |y| \geq \frac{L_0}{R_0}|x|, |y| \leq L_0, |x| \leq R_0\big\}$.  Factor $\frac{1}{2R_0L_0}$ is the normalization constant of the probability density function that ensures that the integral is $1$.

\begin{figure}
\centering
\includegraphics[width=0.5\linewidth]{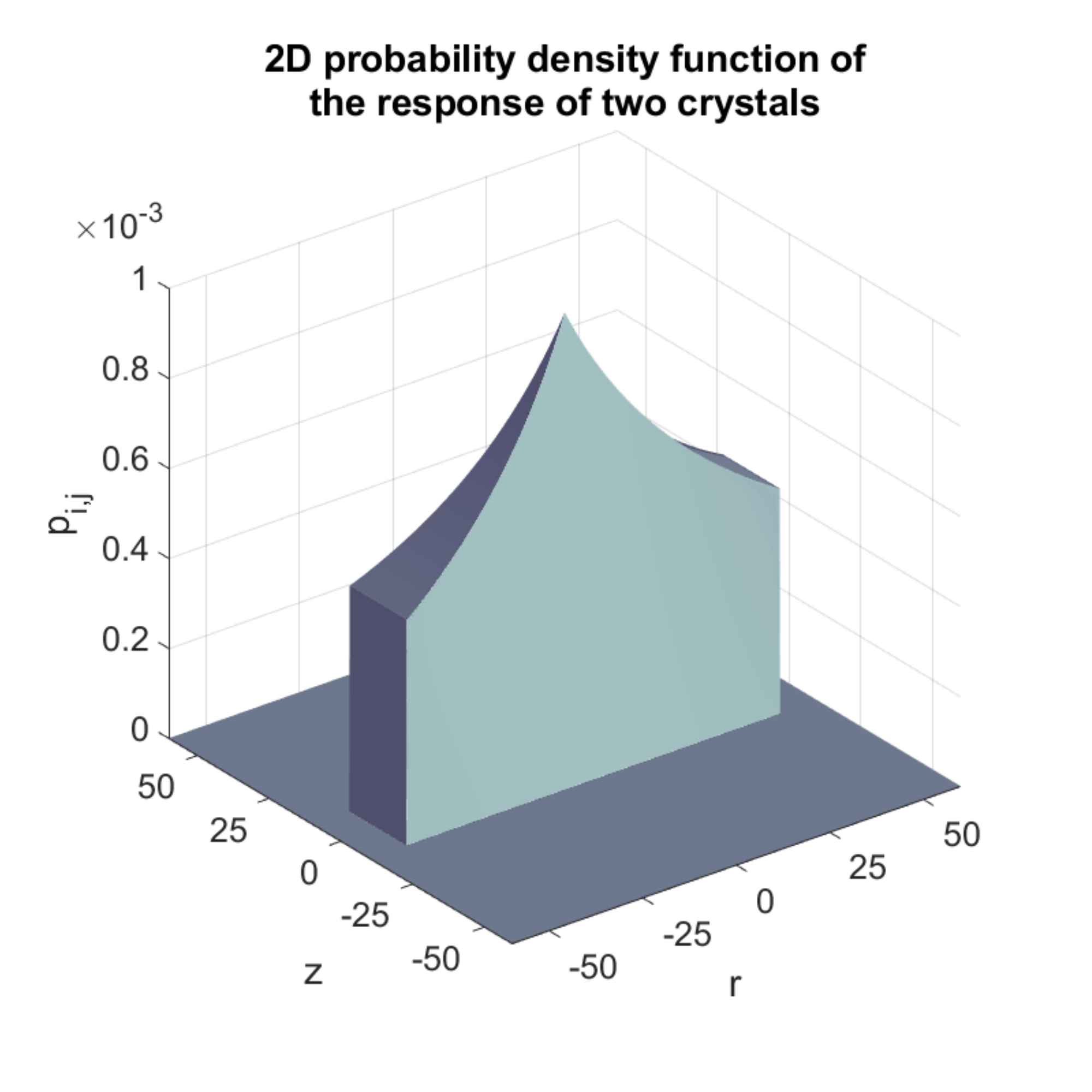}
\caption{The tent-like probability density function of the crystal-to-crystal response ($R_0=50$, $L_0=10$).}
\label{fig:PDF2Dcenter}
\end{figure}

The PDF given by expression (\ref{eq:prob_two_crystals}) is interpreted as the probability that annihilation occurred at the point $T(x,y)$, if detected by a given pair of crystals.
An example of the entire PDF with $R_0=50$ and $L_0=10$ can be seen in Fig. \ref{fig:PDF2Dcenter}.
Notice that the center of the symmetry of the obtained PDF model is set to coordinate $T(0,0)$, which corresponds to the rotation origin. Measurement events in which exactly the opposite crystals on the ring were involved fit that model. In general, there is a displacement between the rotation origin and the PDF center. To model that, a shifted probability density function in $y$ direction by amount $h$ is given by
\begin{equation}
    P_{shift}(x,y;h) = \frac{1}{2R_0L_0}
    \begin{cases} 
        \frac{R_0}{R_0+|x|}, & |y-h| < \frac{L_0}{R_0}|x|, |y-h| \leq L_0, |x| \leq R_0\\
        \frac{R_0^2}{R_0^2-x^2}\frac{L_0-|y-h|}{L_0}, & |y-h| \geq \frac{L_0}{R_0}|x|, |y-h| \leq L_0, |x| \leq R_0\\
        0, & elsewhere
   \end{cases}
\label{eq:prob_two_crystals_shifted}.
\end{equation}
\subsection{Closed-form solution of the crystal-to-crystal response}
During the measurement, the set of crystals rotates around the origin. In our work, we consider that the rotation is $2\pi$ radians, or its multiple. Hence, to calculate the white image we need to integrate PDF-s of all possible crystal pairs while rotating full circle $[0, 2\pi]$. 

In the beginning, we focus our attention on the rotation of the unshifted crystal-to-crystal responses ($h=0$) around the origin. In that purpose, we express $P(x,y)$ in polar coordinates
\begin{equation}
    P_{p}(r,\phi) = \frac{1}{2R_0L_0}
    \begin{cases} 
        \frac{R_0}{R_0+r|\cos(\phi)|}, & (r,\phi) \in A_p \\ 
        \frac{R_0^2}{R_0^2-r^2\cos^2(\phi)}\frac{L_0-r|\sin(\phi)|}{L_0}, & (r,\phi) \in B_p \\ 
        0, & elsewhere
   \end{cases},
\label{eq:prob_two_crystals_polar}
\end{equation}
where $A_p = \Big\{(r,\phi) \in \left[0,+\infty\right>\times \mathbb{R} \;\; \Big| \;\; |\sin(\phi)| < \frac{L_0}{R_0}|\cos(\phi)|,\; r|\sin(\phi)| \leq L_0,\; r|\cos(\phi)| \leq R_0 \Big\}$ and $B_p =\big\{ (r,\phi) \in \left[0,+\infty\right>\times \mathbb{R} \;\; \Big| \;\; |\sin(\phi)| \geq \frac{L_0}{R_0}|\cos(\phi)|,\; r|\sin(\phi)| \leq L_0,\; r|\cos(\phi)| \leq R_0 \big\}$.
Since we have four quadrants symmetry $P(x,y)=P(x,-y)=P(-x,y)=P(-x,-y)$, we can calculate rotation in interval $[0, \frac{\pi}{2}]$ and multiply the result by four:
\begin{equation}
\begin{gathered}
    P_{rot}(r) = \frac{1}{2\pi} \int_{0}^{2\pi} P_p(r,\phi) d\phi = (\textrm{four quadrant symmetry}) = \frac{2}{\pi} \int_{0}^{\frac{\pi}{2}} P_p(r,\phi) d\phi.
    \end{gathered}
\label{eq:rotated}
\end{equation}
The arguments in the last equality belong to the first quadrant solely. Therefore, we can get rid of absolute values in (\ref{eq:prob_two_crystals_polar}) and simplify the integration.

To calculate (\ref{eq:rotated}), we use
\begin{equation}
\begin{gathered}
    \int \frac{1-\frac{r}{L0} \sin(\phi)}{1-\frac{r^2}{R_0^2}\cos^2(\phi)} d\phi =\\
    \begin{cases} 
        \frac{R_0}{2L_0} \log( \frac{t^2 + \frac{R_0+r}{R_0-r} }{t^2 + \frac{R_0-r}{R_0+r} }) + f(r) \Big( \arctan(\sqrt{\frac{R_0+r}{R_0-r}}\cdot t) + \arctan(\sqrt{\frac{R_0-r}{R_0+r}}\cdot t) \Big)  + C, & 0<r\leq R_0\\
        \frac{R_0}{2L_0} \log( \frac{t^2 - \frac{R_0+r}{R_0-r} }{t^2 - \frac{R_0-r}{R_0+r} }) + \frac{f(r)}{2} \log \left| \frac{t^2 - \frac{2R_0}{\sqrt{r^2-R_0^2}}\cdot t -1 }{t^2 + \frac{2R_0}{\sqrt{r^2-R_0^2}}\cdot t -1 } \right| + C, & r>R_0
   \end{cases} \\
    \int \frac{1}{1+k\cos(\phi)} d\phi =
    \begin{cases} 
        \frac{2}{\sqrt{1-k^2}} \arctan( \sqrt{ \frac{1-k}{1+k} } \cdot t) ) + C, & 0<k\leq1\\
        \frac{1}{ \sqrt{k^2-1} } \log \left| \frac{t + \sqrt{ \frac{1+k}{1-k} }}{t - \sqrt{ \frac{1+k}{1-k} }}\right| + C, & k>1
   \end{cases}
\label{eq:integral2}
\end{gathered},
\notag
\end{equation}
where $t = \tan(\frac{\phi}{2})$ and $f(r) = \frac{R_0}{\sqrt{|R_0^2-r^2|}}$. Furthermore, we assume that $L_0<R_0$, i.e. length of the crystal is smaller than the PET scanner radius. After simplification (details can be found in Appendix A), we get the solution
\begin{equation}
    \begin{gathered}
    P_{rot}(r;R_0,L_0) = \frac{1}{R_0 L_0 \pi}\times\\
    \begin{cases} 
        2f(r) \arctan(\sqrt{\frac{R_0-r}{R_0+r}} \frac{C_2}{L_0}) +&\\ \frac{C_1}{2}\log(\frac{L_0^2-C_2(R_0+r) }{ L_0^2-C_2(R_0-r)}) + f(r)(\frac{\pi}{2} - \arctan(\frac{L_0}{\sqrt{R_0^2-r^2}})), & 0 \leq r \leq L_0\\
         \frac{C_1}{2}\log( \frac{R_0+\sqrt{r^2-L_0^2}}{R_0-\sqrt{r^2-L_0^2}} \cdot \frac{D-r}{D+r} ) +  &\\
         2f(r) \arctan(\frac{\sqrt{R_0^2-r^2}}{L_0} \frac{L_0^2+C_2(\sqrt{r^2-L_0^2}+r)}{(\sqrt{r^2-L_0^2}+r)(R_0+r)-C_2(R_0-r)}),   & L_0 < r \leq R_0\\
        f(r) \log(\frac{R_0}{r} \frac{ C_2 +L_0 \sqrt{\frac{r+R_0}{r-R_0}} }{ C_2 -L_0 \sqrt{\frac{r+R_0}{r-R_0}} } ) + \frac{C_1}{2} \log( \frac{R_0+\sqrt{r^2-L_0^2}}{R_0-\sqrt{r^2-L_0^2}} \cdot \frac{D-r}{D+r})+ &\\ \frac{f(r)}{2} g(r),& R_0 < r \leq \sqrt{R_0^2+L_0^2} \\
        0 ,& r > \sqrt{R_0^2+L_0^2}
   \end{cases} \\
   \end{gathered}
\label{eq:expre_around_center}
\end{equation}
where $C_1 = \frac{R_0}{L_0}$, $C_2 = \sqrt{L_0^2+R_0^2}-R_0 = D - R_0$, and
\begin{equation}
\begin{gathered}
g(r) = \scriptstyle{ \log \bigg| \frac{ \sqrt{r^2-R_0^2} + L_0 }{ \sqrt{r^2-R_0^2} - L_0 }}{\frac{  L_0^3R_0 - 2R_0L_0(r+\sqrt{r^2-L_0^2}) +2r^2(r+\sqrt{r^2-L_0^2})\sqrt{r^2-R_0^2} - L_0^2(2r-\sqrt{r^2-L_0^2})\sqrt{r^2-R_0^2}  }{    -L_0^3R_0 + 2R_0L_0(r+\sqrt{r^2-L_0^2}) +2r^2(r+\sqrt{r^2-L_0^2})\sqrt{r^2-R_0^2} - L_0^2(2r+\sqrt{r^2-L_0^2})\sqrt{r^2-R_0^2} }\bigg|}.
\end{gathered}
\notag
\end{equation}

Obtained equation (\ref{eq:expre_around_center}) describes the probability density function of a point source while the measurement system is being rotated a full circle around the origin and the pair of crystals that detected the event were located on the opposite sides of the ring. Even in this very special case (the center of symmetry matches the center of rotation), the final expression is pretty complicated. It motivates us to carefully introduce an approximate model while retaining the accuracy of the exact one.

\section{Approximate solution of the crystal-to-crystal response}
In this Section we investigate approximate solutions for (\ref{eq:expre_around_center}). Once we have an accurate result, we generalize the expression for any shift $h$. 

\subsection{Rotation of Dirac line}

\begin{figure}[H]
\centering
\includegraphics[scale=0.8]{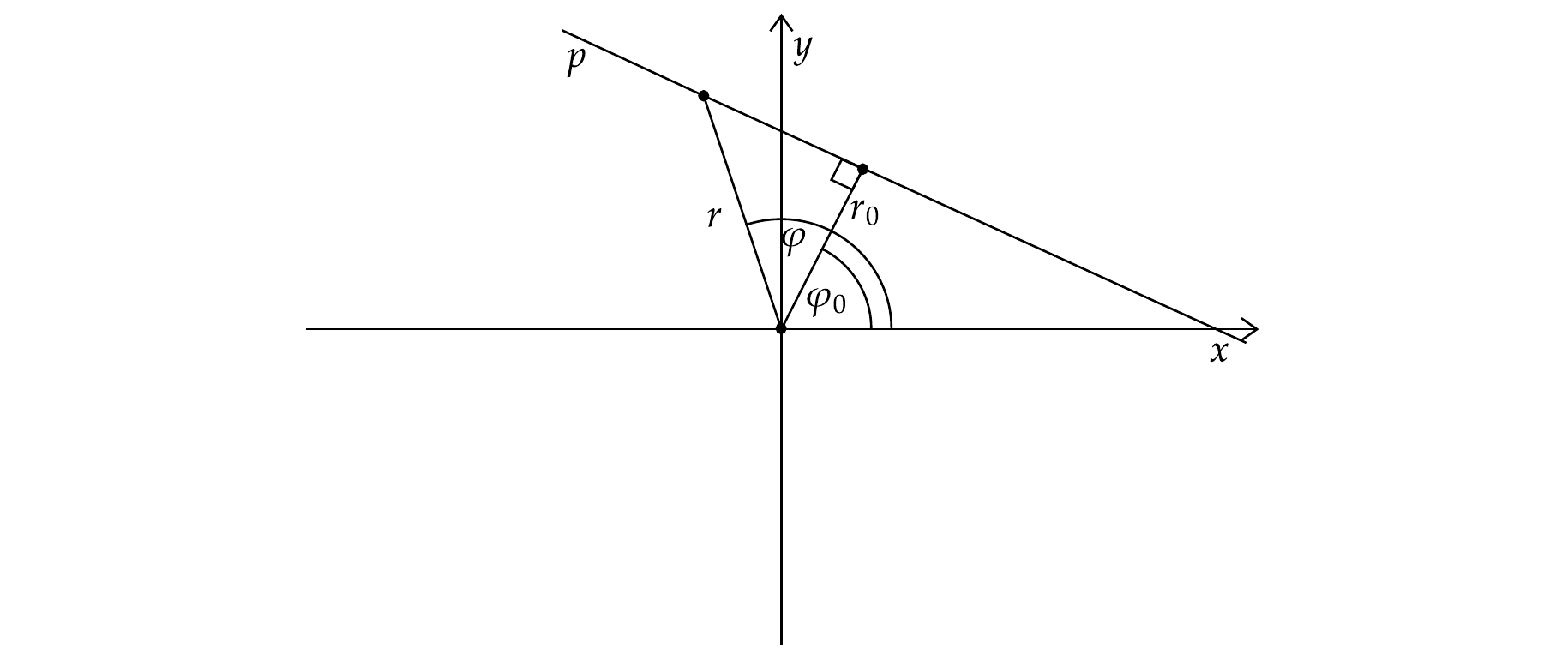}
\caption{Dirac line support}
\label{fig:dirac_line}
\end{figure}

The simplest way to approximate the crystal-to-crystal response is to connect centers of each crystal by a Dirac line. Dirac line can be described as $f_c(x,y;r_0,\varphi_0)= \delta(x\cos(\varphi_0)+y\sin(\varphi_0)-r_0)$ or equivalently in polar coordinates as $f(r,\varphi;r_0,\varphi_0) = \delta(r \cos(\varphi-\varphi_0) - r_0)$, where $r_0$ is the distance between origin and the Dirac line and $\phi_0$ is an angle between the line and x-axis, as shown in Fig. \ref{fig:dirac_line}.

During the measurement, a pair of crystals, as well as our Dirac line is rotating around the origin. Hence, we need to integrate $ \frac{1}{2\pi} \int_0^{2\pi} \delta(r\cos(\varphi - \varphi_0)-r_0) d\varphi_0$. 

To calculate that integral, we apply the well known property $ \int_S \delta(g(x)) dx = \int_S \sum_i \frac{\delta(x-x_i)}{|g'(x_i)|} dx$, where $x_i$ are solutions of equation $g(x)=0$ on some set $S\subseteq \mathbb{R}$. 

In our case,  $S=[0,2\pi]$, $g(\phi_0) = r\,cos(\phi-\phi_0) - r_0$, $|g'(\phi_0)| = r|\sin(\phi-\phi_0)|,\,r\ge0$. Roots of $g$ are $\phi_{0_{1,2}} = \phi \pm \arccos(r/r_0)$. Thus the denominator is $r|sin(\mp\arccos(r_0/r)|=\sqrt{r^2-r_0^2}$. It follows that $\sum_i = 1/\sqrt{r^2-r_0^2}\,\{\delta(\phi_0-\phi_{0_{1}})+\delta(\phi_0-\phi_{0_{2}})\}$. 

After the integration, we get:
\begin{equation}
\hat{P}(r) = \frac{1}{\pi}\frac{1}{\sqrt{r^2-r^2_0}}
\label{eq:dirac_line_sol1},
\end{equation}
since the integral of our two Diracs on interval $[0, 2\pi]$ is $\{1+1\}$. We must take into the account the fact that $P(r)=0$ for $r<r_0$, thus we rewrite (\ref{eq:dirac_line_sol1}) as
\begin{equation}
P(r) = \frac{1}{\pi}\frac{1}{\sqrt{r^2-r^2_0}}\mu(r-r_0),
\label{eq:dirac_line_sol}
\end{equation}
where $\mu(r)$ is Heaviside step function.

\subsection{An approximation of the response between two crystals}

As mentioned before, expression (\ref{eq:expre_around_center}) is pretty long, and therefore impractical for applications. Moreover, it gets even more complicated form in a general, shifted case ($h\neq0$). Hence, we develop an approximate solution with negligible error. 

First, we define a helper function $J : I \rightarrow \mathbb{R}$, $I = \bigg\{ (r,l)\in\mathbb{R}^2 \;\; \Big| \;\; r\geq0,\;l\geq0\;\bigg\} $;
\begin{equation}
J(r,l) = \frac{1}{\pi}\frac{1}{\sqrt{r^2-l^2}}\mu(r-l)
\label{eq:jezgra}.
\end{equation}
$J(r,l)$ is the result of integration of full-circle rotation of the Dirac-line whose distance from the origin (i.e. center of rotation) is $l$.

\begin{figure}[H]
\includegraphics[width=0.5\linewidth]{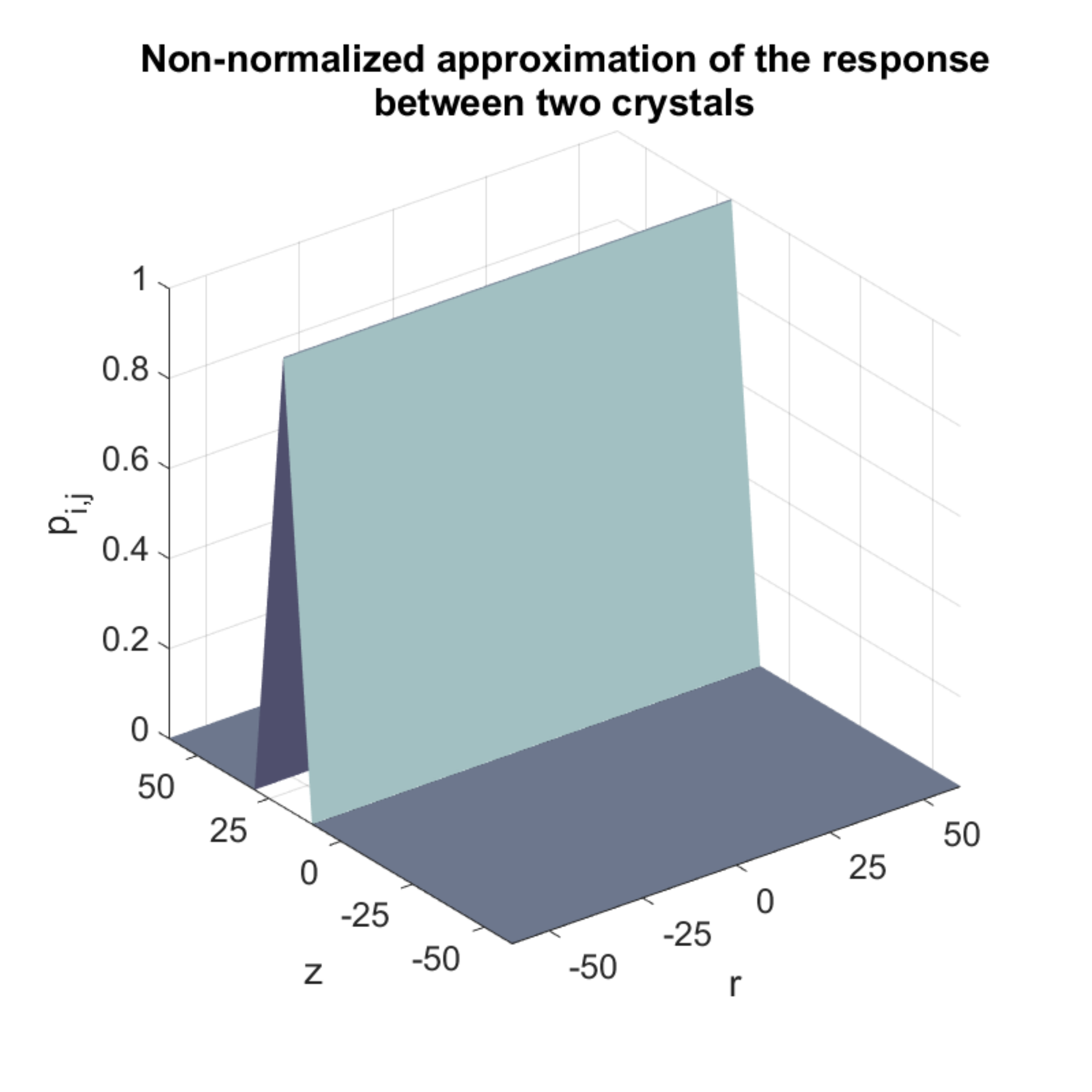}
\centering
\caption{3D visualisation of infinite length triangular crystal-to-crystal response mode, for $h = 20, L_0 = 10$.}
\label{fig:aproxfunction3dplot}
\end{figure}

\begin{figure}[H]
\centering
\begin{subfigure}{0.48\textwidth}
  \centering
  \includegraphics[width=\linewidth]{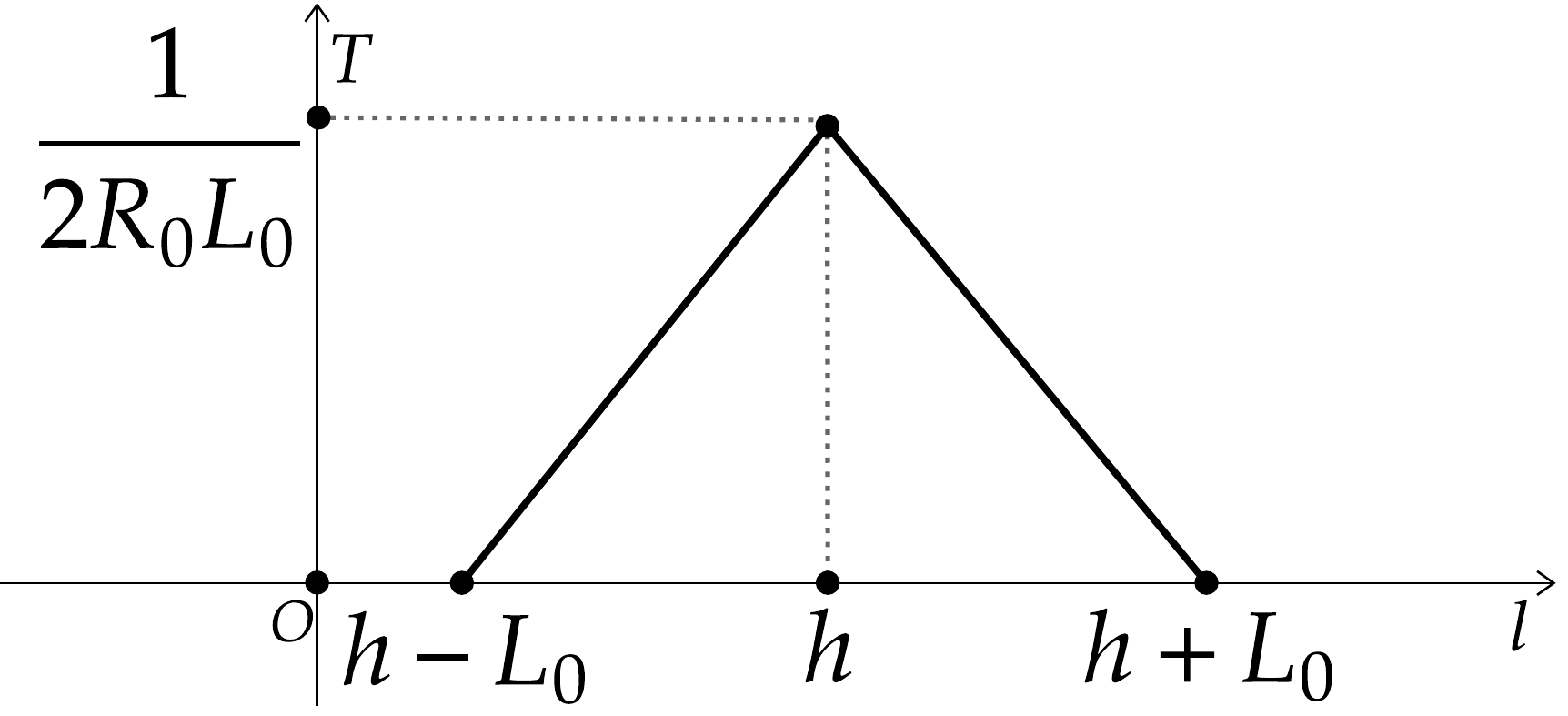}
  \caption{$h\geq L_0$}
  \label{fig:trokutvelikih}
\end{subfigure}
\begin{subfigure}{0.48\textwidth}
  \centering
  \includegraphics[width=\linewidth]{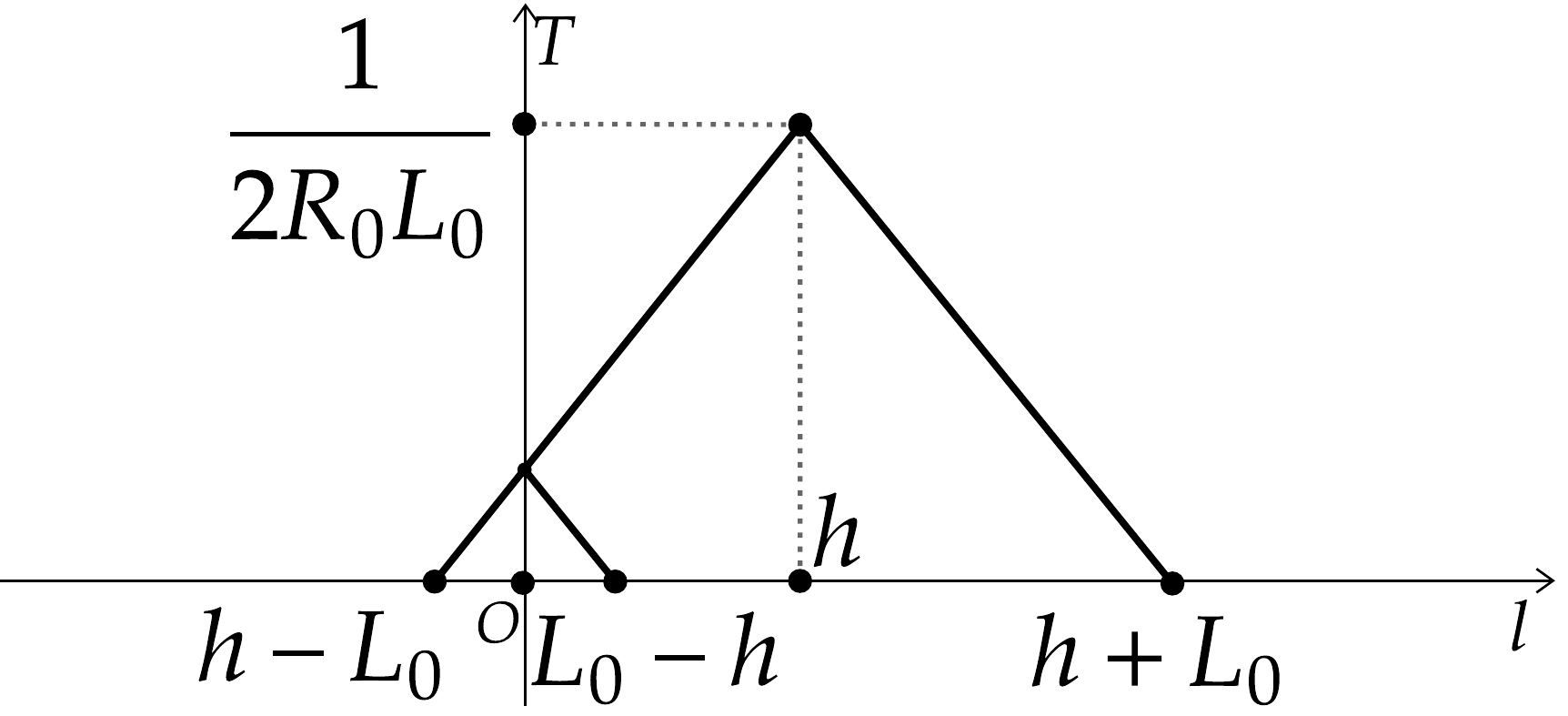}
  \caption{$h<L_0$}
  \label{fig:trokumalih}
\end{subfigure}
\centering
\captionof{figure}{Triangle function - two cases}
\label{fig:casestriangle}
\end{figure}

The idea is to find a proper weight function $T$ such that the expression
\begin{equation}
\displaystyle \int_{0}^{+\infty} T(l)J(r,l)dl
\label{eq:hope}
\end{equation}
is a good approximation of (\ref{eq:expre_around_center}). A reasonable choice for $T(l)$ is a 2D triangular function, as shown in Fig \ref{fig:aproxfunction3dplot}. The $l$ parameter enables introduction of shifts.
Explicit expression for T(l) is
\begin{equation}
T(l) = \frac{1}{2L_0R_0}
\begin{cases}
1 - \frac{|l-h|}{L_0} &, h-L_0 \leq l \leq h+L_0 \\
0 &, elsewhere
\end{cases}.
\label{eq:tezine}
\end{equation}
Calculation of (\ref{eq:hope}) is separated in two cases (Fig. \ref{fig:casestriangle}).

First case is for values $h \geq L_0$ (Fig. \ref{fig:trokutvelikih}). The calculation is straight-forward:

\begin{equation}
\centering
\begin{gathered}
P_a(r;h,R_0,L_0) = \displaystyle \int_{0}^{+\infty} T(l)J(r,l)dl =\\  \frac{1}{2L_0R_0} \bigg(  \displaystyle \int_{h-L_0}^{h} \frac{L_0+l-h}{L_0}J(r,l)dl + \displaystyle \int_{h}^{h+L_0} \frac{L_0-l+h}{L_0}J(r,l)dl \bigg)= \\ \frac{1}{2L_0R_0} Re\{ \frac{1}{\pi L_0} \Big( \left( L_0+h \right) \arcsin\left(\frac{L_0+h}{r}\right)  -2\cdot h\cdot \arcsin\left(\frac{h}{r}\right) +\\ \left( L_0-h \right) \arcsin\left(\frac{L_0-h}{r}\right) + \sqrt{r^2-(L_0+h)^2} - 2 \sqrt{r^2-h^2} + \sqrt{r^2-(L_0-h)^2}\Big)\}
\end{gathered}
\label{eq:proracunt1}
\end{equation}
We assume complex functions. The real part of the inverse sine function is $Re\{ arcsin(x) \} = \begin{cases} -\frac{\pi}{2}&, x<-1\\ arcsin(x)&, \abs{x}\leq 1\\ \frac{\pi}{2}&,x>1\end{cases} $, and the real part of the square root is $Re\{ \sqrt{x} \} = \begin{cases} \sqrt{x}&, x\geq 0\\ 0&,x<0\end{cases} $. 

In somewhat similar fashion, we can calculate the solution for the second case, i.e. when $h<L_0$. The only difference is that we must mirror the negative part of $T(l)$  around the x axis as shown in Fig. \ref{fig:trokumalih}. This is due the definition of $J(r,l)$, since it is defined on $[0,+\infty>^2$. 
\begin{equation}
\begin{gathered}
P_a(r;h,R_0,L_0) = \displaystyle \int_{0}^{+\infty} T(l)J(r,l)dl = \frac{1}{2L_0R_0} \bigg( \displaystyle \int_{0}^{h} \frac{L_0+l-h}{L_0}J(r,l)dl +\\ \displaystyle \int_{0}^{L_0-h} \frac{L_0-l-h}{L_0}J(r,l)dl +  \displaystyle \int_{h}^{h+L_0} \frac{L_0-l+h}{L_0}J(r,l)dl \bigg)= \\  \frac{1}{2L_0R_0} Re\{ \frac{1}{\pi L_0} \Big( \left( L_0+h \right) \arcsin\left(\frac{L_0+h}{r}\right)  -2\cdot h\cdot \arcsin\left(\frac{h}{r}\right) +\\ \left( L_0-h \right) \arcsin\left(\frac{L_0-h}{r}\right) + \sqrt{r^2-(L_0+h)^2} - 2 \sqrt{r^2-h^2} + \sqrt{r^2-(L_0-h)^2}\Big)\}
\end{gathered}
\label{eq:proracunt2}
\end{equation}
Obtained formulae in (\ref{eq:proracunt1}) and  (\ref{eq:proracunt2}) are equal! Therefore, we have only one expression $P_a(r;h,R_0,L_0)$ for any given $h$ (for any given shift in $y$ direction, namely, for any distance from the origin) over the entire plain.
Validation of the approximate solution for the unshifted case ($h=0$) is given in Appendix C.

\subsection{Additional simplification with constant response between two crystals}

\begin{figure}[H]
\centering
\begin{subfigure}{0.48\textwidth}
  \centering
  \includegraphics[width=\linewidth]{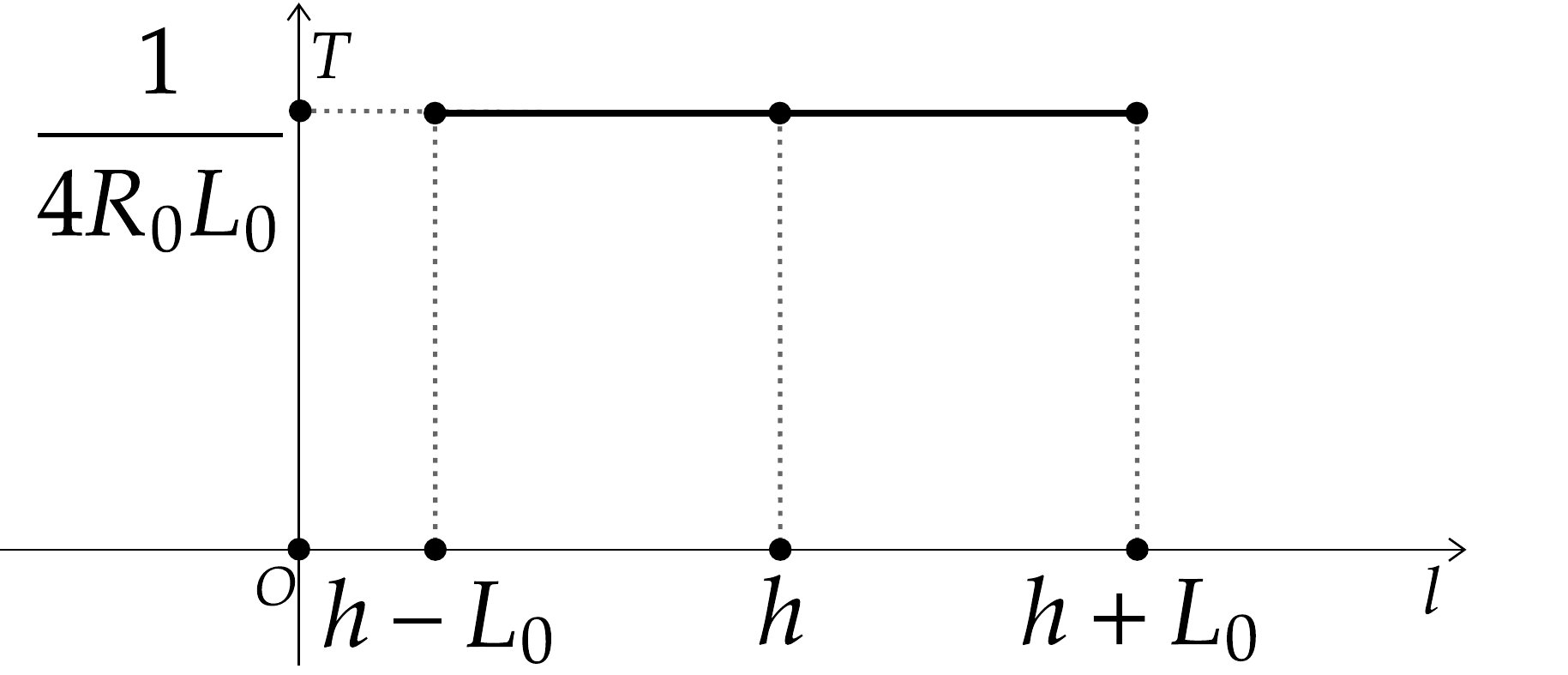}
  \caption{$h\geq L_0$}
  \label{fig:konstanta1}
\end{subfigure}
\begin{subfigure}{0.48\textwidth}
  \centering
  \includegraphics[width=\linewidth]{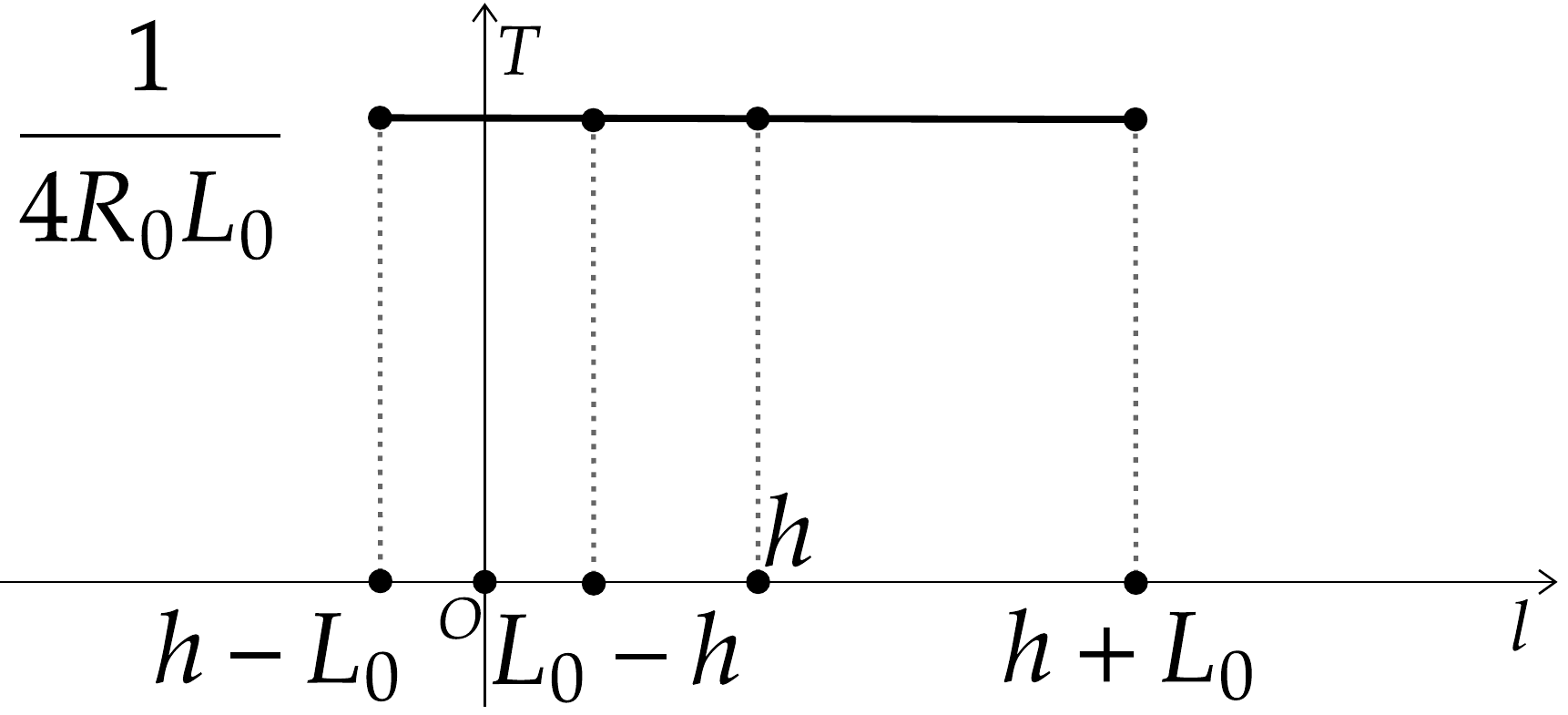}
  \caption{$h<L_0$}
  \label{fig:konstanta2}
\end{subfigure}
\centering
\caption{Constant function - two cases}
\label{fig:casesconstant}
\end{figure}

Another, simpler approximation, can be done if we use a rectangular window as weight function:
\begin{equation}
T(l) = \frac{1}{4L_0R_0}
\begin{cases}
1, & h-L_0 \leq l \leq h+L_0 \\
0, & elsewhere
\end{cases}.
\label{eq:tezine2}
\end{equation}
The calculation of $P_a(r;h,R_0,L_0)$ is done in the same way as in the triangle case (based on Fig. \ref{fig:casesconstant}). The result is used for additional control, and can be used for future generalization to more dimensions. The solution is:
\begin{equation}
\begin{gathered}
P_{a_2}(r;h,R_0,L_0) = \frac{1}{ 4 L_0 R_0 \pi} Re\bigg\{ \arcsin(\frac{h+L_0}{r}) - \arcsin(\frac{h-L_0}{r})\bigg\}
\end{gathered}
\label{eq:ravankrov}.
\end{equation}

\subsection{Comparison of approximations}

\begin{figure}[H]
\centering
\begin{subfigure}{0.32\textwidth}
\includegraphics[width=\linewidth]{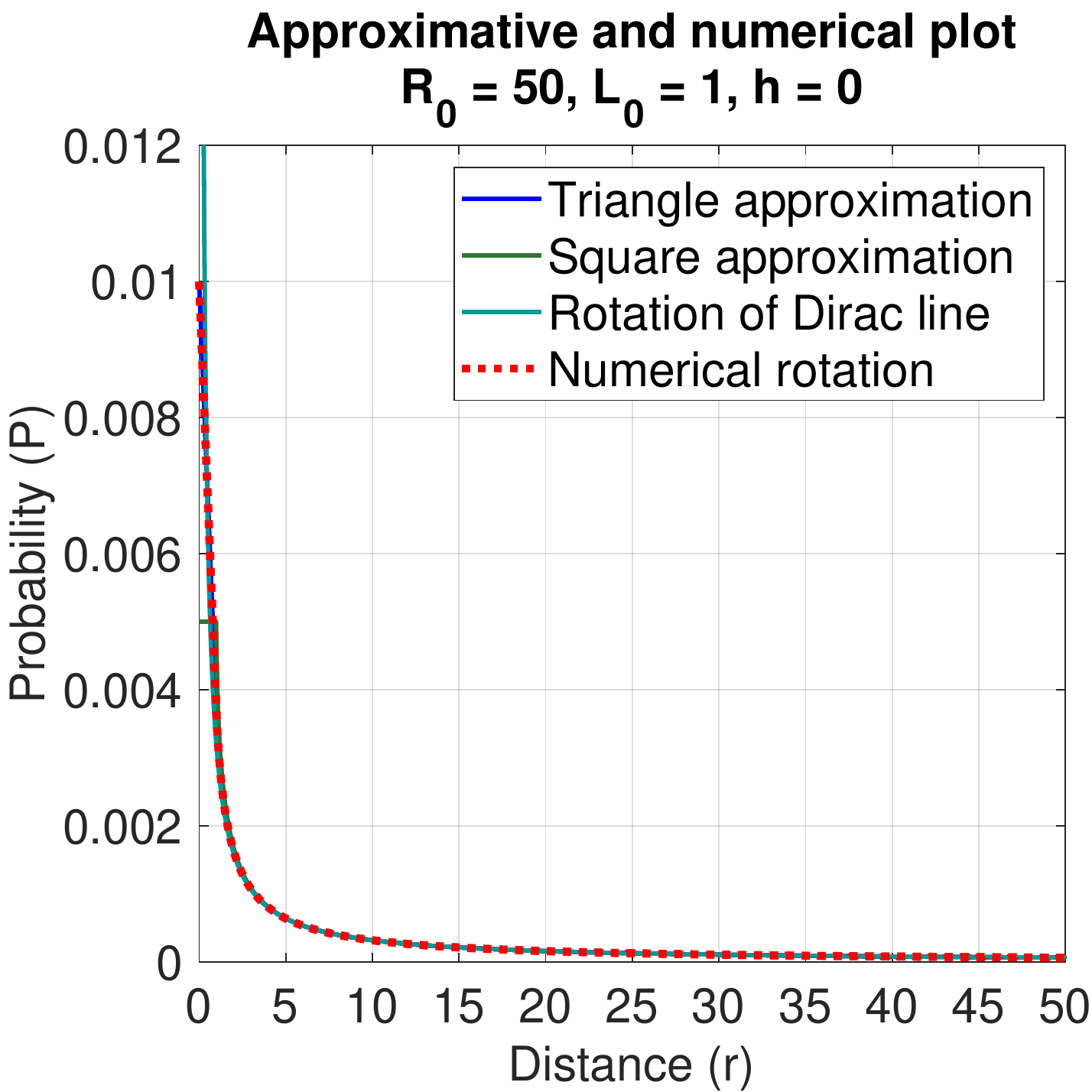}
\caption{$R_0=50$, $L_0=1$, $h=0$}
\end{subfigure}
\begin{subfigure}{0.32\textwidth}
\includegraphics[width=\linewidth]{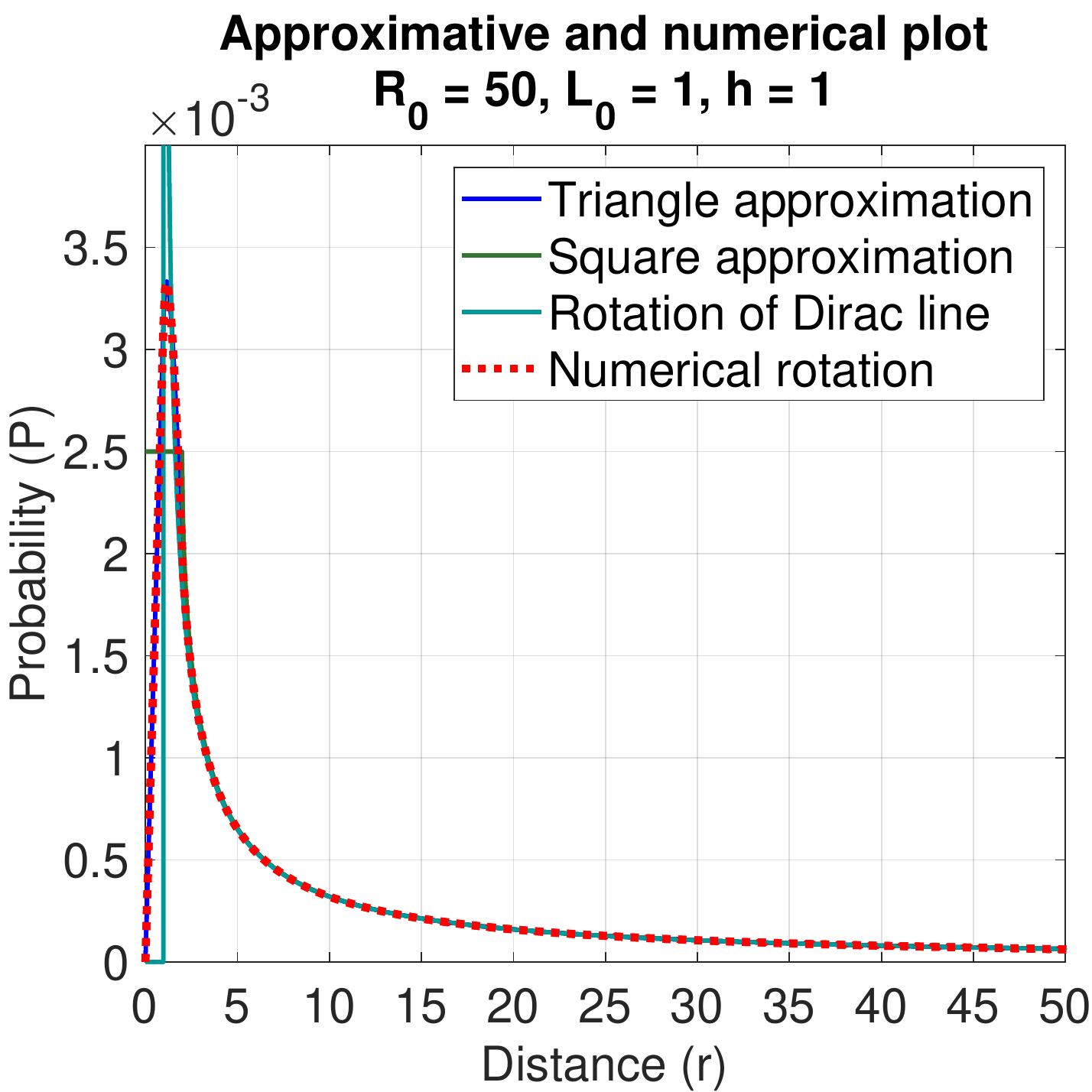}
\caption{$R_0=50$, $L_0=1$, $h=1$}
\end{subfigure}
\begin{subfigure}{0.32\textwidth}
\includegraphics[width=\linewidth]{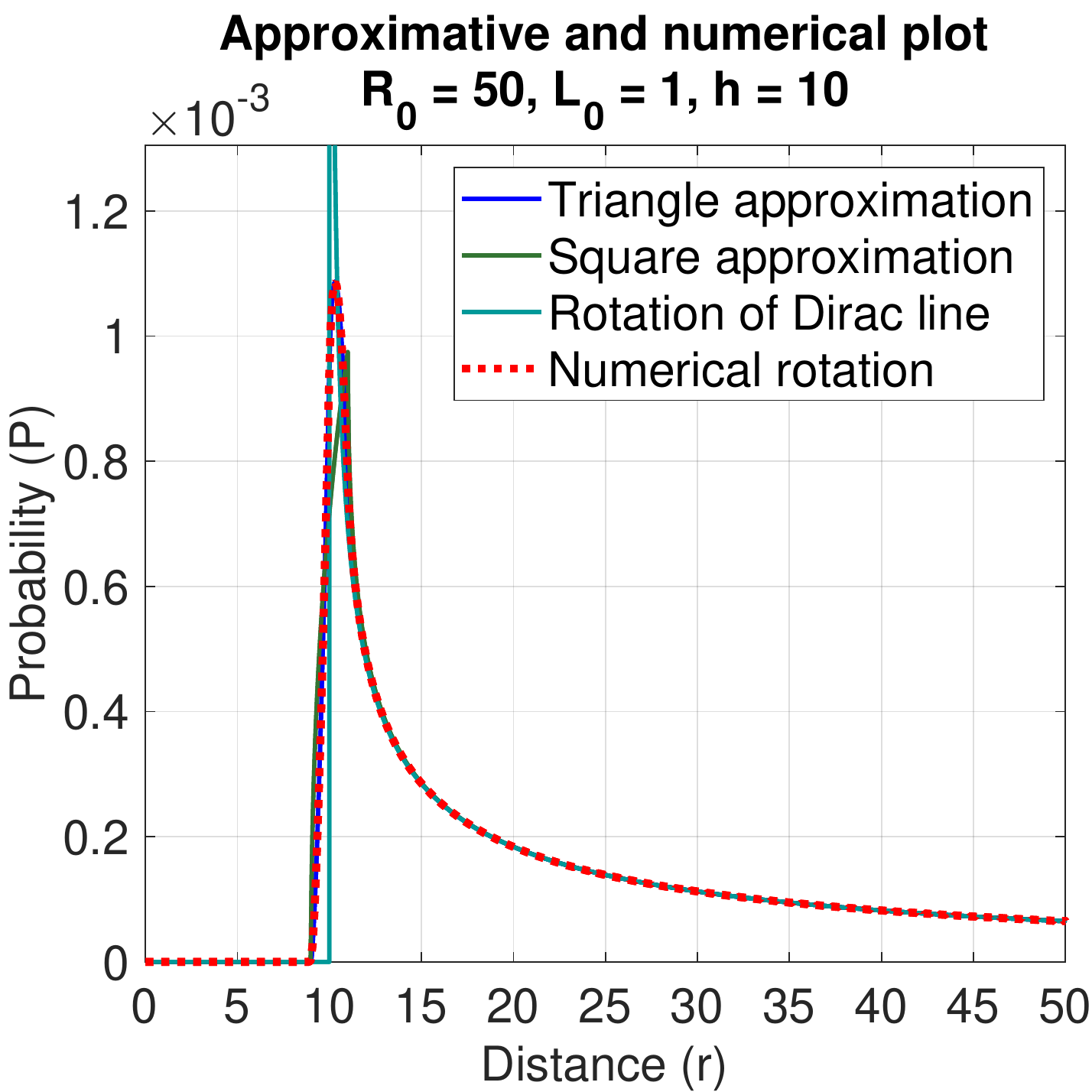}
\caption{$R_0=50$, $L_0=1$, $h=10$}
\end{subfigure}
\medskip
\centering
\begin{subfigure}{0.32\textwidth}
\includegraphics[width=\linewidth]{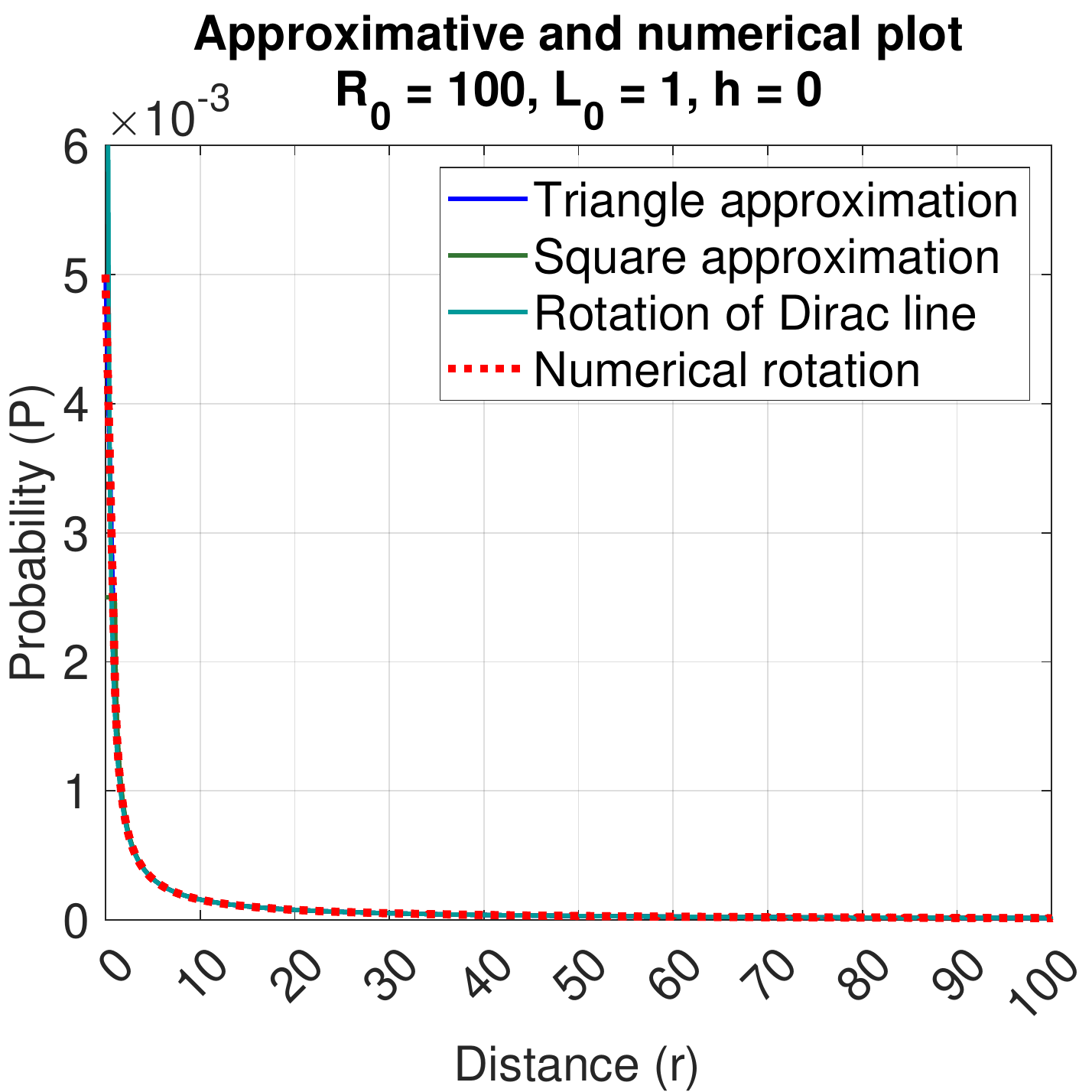}
\caption{$R_0=100$, $L_0=1$, $h=0$}
\end{subfigure}
\begin{subfigure}{0.32\textwidth}
\includegraphics[width=\linewidth]{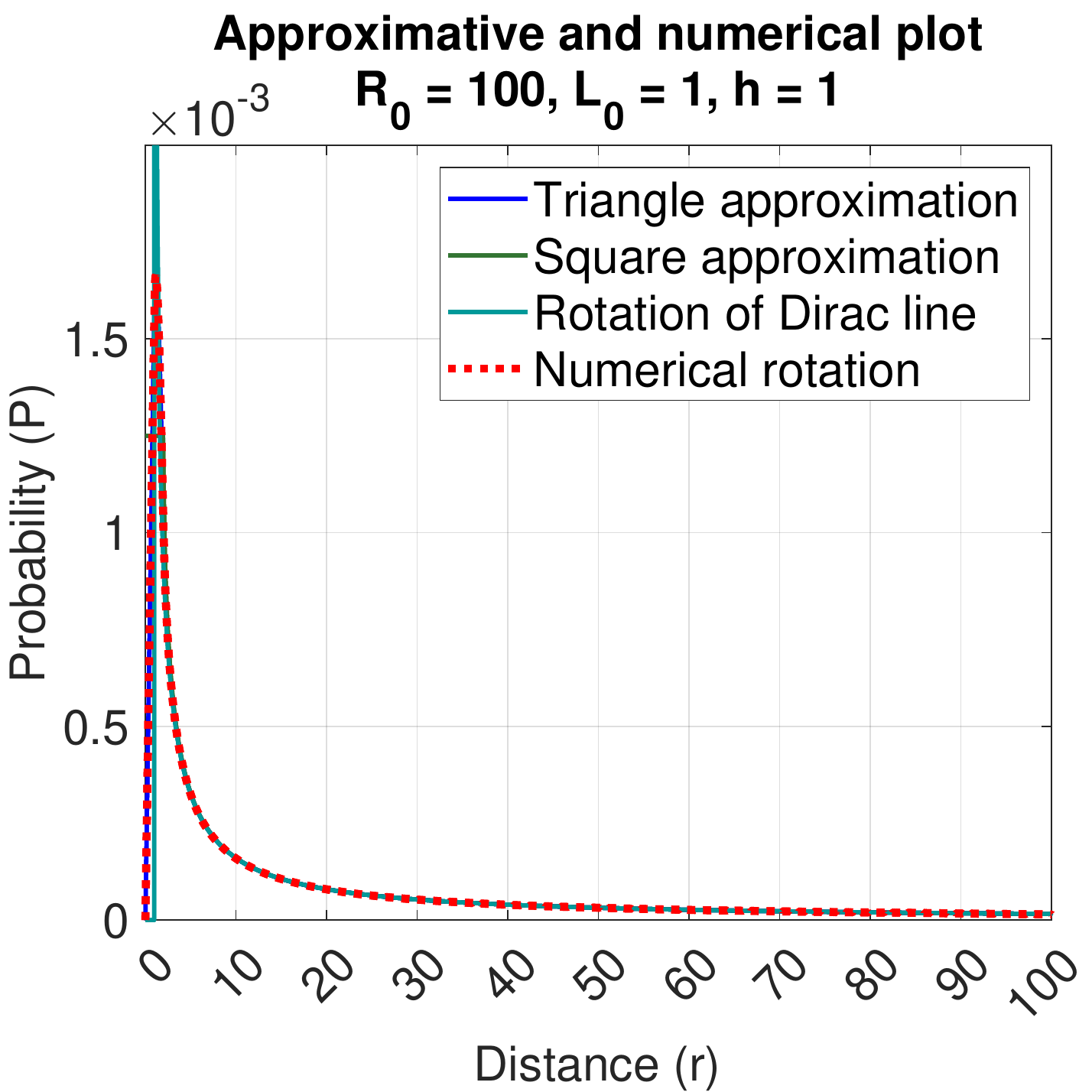}
\caption{$R_0=100$, $L_0=1$, $h=1$}
\end{subfigure}
\begin{subfigure}{0.32\textwidth}
\includegraphics[width=\linewidth]{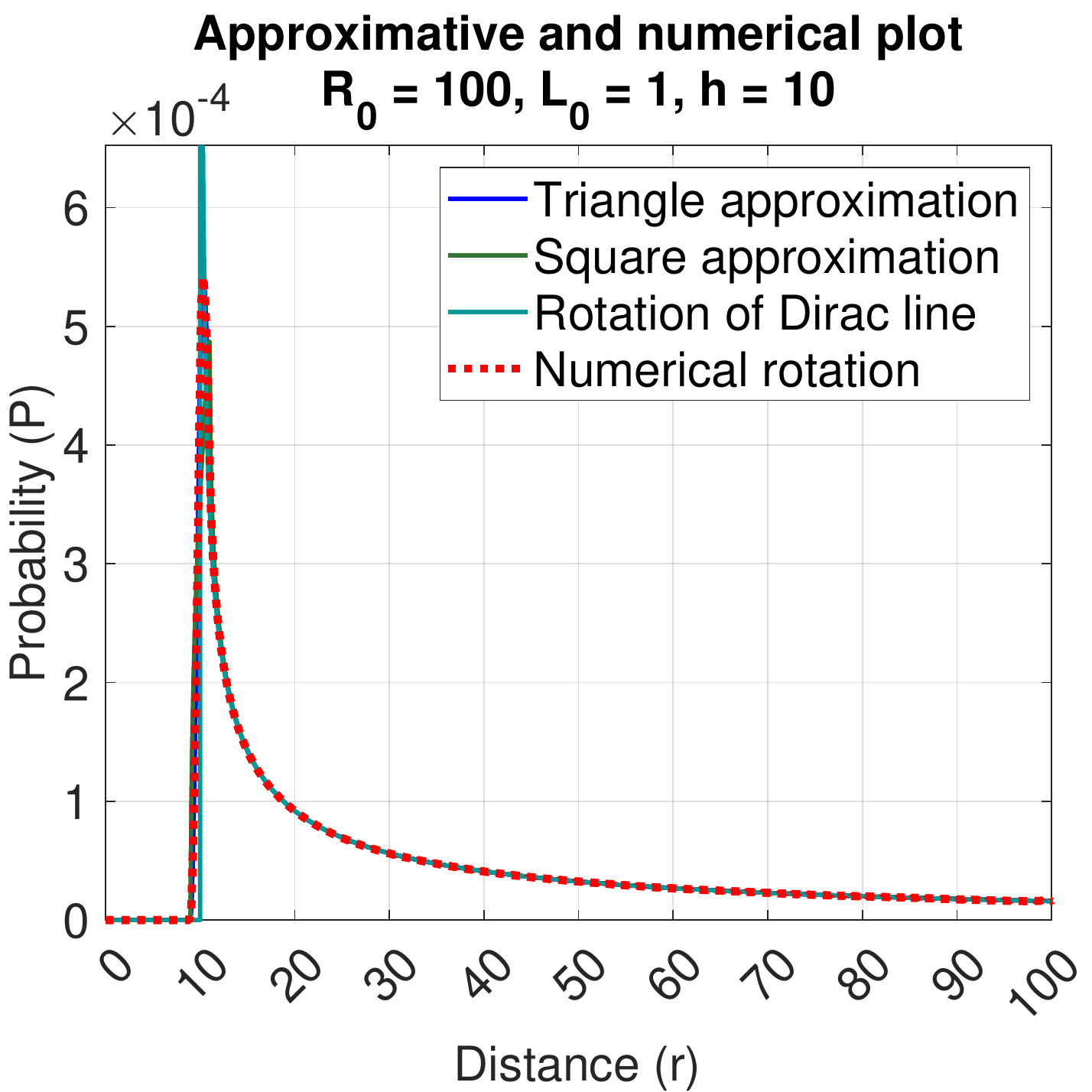}
\caption{$R_0=100$, $L_0=1$, $h=10$}
\end{subfigure}
\caption{Rotated crystal-to-crystal responses and its approximations. We compare triangle approximation, square approximation, and rotation of the Dirac line with the numerical integration of the exact crystal-to-crystal responses (taken as the reference). The triangle approximation fits the best.} 
\label{fig:comperison}
\end{figure}

To confirm our results, we compare all previously described approximations and numerically integrated crystal-to-crystal responses, which can be expressed as $$P_{rot_h}(r; h) =\frac{1}{2\pi} \int_0^{2\pi} P_{shift}(r \cos{\phi},r \cos{\phi};h)\, d\phi$$. In Fig. \ref{fig:comperison}, we can see all the approximations together with numerically integrated crystal-to-crystal responses for different values of $h$. The latter is considered as the reference. There were two geometric configuration choices: $R_0=50$ and $L_0=1$; $R_0=100$ and $L_0=1$. In addition, we estimate the root mean square error (RMSE) for each approximation and for both configurations. Since the rotation of pure Dirac line diverges at $r=h$, we estimated the RMSE on a bit shortened interval $\left[h+0.1, R_0 \right]$. Values of the RMSE where calculated with discrete step $h_{step}=0.1$. Maximum RMSE values for all $h$ are presented in Tab. \ref{tab:RMSE}. As expected, approximation with triangular window has the lowest RMSE. Furthermore, the more elongated configuration with $R_0=100$ and $L_0=1$ has lower RMSE when compared to the configuration with $R_0=50$ and $L_0=1$. This is expected since $\frac{1}{100} < \frac{1}{50}$, and we have proven that approximation holds as $\frac{L_0}{R_0}\rightarrow 0$ (see Appendix C). Therefore, smaller the ratio, better the approximation.
\begin{table}[b]
\centering
\begin{tabular}{|c|c|c|c|}
    \hline
     & Dirac line & Square window approx. & Triangle window approx. \\
     \hline 
     RMSE ($R_0=50$)\;\;  & $2.39 \cdot 10^{-4}$ & $3.45 \cdot 10^{-5}$ & $8.38 \cdot 10^{-7}$ \\
     RMSE ($R_0=100$) & $5.87 \cdot 10^{-5}$ & $8.58 \cdot 10^{-6}$ & $7.25 \cdot 10^{-7}$ \\
    \hline
\end{tabular}
\captionof{table}{RMSE for different approximations and geometrical configurations\label{tab:RMSE}}
\end{table}

\section{White image and Expectation-Maximization algorithm}

The white image describes the probability of detecting uniformly distributed point sources over the entire measurement area of the PET scanner, including rotation of all crystals around the object that is being measured.

One way to generate the white image is MC simulation. Downside of MC simulation is a long execution time if the number of simulated events is high, as well as the noisy result.

\begin{figure}[H]
\includegraphics[width=1\linewidth]{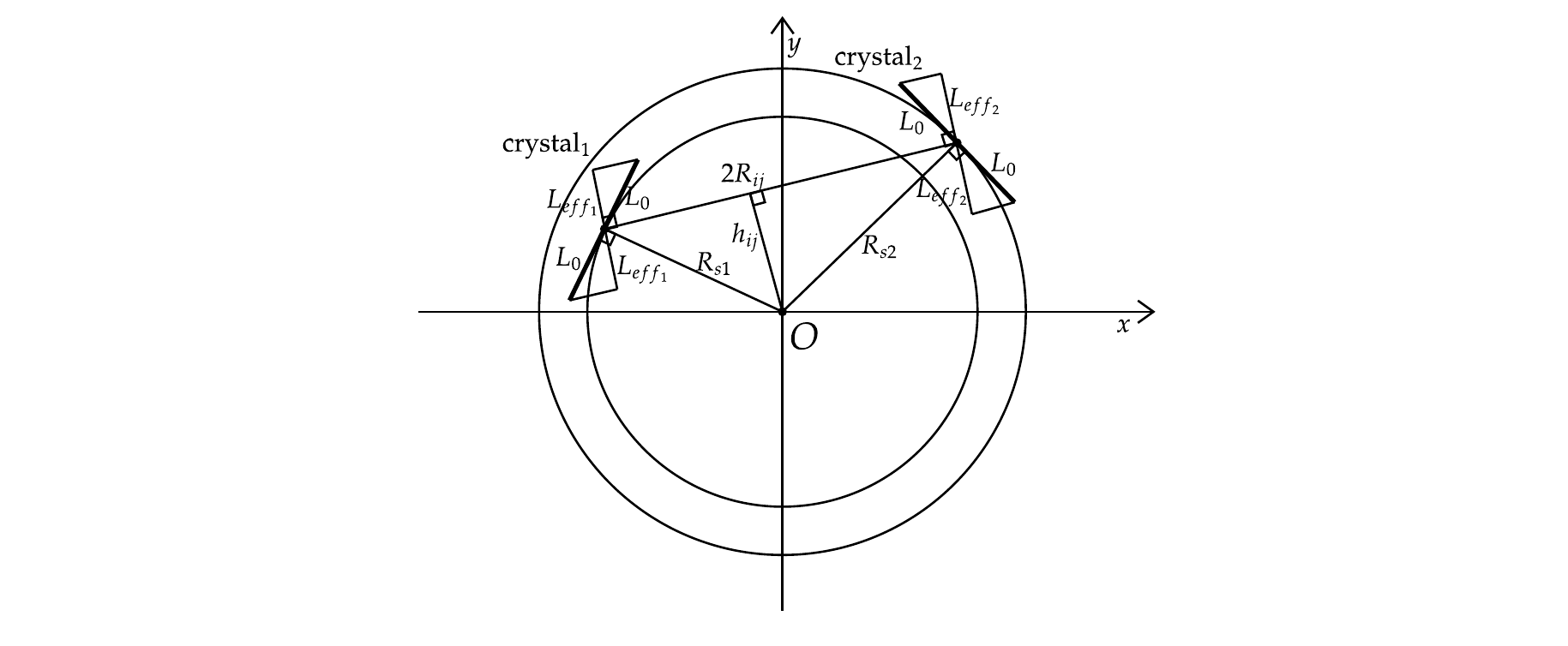}
\centering
\caption{Effective length of a crystal}
\label{fig:Awhitesurfaceparametars}
\end{figure}

Another way to calculate the white image is to connect all possible pairs of crystals under all possible angles of rotation using their PDF-s, and solve it analytically. We can express it in polar coordinates as:
\begin{equation}
I_{WI}(r) = \frac{1}{N_p\sum_{i,j} w_{ij}}\displaystyle \sum_{i,j} w_{ij}\, P_a(r;h_{ij},R_{ij},L_{ij}), 
\label{eq:WI}
\end{equation}
assuming that we are using triangular crystal-to-crystal response approximation. $N_p$ denotes the number of all possible crystal pairs, $h_{ij}$ is the distance from the center of rotation to the line that is connecting centers of crystals $i$ and $j$ that makes a pair, $R_{ij}$ is half of the distance between the two crystals, and $L_{ij}$ is half of the effective length of the crystal, as shown in Fig. \ref{fig:Awhitesurfaceparametars}. 
Weight $w_{ij}$ models the contribution of each PDF $P_a(r;h_{ij},R_{ij},L_{ij})$ to the entire white image. It depends on distances between crystals $R_{ij}$ and crystal lengths $L_{ij}$. Taking both into the account, it can be shown that the $w_{ij}$ weight is equal to $L_{ij}^2$ (see Appendix D). For PET scanners with multiple rings, $L_{ij}$ is calculated by averaging the effective lengths. In our example, Raytest ClearPET has two rings of detectors located at radii $R_{s1}$ and $R_{s2}$. If the detection occurs by some pair of crystals $(i,j)$, and concomitant radii are $R_{si}$ and $R_{sj}$, the effective length of each crystal is $L_{eff_k} = \frac{L_0}{2}\sqrt{1-\frac{h_{ij}^2}{R_{sk}^2}}$, where $k\in\{i,j\}$. Hence, the effective length is the average of $L_{eff_i}$ and $L_{eff_j}$.

\begin{figure}[h]
\centering
\begin{subfigure}{0.3\textwidth}
  \centering
  \includegraphics[width=1\linewidth]{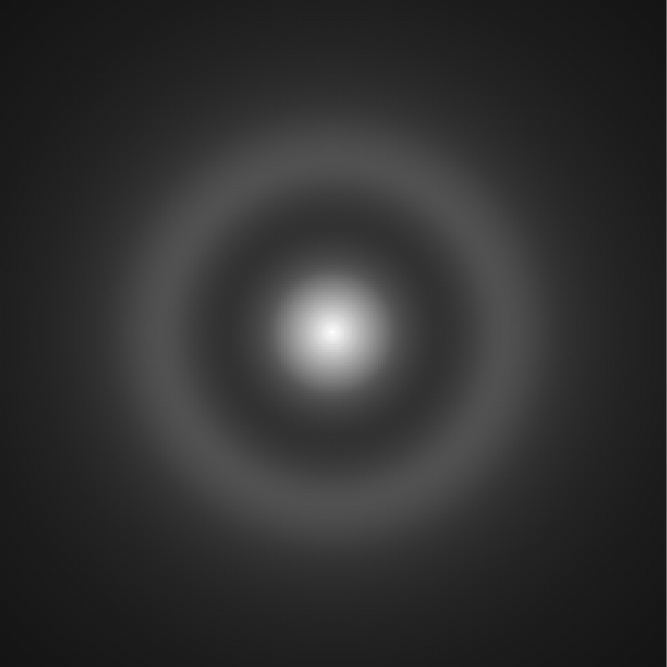}
  \caption{White image obtained by expression (\ref{eq:WI})}
  \label{fig:WI_EX}
\end{subfigure}
\begin{subfigure}{0.3\textwidth}
  \centering
  \includegraphics[width=1\linewidth]{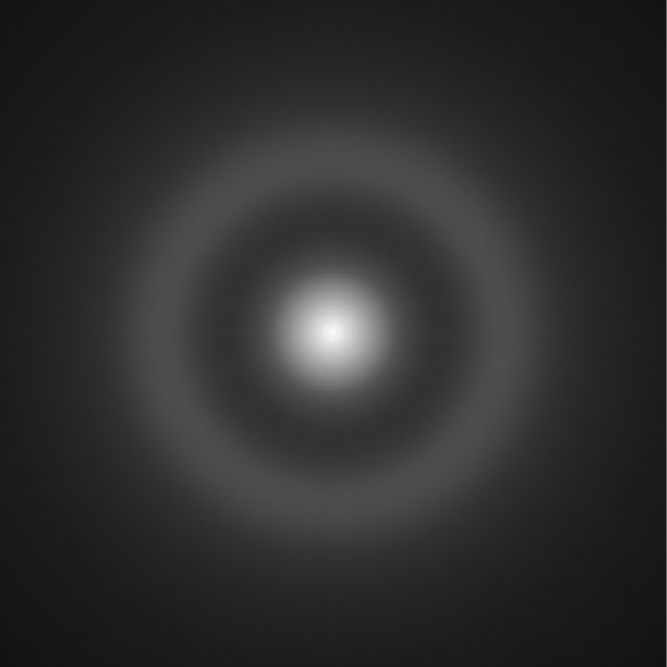}
  \caption{White image obtained via the MC simulation}
  \label{fig:WI_MC}
\end{subfigure}
\begin{subfigure}{0.3\textwidth}
  \centering
  \includegraphics[width=1\linewidth]{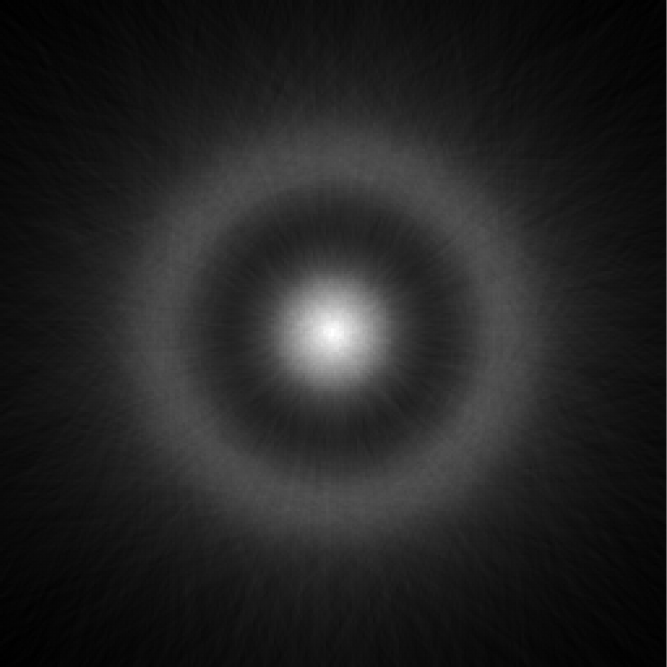}
  \caption{White image obtained by a real calibration measurement}
  \label{fig:WI_RM}
\end{subfigure}
\centering
\caption{White image comparison proves our analytic results}
\label{fig:WI}
\end{figure}

To check our calculation, we perform the MC simulation and compare the result with the expression (\ref{eq:WI}). In Fig. \ref{fig:WI} we can see several images: Fig. \ref{fig:WI_EX} depicts the result obtained by our closed-form expression and in Fig. \ref{fig:WI_MC} the one obtained by the MC simulation is depicted. Additionally, Fig. \ref{fig:WI_RM} depicts real measurement obtained by using the calibration phantom with the uniformly distributed radioactive tracer. Although we are approximating the crystal-to-crystal response with a triangular PDF, the differences between the analytic, synthetic and real models are indistinguishable. 

\begin{figure}[H]
\centering
\begin{subfigure}{0.48\textwidth}
  \centering
  \includegraphics[width=0.7\linewidth]{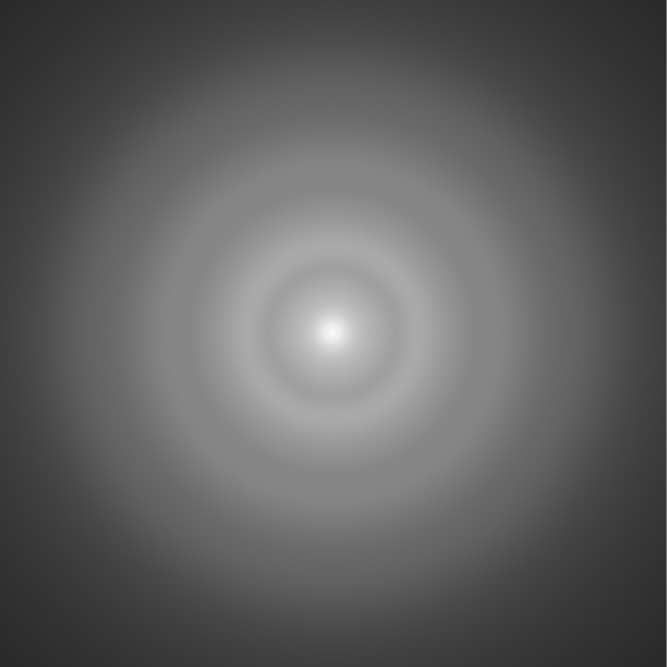}
  \caption{Intersection with eight active sectors}
  \label{fig:WI_type1}
\end{subfigure}
\begin{subfigure}{0.48\textwidth}
  \centering
  \includegraphics[width=0.7\linewidth]{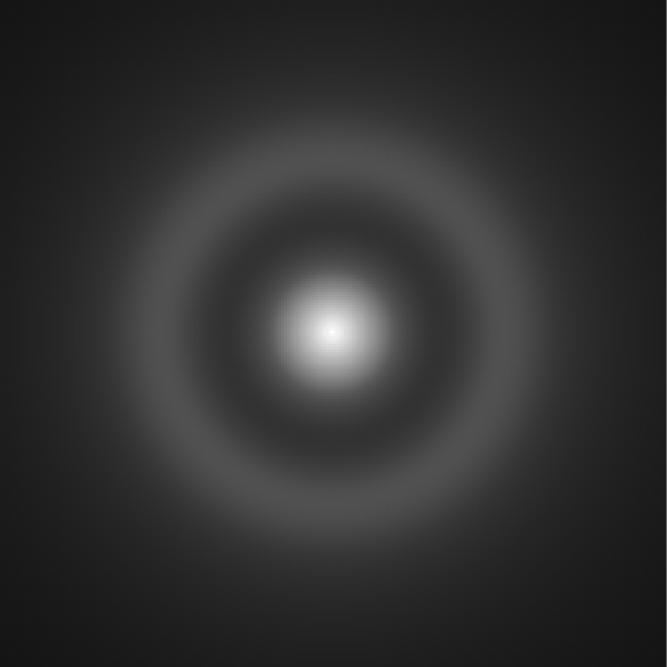}
  \caption{Intersection with four active sectors}
  \label{fig:WI_type2}
\end{subfigure}
\centering
\caption{White images for two different intersections, and geometric configurations of the ClearPET scanner.}
\label{fig:bothwhiteimages}
\end{figure}

Usually, the ClearPET scanner consists of 20 sectors that are covering the entire $2\pi$ ring. We focus on our incomplete ClearPET scanner that contains only eight sectors. Four sectors are stacked together and the other four are located on the opposite side of the ring. For details regarding ClearPET and its geometry see \cite{Matulic2021}. Even and odd sectors are mutually shifted in the axial direction. Hence, there are two different crystal configurations, depending on the position of the observed axial intersection. In Fig. \ref{fig:bothwhiteimages} we can see two white images, corresponding to the two intersections. The left one corresponds to the position with eight active sectors while the right one corresponds to the position with only four active sectors. 

A well-know expression for MLEM image reconstruction is $ \displaystyle I_j^{(n+1)} = \frac{I_j^{(n)}}{\sum_i a_{ij}} \sum_i a_{ij} \frac{p_i}{\sum_{\hat{j}} a_{i\hat{j}} I_{\hat{j}}^{(n)} }$ where $A=\left( a_{ij} \right)$ is the system matrix, $p_i$ are the measurements and $I_{j}^{n}$ is $j$-th pixel of the reconstructed image in $n$-th iteration. The MLEM algorithm can be rewritten in a matrix form as:
\begin{equation}
I^{(n+1)} = I^{(n)} \odot \big( A^{*}\cdot (S \oslash (A\cdot I^{(n)}) \big) \oslash \big(A^{*}\cdot\mathbb{1}).
\end{equation}

In the latter equation, we consider the reconstructed image reshaped as vector $I^{(n)}$. Symbol $\odot$ denotes element by element multiplication, symbol $\oslash$ denotes element by element division, and by $\cdot$ we denoted matrix multiplication. Matrix $S$ denotes sinogram data obtained by measurement, and vector $\mathbb{1}$ is an all-ones vector.

We modify the MLEM reconstruction in the following way:
\begin{itemize}
  \item We substitute $A^{*}\cdot\mathbb{1}$ with our white image model $I_{WI}$.
  \item We replace $A\cdot I^{(n)}$ with Radon transform operator $\mathcal{R}(I^{(n)})$. This is a ray-driven projection. It can be easily implemented, so there is no need for building a large system matrix.
  \item For each measurement, we add dithering to center-of-crystals connecting lines. In that way, we decrease the effects of quantization. The resulting lines are collected in the sinogram data matrix $S$.
  \item The entire sub-expression  $ A^{*}\cdot (S \oslash (A\cdot I^{(n)}) $ can be replaced with Radon transform $R$ and its adjoint operator $R^{*}$ as $\mathcal{R^{*}}\big(S \oslash \mathcal{R}(I^{(n)})\big)$. Adjoint operator $R^{*}$ of the Radon transform is a ray-driven back-projection and can easily be implemented without the need for having full system matrix. 
\end{itemize}
In summary, we get the following iterative formula:
\begin{equation}
I^{(n+1)} = I^{(n)} \odot \mathcal{R^{*}}\big(S \oslash \mathcal{R}(I^{(n)})\big) \oslash I_{WI}.
\label{eq:iterative}
\end{equation}
Realization of the proposed algorithm can be found at: \url{https://github.com/tm2005/Analytical_WI_Model}.

\begin{figure}[H]
\centering
\begin{subfigure}{0.48\textwidth}
  \centering
  \includegraphics[width=0.7\linewidth]{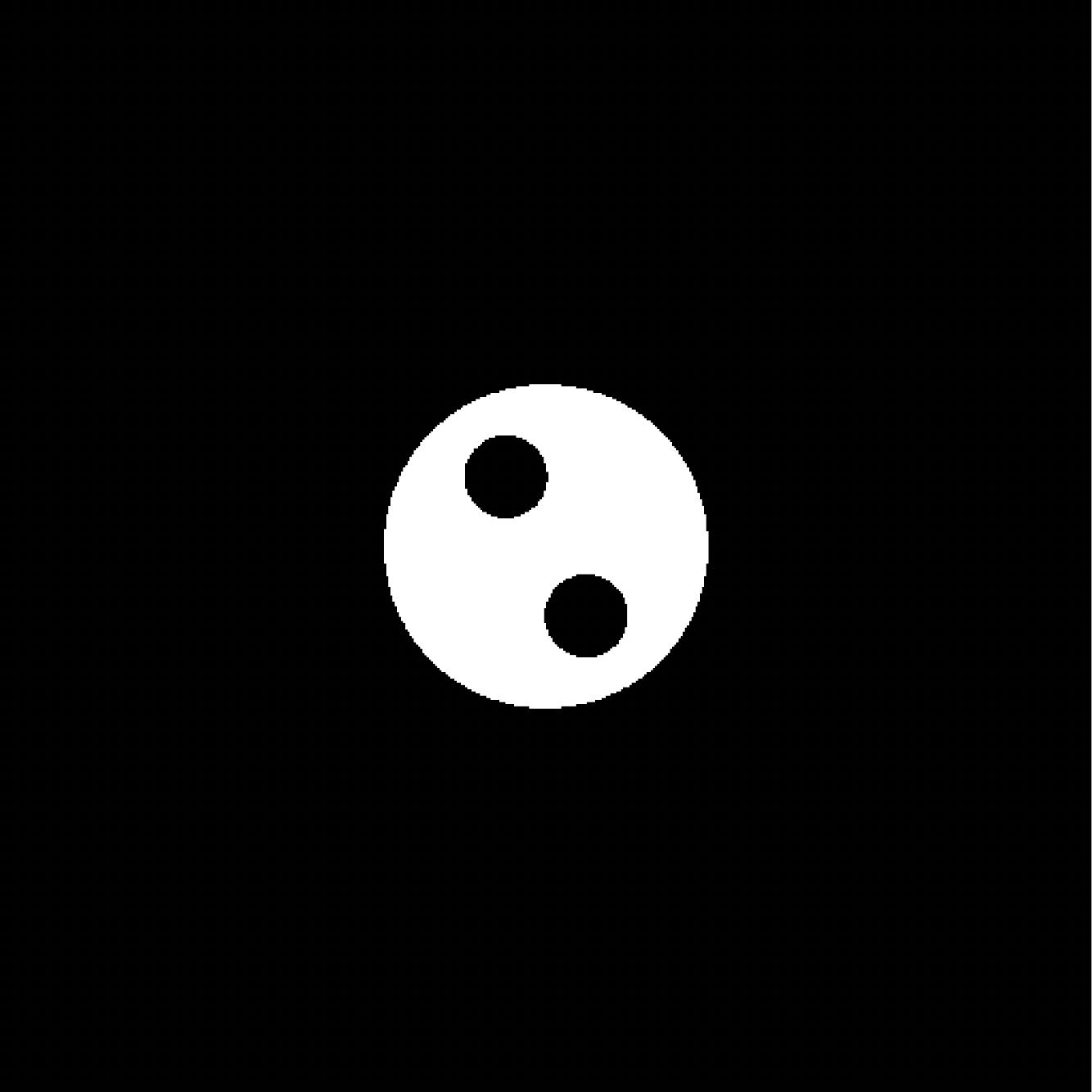}
  \caption{Intersection with big cylinder and two cylindrical holes}
  \label{fig:phantom_five}
\end{subfigure}
\begin{subfigure}{0.48\textwidth}
  \centering
  \includegraphics[width=0.7\linewidth]{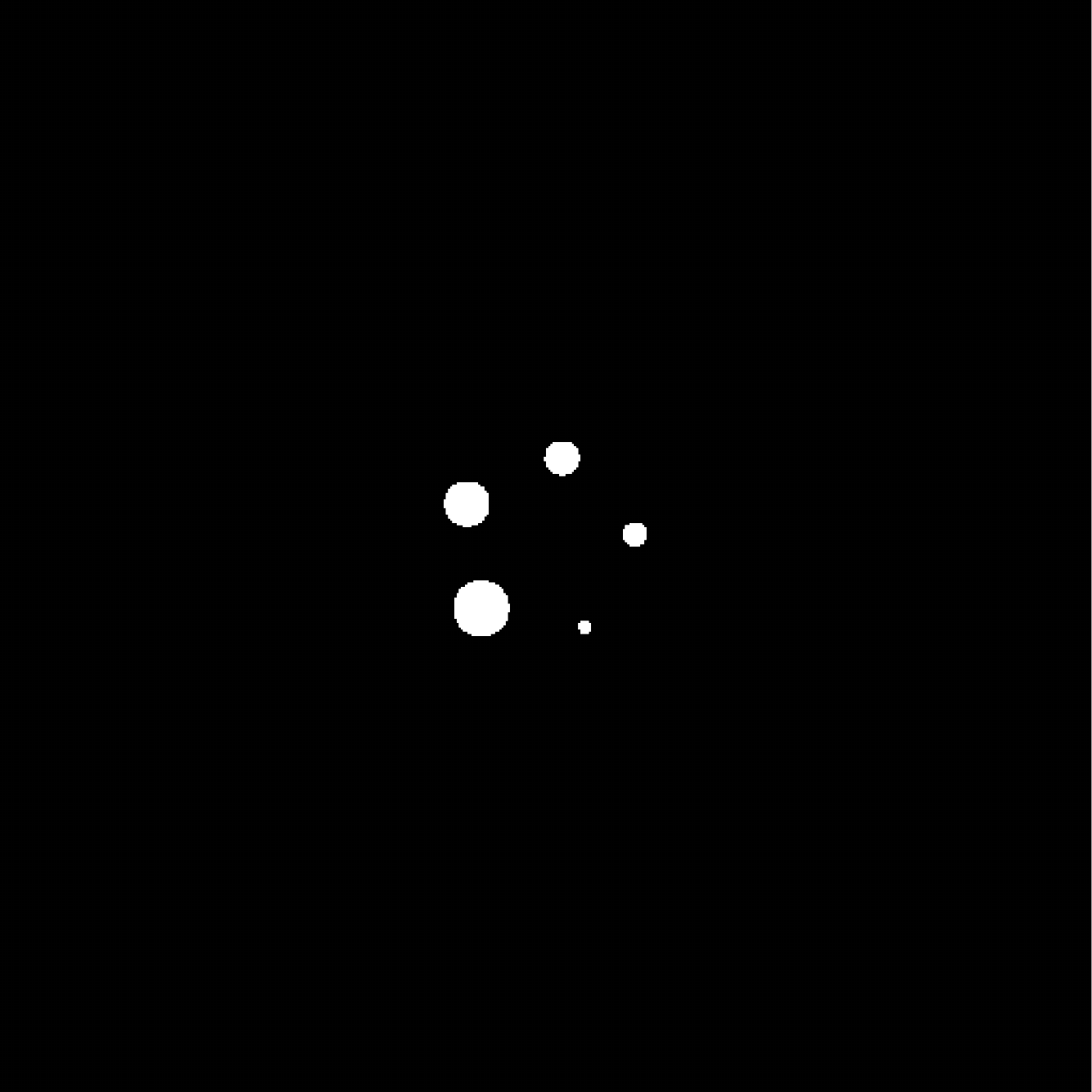}
  \caption{Intersection with five small cylinders of different size}
  \label{fig:phantom_two}
\end{subfigure}
\centering
\caption{Two different intersections of NEMA NU 4-2008 phantom. Cylinders are filled by the radioactive tracer, except for the holes.}
\label{fig:phantoms}
\end{figure}

\section{Results}
To confirm our method, we perform reconstruction on synthetic data, as well as on the real measurements. Synthetic measurements correspond to our partially incomplete ClearPET scanner, as described in the previous section. 

In Fig. \ref{fig:phantoms}, we can see the ground truth images. They correspond to two different axial intersections of the NEMA NU 4-2008 phantom. 

\begin{figure}[H]
\centering
\begin{subfigure}{0.48\textwidth}
  \centering
  \includegraphics[width=0.7\linewidth]{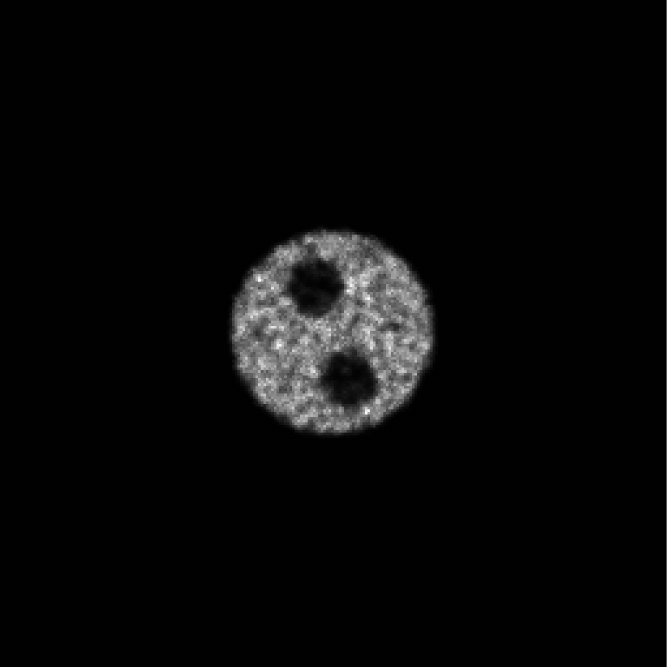}
  \caption{Reconstruction using eight active sectors for intersections with two cylindrical holes.}
  \label{fig:synt_type1_two}
\end{subfigure}
\begin{subfigure}{0.48\textwidth}
  \centering
  \includegraphics[width=0.7\linewidth]{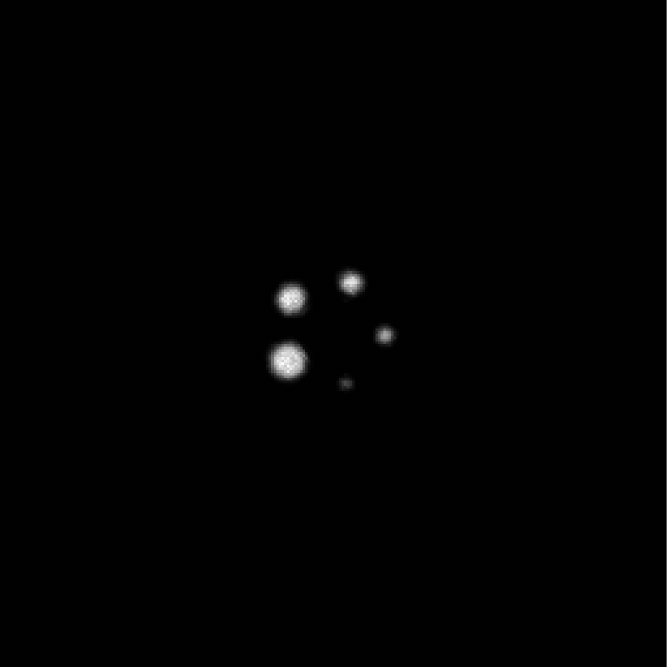}
  \caption{Reconstruction using eight active sectors for intersections with five cylinders.}
  \label{fig:synt_type1_five}
\end{subfigure}
\begin{subfigure}{0.48\textwidth}
  \centering
  \includegraphics[width=0.7\linewidth]{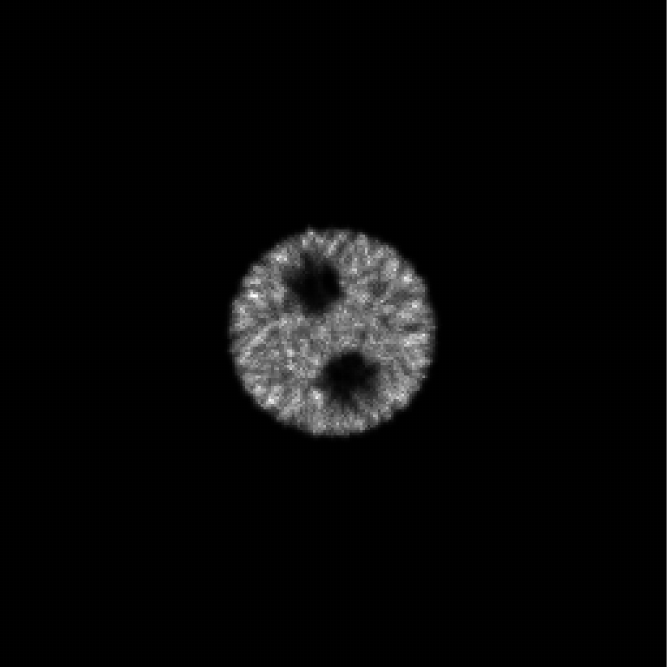}
  \caption{Reconstruction using four active sectors for intersections with two cylindrical holes.}
  \label{fig:synt_type2_two}
\end{subfigure}
\begin{subfigure}{0.48\textwidth}
  \centering
  \includegraphics[width=0.7\linewidth]{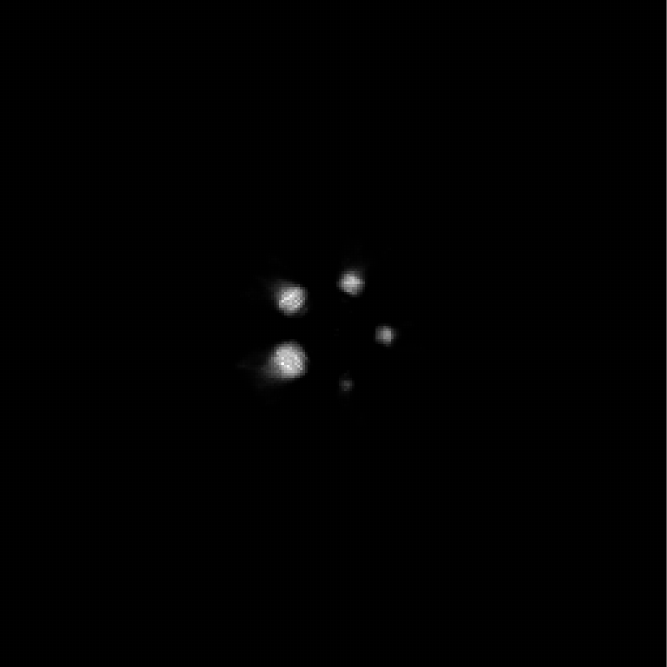}
  \caption{Reconstruction using four active sectors for intersections with five cylinders.}
  \label{fig:synt_type2_five}
\end{subfigure}
\centering
\caption{Reconstruction of two different intersections of NEMA NU 4-2008 using two different crystal configurations on synthetic data. Cylinders are filled by the radioactive tracer, except for the holes.}
\label{fig:synt_recon}
\end{figure}

Images reconstructed with the proposed algorithm on synthetic data can be seen in Fig. \ref{fig:synt_recon}. Reconstructions in Fig \ref{fig:synt_type1_two} and Fig. \ref{fig:synt_type1_five} are obtained after simulation with eight sectors active, while in Fig. \ref{fig:synt_type2_two} and Fig. \ref{fig:synt_type2_five} there were only four active sectors.

\begin{figure}[H]
\centering
\begin{subfigure}{0.32\textwidth}
  \centering
  \includegraphics[width=\linewidth]{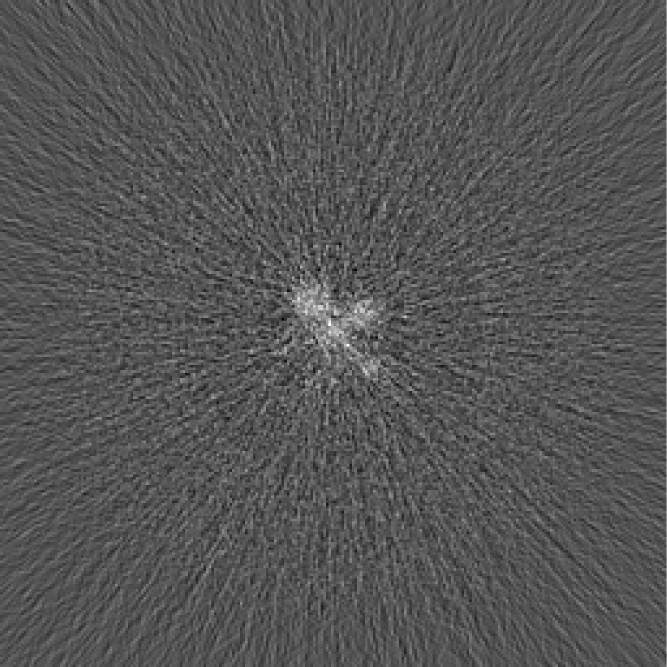}
  \caption{Filtered back-projection algorithm using eight active sector for intersections containing five cylinders.}
  \label{fig:real_five_FBP}
\end{subfigure}
\begin{subfigure}{0.32\textwidth}
  \centering
  \includegraphics[width=\linewidth]{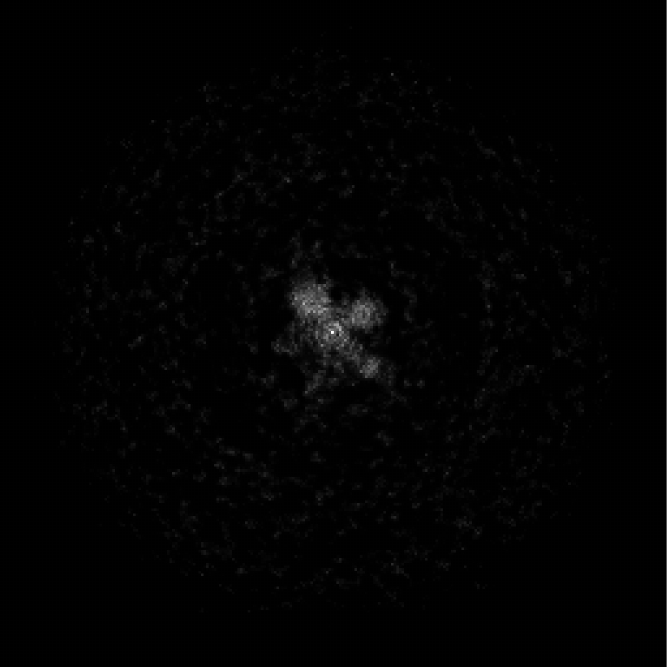}
  \caption{MLEM algorithm (no compensation) using eight active sector for intersections containing five cylinders.}
  \label{fig:real_five_EM_nocomp}
\end{subfigure}
\begin{subfigure}{0.32\textwidth}
  \centering
  \includegraphics[width=\linewidth]{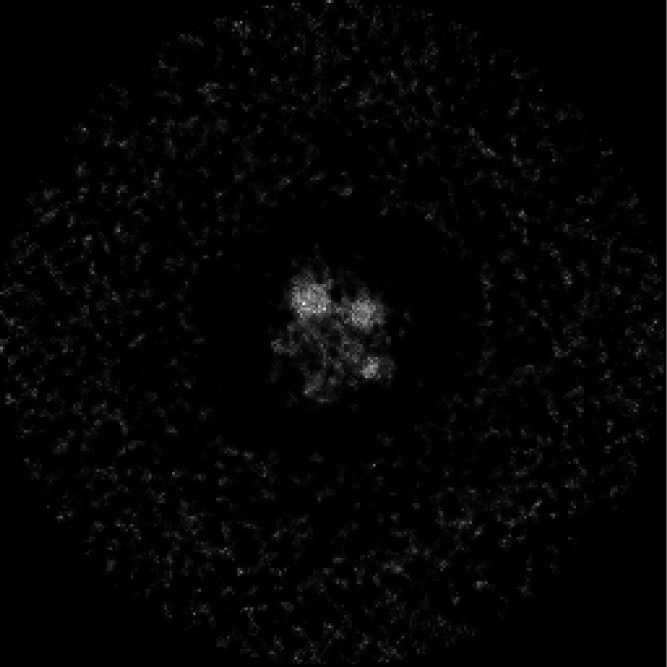}
  \caption{Proposed reconstruction method using eight active sector for intersections containing five cylinders.}
  \label{fig:real_five_EM}
\end{subfigure}
\begin{subfigure}{0.32\textwidth}
  \centering
  \includegraphics[width=\linewidth]{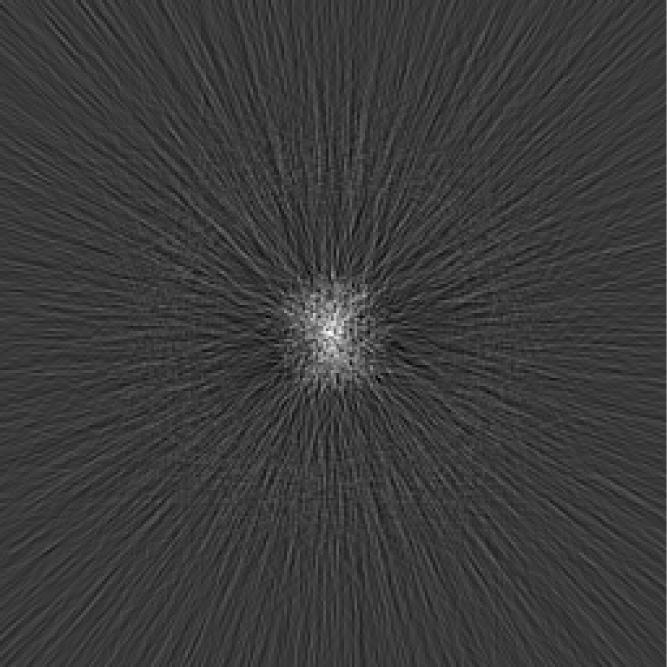}
  \caption{Filtered back-projection algorithm using eight active sector for intersections containing two cylindrical holes.}
  \label{fig:real_two_FBP}
\end{subfigure}
\begin{subfigure}{0.32\textwidth}
  \centering
  \includegraphics[width=\linewidth]{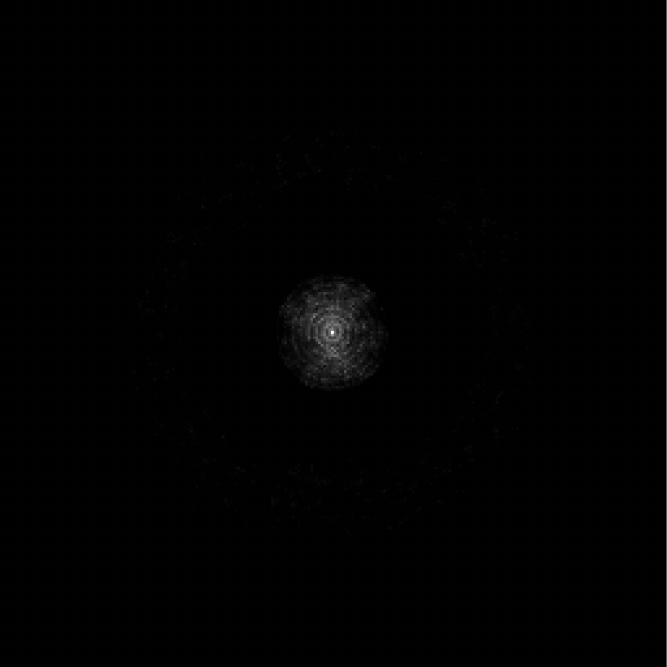}
  \caption{MLEM algorithm (no compensation) using eight active sector for intersections containing two cylindrical holes.}
  \label{fig:real_two_EM_nocomp}
\end{subfigure}
\begin{subfigure}{0.32\textwidth}
  \centering
  \includegraphics[width=\linewidth]{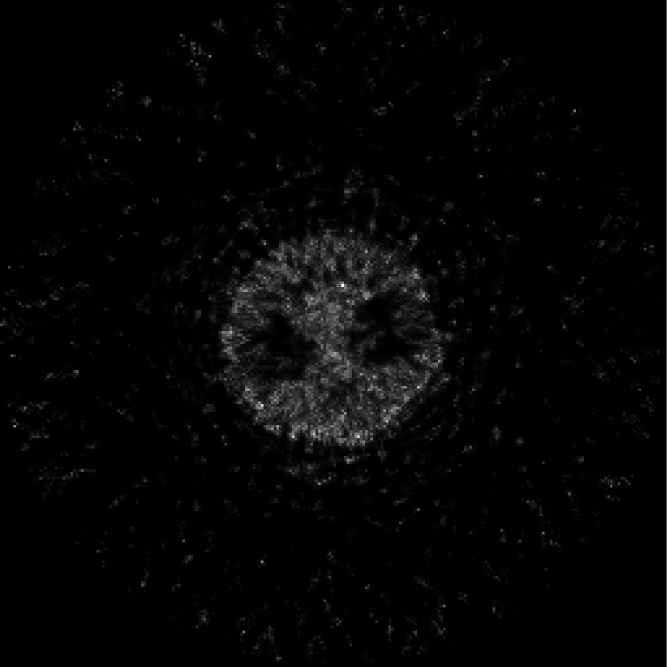}
  \caption{Proposed reconstruction method using eight active sector for intersections containing two cylindrical holes.}
  \label{fig:real_two_EM}
\end{subfigure}
\centering
\caption{Reconstruction of two different intersections of the NEMA NU 4-2008 phantom using two different crystal configurations on real measurements.}
\label{fig:real_recon}
\end{figure}

In Fig. \ref{fig:real_recon} we can see: reconstructed images using filtered back-projection (Fig. \ref{fig:real_five_FBP} and \ref{fig:real_two_FBP}), MLEM algorithm implemented as a series of projections and back-projections without taking into the account white-image compensation (Fig. \ref{fig:real_five_EM_nocomp} and \ref{fig:real_two_EM_nocomp}), and finally the proposed algorithm (Fig. \ref{fig:real_five_EM} and \ref{fig:real_two_EM}), all conducted on the real data measured by our Raytest ClearPET. In the first row of Fig. \ref{fig:real_recon}, there were eight active sectors. In the second row, there were only four active sectors. Same measurement data, as well as the dithering approach were used in all reconstructions.
Reconstruction based on the real measurements without the white image compensation suffers from significant drawbacks. In Fig. \ref{fig:real_five_FBP} and \ref{fig:real_five_EM_nocomp}, we can recognize five circles (five radiotracer filled cylinders), but an annoying central artifact occurs in both images. This artifact can be easily explained using our white image model. If we take a closer look of Fig. \ref{fig:WI_type1} and consider its probabilistic interpretation, we can see that many events occur in the region of the highest probability, i.e. many lines of the response pass through the center of the rotation. It causes the central artifact. Additionally, ring-like artifacts around the origin can also be seen in Fig. \ref{fig:real_five_EM_nocomp}. They are caused by quantization effects due to finite crystal lengths and distances between the adjacent ones. Reconstruction conducted using the proposed algorithm (Fig. \ref{fig:real_five_EM}) does not suffer from any of the previously mentioned artifacts. Some noise is still present, due to scatter and random events that were not included in our model. 

Similar observations are revealed in the second row of Fig. \ref{fig:real_recon}. Captured object is completely unrecognizable in Fig. \ref{fig:real_two_FBP} and \ref{fig:real_two_EM_nocomp}. This can be explained in a probabilistic manner, too. Reconstructions without valid correction based on the scanner geometry resemble the distribution explained by the white image model. Uncompensated reconstructions correspond to the highest probability areas of the white image, as clearly visible in Fig. \ref{fig:WI_type2}, including some ring-like artifacts around the origin due to quantization. Proposed reconstruction is depicted in Fig. \ref{fig:real_two_EM}. Despite some scatter and noise, it clearly overperforms the uncompensated methods.

All obtained images have resolution of $256 \times 256$ pixels. Reconstructions obtained by iterative methods were halted after 50 iterations. Time to produce such images by the proposed algorithm, where the measurement data contained about 50000 coincidences per intersection, was approximately 1.3 seconds on Intel i9-11900K with 64 GB of RAM. Since ClearPET has 48 possible axial intersections, a bit over a minute was needed for a full high-quality reconstruction.

\section{Conclusion}

The first goal of our research was to develop a precise white image model. Such a compensation model is necessary for overcoming physical limitations of the real PET scanner. We started from a closed-form expression of crystal-to-crystal response, under the condition that the distance between two crystals is larger than a size of a crystal. Moreover, we developed a closed-form solution that describes full-circle rotation of such a response, under the condition that the center of rotation matches the center of the response. The rotation of such objects resulted in a long and unpractical expression, but was used as the reference for approximate solutions. We proposed a few, in such a way that the error was negligible when compared to the exact solution. The winner was the triangular-shaped response, as described in Appendix C. It lead to simple expression that can be easily implemented in the reconstruction algorithm. To the best of our knowledge, such a detailed mathematical model of PET camera, as described in Appendices A, B, C, and D, was not yet reported.

Our second goal was to create a white image model that is used for compensation. It is a weighted sum of all responses, by taking into account all possible pairs of crystals and rotating their responses (integrating over $2\pi$ angle). We have shown that our white image model excellently corresponds to the one obtained via MC simulation, as well as to the real calibration measurement. As a result, we were able to generate the white image model on the fly, due to the simplicity of the obtained expression. 

Finally, we utilized the white image model for the fast PET image reconstruction. We modified the MLEM algorithm in such a way that we do not need a large system matrix. The MLEM reconstruction algorithm is conducted as a series of ray-driven projections and back-projections. The compensation is done in each step of the MLEM algorithm as an element-wise division using our white image model. It replaces the entire system model, with memory demands comparable to the reconstructed image size. Finally, we have checked our approach on synthetic and real data. For the real-world acquisition, we used the Raytest ClearPET camera, calibration phantom and test NEMA NU 4-2008 phantom. We have shown that the images obtained by the proposed algorithm have successfully reconstructed the measured object, and clearly outperformed the uncompensated methods. 

Our approach does not take into the account scatter and random events in measurements, which remains a challenge for the future work. Furthermore, we developed several simplified response models and proved their accuracy, thus paving the way for extending proposed approach to full 3D reconstruction schemes. Instead of the white image model, it will utilize a 4D compensation subspace. Since our approach is based on ray-driven projections and back-projections instead of the full system operator, it is expected to be very useful for multi-dimensional applications.

\section*{Acknowledgements}


\bibliographystyle{plain}
\bibliography{refer.bib}

\begin{thebibliography}{10}

\bibitem{Berker2018a}
Y.~Berker, J.~Maier, and M.~Kachelries.
\newblock Deep scatter estimation in {PET}: Fast scatter correction using a
  convolutional neural network.
\newblock In {\em 2018 {IEEE} Nuclear Science Symposium and Medical Imaging
  Conference Proceedings ({NSS}/{MIC})}. {IEEE}, November 2018.

\bibitem{delaPrieta2006}
R.~de~la Prieta, J.~A. Hernandez, E.~Schiavi, and N.~Malpica.
\newblock Analytical geometric model for photon coincidence detection in 3d
  {PET}.
\newblock In {\em 2006 {IEEE} Nuclear Science Symposium Conference Record}.
  {IEEE}, 2006.

\bibitem{Fati2019}
I.~Fatwasauri and M.~Rizkinia.
\newblock Compressive sensing image reconstruction with total variation and
  l2,1 norm for microwave imaging.
\newblock In {\em 2019 {IEEE} International Conference on Innovative Research
  and Development ({ICIRD})}. {IEEE}, June 2019.

\bibitem{Gong2019b}
K.~Gong, C.~Catana, J.~Qi, and Q.~Li.
\newblock {PET} image reconstruction using deep image prior.
\newblock {\em {IEEE} Transactions on Medical Imaging}, 38(7):1655--1665, July
  2019.

\bibitem{Gong2019a}
K.~Gong, J.~Guan, K.~Kim, X.~Zhang, J.~Yang, Y.~Seo, G.~E. Fakhri, J.~Qi, and
  Q.~Li.
\newblock Iterative {PET} image reconstruction using convolutional neural
  network representation.
\newblock {\em {IEEE} Transactions on Medical Imaging}, 38(3):675--685, March
  2019.

\bibitem{Han2017a}
X.~Han.
\newblock {MR}-based synthetic {CT} generation using a deep convolutional
  neural network method.
\newblock {\em Medical Physics}, 44(4):1408--1419, March 2017.

\bibitem{Hanif2013}
A.~Hanif, A.~B. Mansoor, and T.~Ejaz.
\newblock A new approach to radionuclide imaging using compressed sensing.
\newblock {\em The Imaging Science Journal}, 61(6):503--508, July 2013.

\bibitem{Harrison}
R.~L. Harrison, S.~D. Vannoy, D.~R. Haynor, S.~B. Gillispie, M.~S. Kaplan, and
  T.~K. Lewellen.
\newblock Preliminary experience with the photon history generator module of a
  public-domain simulation system for emission tomography.
\newblock In {\em 1993 {IEEE} Conference Record Nuclear Science Symposium and
  Medical Imaging Conference}. {IEEE}.

\bibitem{Herraiz2006}
J.~L. Herraiz, S.~Espa{\~{n}}a, J.~J. Vaquero, M.~Desco, and J.~M.
  Ud{\'{\i}}as.
\newblock {FIRST}: Fast iterative reconstruction software for ({PET})
  tomography.
\newblock {\em Physics in Medicine and Biology}, 51(18):4547--4565, August
  2006.

\bibitem{Hu2007}
Z.~Hu, W.~Wang, E.~E. Gualtieri, Y.~L. Hsieh, J.~S. Karp, S.~Matej, M.~J.
  Parma, C.~H. Tung, E.~S. Walsh, M.~Werner, and D.~Gagnon.
\newblock An {LOR}-based fully-3d {PET} image reconstruction using a blob-basis
  function.
\newblock In {\em 2007 {IEEE} Nuclear Science Symposium Conference Record}.
  {IEEE}, 2007.

\bibitem{Hudson1994}
H.~M. Hudson and R.~S. Larkin.
\newblock Accelerated image reconstruction using ordered subsets of projection
  data.
\newblock {\em {IEEE} Transactions on Medical Imaging}, 13(4):601--609, 1994.

\bibitem{Ida2019a}
I.~Häggströma, R.~Schmidtlein, G.~Campanella, and T.~J.Fuchs.
\newblock Deeppet: A deep encoder-decoder network for directly solving the pet
  reconstruction inverse problem.
\newblock 2018.

\bibitem{Iriarte2016}
A.~Iriarte, R.~Marabini, S.~Matej, C.O.S. Sorzano, and R.M. Lewitt.
\newblock System models for {PET} statistical iterative reconstruction: A
  review.
\newblock {\em Computerized Medical Imaging and Graphics}, 48:30--48, March
  2016.

\bibitem{Jan2004}
S.~Jan, G.~Santin, D~Strul, S~Staelens, K~Assi{\'{e}}, D~Autret, S~Avner,
  R~Barbier, M~Bardi{\`{e}}s, P~M Bloomfield, D~Brasse, V~Breton,
  P~Bruyndonckx, I~Buvat, A~F Chatziioannou, Y~Choi, Y~H Chung, C~Comtat,
  D~Donnarieix, L~Ferrer, S~J Glick, C~J Groiselle, D~Guez, P-F Honore,
  S~Kerhoas-Cavata, A~S Kirov, V~Kohli, M~Koole, M~Krieguer, D~J van~der Laan,
  F~Lamare, G~Largeron, C~Lartizien, D~Lazaro, M~C Maas, L~Maigne, F~Mayet,
  F~Melot, C~Merheb, E~Pennacchio, J~Perez, U~Pietrzyk, F~R Rannou, M~Rey, D~R
  Schaart, C~R Schmidtlein, L~Simon, T~Y Song, J-M Vieira, D~Visvikis, R~Van
  de~Walle, E~Wieërs, and C~Morel.
\newblock {GATE}: a simulation toolkit for {PET} and {SPECT}.
\newblock {\em Physics in Medicine and Biology}, 49(19):4543--4561, September
  2004.

\bibitem{Lee2018ab}
K.~Lee, Y.~Wu, and Y.~Bresler.
\newblock Near-optimal compressed sensing of a class of sparse low-rank
  matrices via sparse power factorization.
\newblock {\em {IEEE} Transactions on Information Theory}, 64(3):1666--1698,
  March 2018.

\bibitem{Liu2018a}
F.~Liu, H.~Jang, R.~Kijowski, T.~Bradshaw, and A.~B. McMillan.
\newblock Deep learning {MR} imaging{\textendash}based attenuation correction
  for {PET}/{MR} imaging.
\newblock {\em Radiology}, 286(2):676--684, February 2018.

\bibitem{Lougovski2015}
A.~Lougovski, F.~Hofheinz, J.~Maus, G.~Schramm, and J.~van~den Hoff.
\newblock On the relation between kaiser{\textendash}bessel blob and tube of
  response based modelling of the system matrix in iterative {PET} image
  reconstruction.
\newblock {\em Physics in Medicine and Biology}, 60(10):4209--4224, May 2015.

\bibitem{Lougovski2014}
A.~Lougovski, F.~Hofheinz, J.~Maus, G.~Schramm, E.~Will, and J.~van~den Hoff.
\newblock A volume of intersection approach for on-the-fly system matrix
  calculation in 3d {PET} image reconstruction.
\newblock {\em Physics in Medicine and Biology}, 59(3):561--577, January 2014.

\bibitem{Matulic2021}
T.~Matuli{\'{c}}, R.~Bagari{\'{c}}, and D.~Ser{\v{s}}i{\'{c}}.
\newblock Enhanced reconstruction for {PET} scanner with a narrow field of view
  by using backprojection method.
\newblock In {\em 2021 44th International Convention on Information,
  Communication and Electronic Technology ({MIPRO})}. {IEEE}, September 2021.

\bibitem{Moehrs2008}
S.~Moehrs, M.~Defrise, N.~Belcari, A.~Del Guerra, A.~Bartoli, S.~Fabbri, and
  G.~Zanetti.
\newblock Multi-ray-based system matrix generation for 3d {PET} reconstruction.
\newblock {\em Physics in Medicine and Biology}, 53(23):6925--6945, November
  2008.

\bibitem{Nie2017a}
D.~Nie, R.~Trullo, Jun Lian, Caroline Petitjean, Su~Ruan, Qian Wang, and
  Dinggang Shen.
\newblock Medical image synthesis with context-aware generative adversarial
  networks.
\newblock In {\em Medical Image Computing and Computer Assisted Intervention -
  {MICCAI} 2017}, pages 417--425. Springer International Publishing, 2017.

\bibitem{Panin2006}
V.~Y. Panin, F.~Kehren, C.~Michel, and M.~Casey.
\newblock Fully 3-d {PET} reconstruction with system matrix derived from point
  source measurements.
\newblock {\em {IEEE} Transactions on Medical Imaging}, 25(7):907--921, July
  2006.

\bibitem{Qi1998}
J.~Qi, R.~M. Leahy, S.~R. Cherry, A.~Chatziioannou, and T.~H. Farquhar.
\newblock High-resolution 3d bayesian image reconstruction using the {microPET}
  small-animal scanner.
\newblock {\em Physics in Medicine and Biology}, 43(4):1001--1013, April 1998.

\bibitem{Qian2017a}
H.~Qian, X.~Rui, and S.~Ahn.
\newblock Deep learning models for {PET} scatter estimations.
\newblock In {\em 2017 {IEEE} Nuclear Science Symposium and Medical Imaging
  Conference ({NSS}/{MIC})}. {IEEE}, October 2017.

\bibitem{Rahmim2005}
A.~Rahmim, J.~C. Cheng, S.~Blinder, M.~L. Camborde, and V.~Sossi.
\newblock Statistical dynamic image reconstruction in state-of-the-art
  high-resolution {PET}.
\newblock {\em Physics in Medicine and Biology}, 50(20):4887--4912, October
  2005.

\bibitem{Ralai2020}
I.~Rala{\v{s}}i{\'{c}}, D.~Ser{\v{s}}i{\'{c}}, and S.~{\v{S}}egvi{\'{c}}.
\newblock Perceptual autoencoder for compressive sensing image reconstruction.
\newblock {\em Informatica}, pages 561--578, 2020.

\bibitem{Ralasic2018a}
I.~Rala{\v{s}}i{\'{c}}, A.~Tafro, and D.~Ser{\v{s}}i{\'{c}}.
\newblock Statistical compressive sensing for efficient signal reconstruction
  and classification.
\newblock In {\em 2018 4th International Conference on Frontiers of Signal
  Processing ({ICFSP})}. {IEEE}, September 2018.

\bibitem{Sarrut2021}
D.~Sarrut, M.~Ba{\l}a, Manuel Bardi{\`{e}}s, Julien Bert, Maxime Chauvin,
  Konstantinos Chatzipapas, Mathieu Dupont, Ane Etxebeste, Louise~M Fanchon,
  S{\'{e}}bastien Jan, Gunjan Kayal, Assen~S Kirov, Pawe{\l} Kowalski, Wojciech
  Krzemien, Joey Labour, Mirjam Lenz, George Loudos, Brahim Mehadji, Laurent
  M{\'{e}}nard, Christian Morel, Panagiotis Papadimitroulas, Magdalena Rafecas,
  Julien Salvadori, Daniel Seiter, Mariele Stockhoff, Etienne Testa, Carlotta
  Trigila, Uwe Pietrzyk, Stefaan Vandenberghe, Marc-Antoine Verdier, Dimitris
  Visvikis, Karl Ziemons, Milan Zvolsk{\'{y}}, and Emilie Roncali.
\newblock Advanced monte carlo simulations of emission tomography imaging
  systems with {GATE}.
\newblock {\em Physics in Medicine {\&} Biology}, 66(10):10TR03, May 2021.

\bibitem{Seri2016}
D.~Ser{\v{s}}i{\'{c}}, A.~Sovi{\'{c}} Kr{\v{z}}i{\'{c}}, and C.~S. Menoni.
\newblock Relative intersection of confidence intervals rule for sharper
  restoration of soft x-ray images.
\newblock {\em Applied Optics}, 55(31):8932, October 2016.

\bibitem{Seri2014}
D.~Ser{\v{s}}i{\'{c}}, A.~Sovi{\'{c}}, and C.~S. Menoni.
\newblock Restoration of soft x-ray laser images of nanostructures.
\newblock {\em Optics Express}, 22(11):13846, May 2014.

\bibitem{Shepp1982}
L.~A. Shepp and Y.~Vardi.
\newblock Maximum likelihood reconstruction for emission tomography.
\newblock {\em {IEEE} Transactions on Medical Imaging}, 1(2):113--122, October
  1982.

\bibitem{Siddon1985}
R.~L. Siddon.
\newblock Fast calculation of the exact radiological path for a
  three-dimensional {CT} array.
\newblock {\em Medical Physics}, 12(2):252--255, March 1985.

\bibitem{Tafro2019a}
A.~Tafro and D.~Ser{\v{s}}i{\'{c}}.
\newblock Iterative algorithms for gaussian mixture model estimation in 2d
  {PET} imaging.
\newblock In {\em 2019 11th International Symposium on Image and Signal
  Processing and Analysis ({ISPA})}. {IEEE}, September 2019.

\bibitem{Terstegge}
A.~Terstegge, S.~Weber, H.~Herzog, H.W. Muller-Gartner, and H.~Halling.
\newblock High resolution and better quantification by tube of response
  modelling in 3d {PET} reconstruction.
\newblock In {\em 1996 {IEEE} Nuclear Science Symposium. Conference Record}.
  {IEEE}.

\bibitem{Thompson1992}
C.~J. Thompson, J.~Moreno-Cantu, and Y.~Picard.
\newblock {PETSIM}: Monte carlo simulation of all sensitivity and resolution
  parameters of cylindrical positron imaging systems.
\newblock {\em Physics in Medicine and Biology}, 37(3):731--749, March 1992.

\bibitem{Tomic2012}
M.~Tomi{\'{c}} and D.~Ser{\v{s}}i{\'{c}}.
\newblock Adaptive edge-preserving denoising by point-wise wavelet basis
  selection.
\newblock {\em {IET} Signal Processing}, 6(1):1, 2012.

\bibitem{Tomic2013}
M.~Tomi{\'{c}} and D.~Ser{\v{s}}i{\'{c}}.
\newblock Point-wise adaptive wavelet transform for signal denoising.
\newblock {\em Informatica}, 24(4):637--656, January 2013.

\bibitem{Vlasic2022}
T.~Vla{\v{s}}i{\'{c}} and D.~Ser{\v{s}}i{\'{c}}.
\newblock Sampling and reconstruction of sparse signals in shift-invariant
  spaces: Generalized shannon's theorem meets compressive sensing.
\newblock {\em {IEEE} Transactions on Signal Processing}, 70:438--451, 2022.

\bibitem{Wang2018a}
Y.~Wang, B.~Yu, Lei Wang, Chen Zu, David~S. Lalush, Weili Lin, Xi~Wu, Jiliu
  Zhou, Dinggang Shen, and Luping Zhou.
\newblock 3d conditional generative adversarial networks for high-quality {PET}
  image estimation at low dose.
\newblock {\em {NeuroImage}}, 174:550--562, July 2018.

\bibitem{Wolterink2017a}
J.~M. Wolterink, A.~M. Dinkla, M.~H.~F. Savenije, Peter~R. Seevinck, Cornelis
  A.~T. van~den Berg, and Ivana I{\v{s}}gum.
\newblock Deep {MR} to {CT} synthesis using unpaired data.
\newblock In {\em Simulation and Synthesis in Medical Imaging}, pages 14--23.
  Springer International Publishing, 2017.

\bibitem{Xie2020a}
Z.~Xie, R.~Baikejiang, T.~Li, X.~Zhang, K.~Gong, M.~Zhang, W.~Qi, E.~Asma, and
  J.~Qi.
\newblock Generative adversarial network based regularized image reconstruction
  for {PET}.
\newblock {\em Physics in Medicine {\&} Biology}, 65(12):125016, June 2020.

\bibitem{Yamaya2008}
T.~Yamaya, E.~Yoshida, T.~Obi, H.~Ito, K.~Yoshikawa, and H.~Murayama.
\newblock First human brain imaging by the {jPET}-d4 prototype with a
  pre-computed system matrix.
\newblock {\em {IEEE} Transactions on Nuclear Science}, 55(5):2482--2492,
  October 2008.

\bibitem{Yang2020GAN}
X.~Yang, M.~Kahnt, D.~Br\"{u}ckner, A.~Schropp, Y.~Fam, J.~Becher, J.~D.
  Grunwaldt, T.~L. Sheppard, and C.~G. Schroer.
\newblock Tomographic reconstruction with a generative adversarial network.
\newblock {\em Journal of Synchrotron Radiation}, 27(2):486--493, February
  2020.

\bibitem{Zhang2017a}
K.~Zhang, W.~Zuo, Y.~Chen, D.~Meng, and L.~Zhang.
\newblock Beyond a gaussian denoiser: Residual learning of deep {CNN} for image
  denoising.
\newblock {\em {IEEE} Transactions on Image Processing}, 26(7):3142--3155, July
  2017.

\bibitem{Zhu2013a}
Z.~Zhu, K.~Wahid, Paul Babyn, David Cooper, Isaac Pratt, and Yasmin Carter.
\newblock Improved compressed sensing-based algorithm for sparse-view {CT}
  image reconstruction.
\newblock {\em Computational and Mathematical Methods in Medicine}, 2013:1--15,
  2013.

\end{thebibliography}
\clearpage
\appendix

\section{Rotation of the exact crystal-to-crystal response}

\begin{figure}[H]
\begin{subfigure}{0.32\textwidth}
  \centering
  \includegraphics[width=1\linewidth]{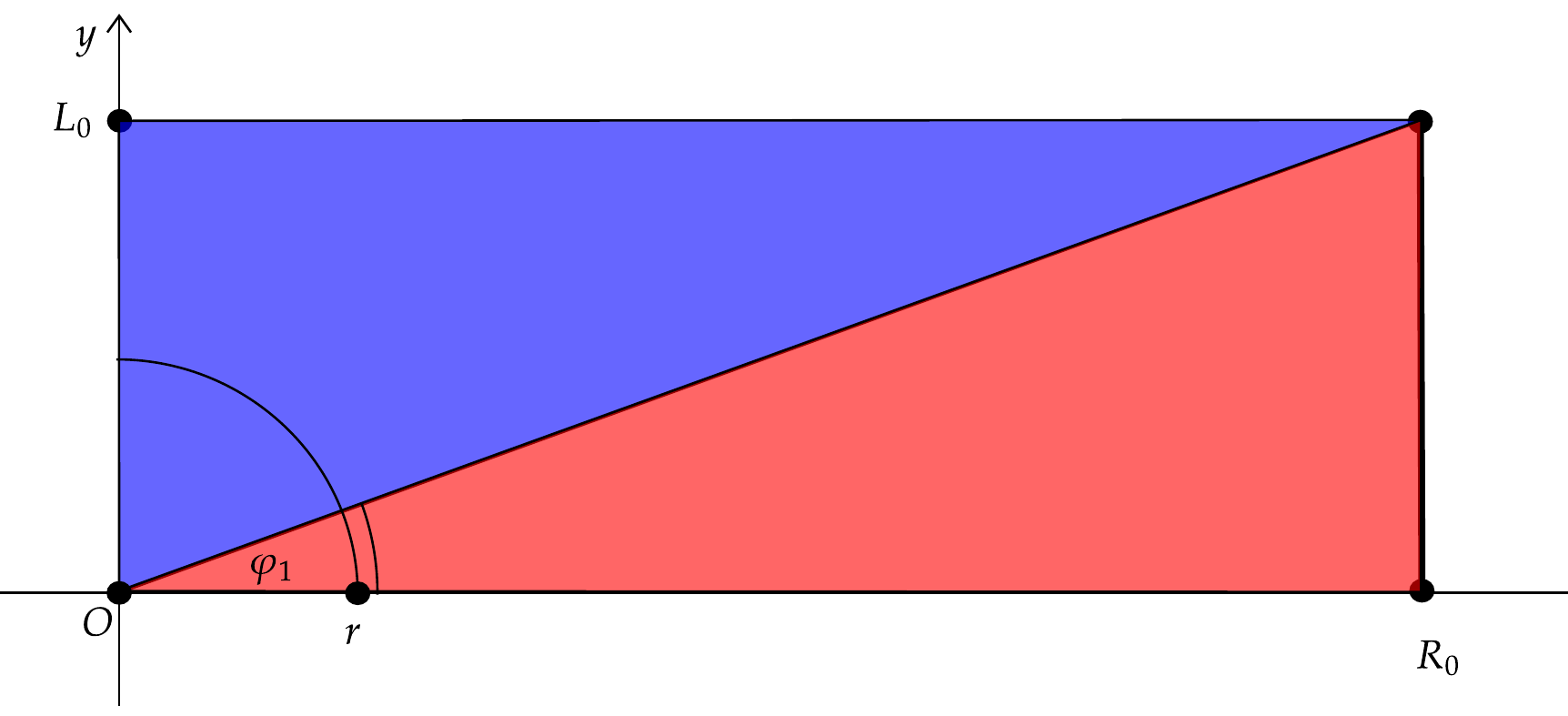}  
  \caption{$r \in \left[0,L_0\right]$}
  \label{fig:case1}
\end{subfigure}
\begin{subfigure}{0.32\textwidth}
  \centering
  \includegraphics[width=1\linewidth]{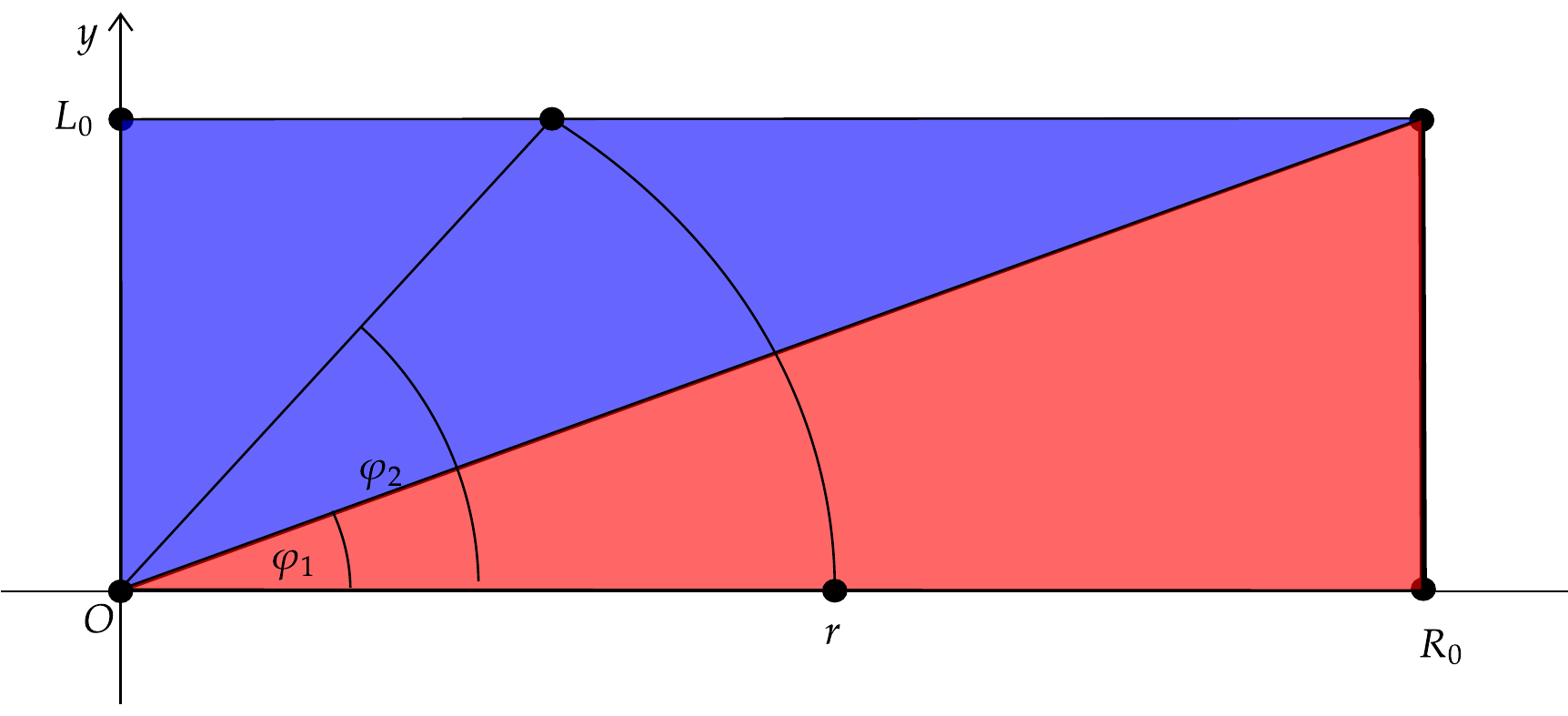}  
  \caption{$r \in \left<L_0,R_0\right]$}
  \label{fig:case2}
\end{subfigure}
\begin{subfigure}{0.32\textwidth}
  \centering
  \includegraphics[width=1\linewidth]{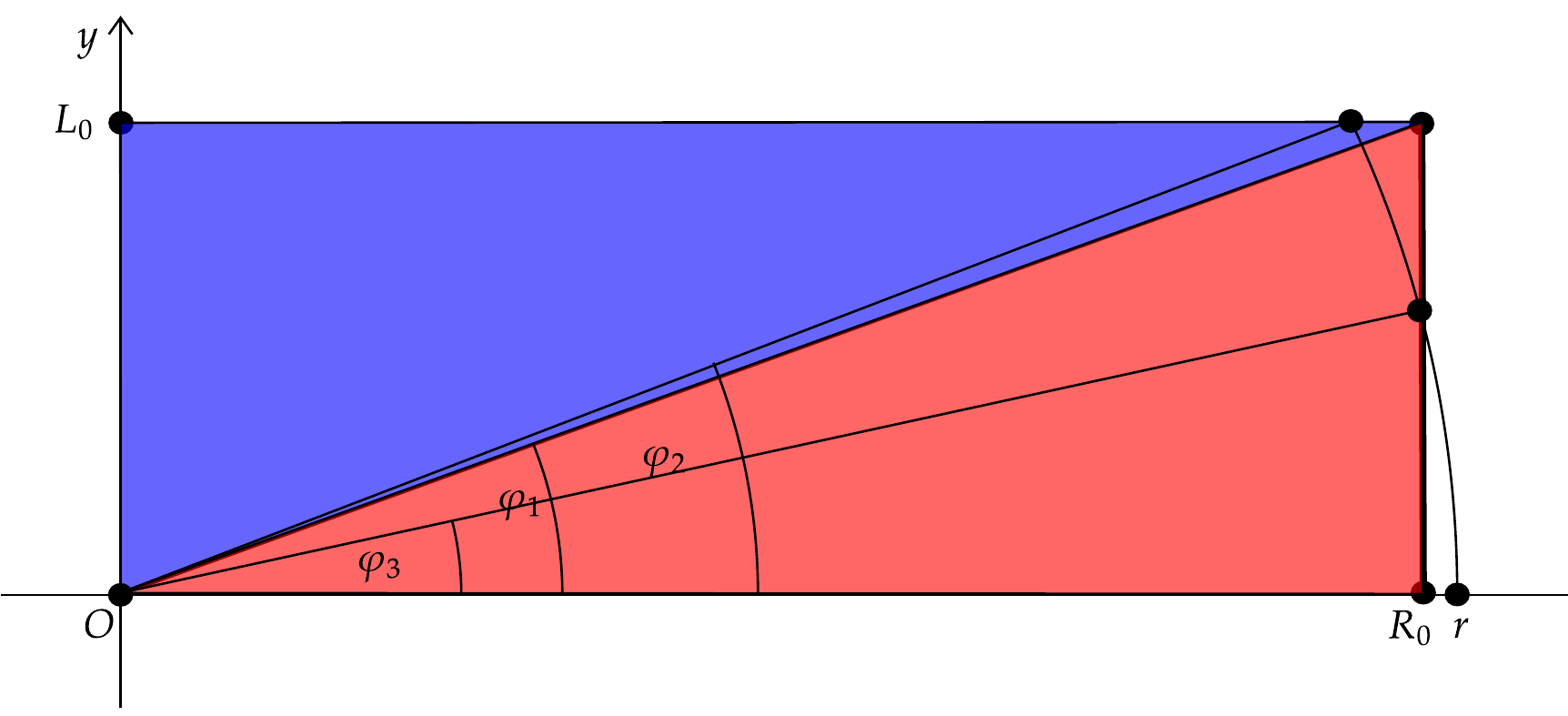}  
  \caption{$r \in \left<R_0,\sqrt{R_0^2+L_0^2}\right]$}
  \label{fig:case3}
\end{subfigure}
\caption{Nontrivial cases when calculating $I(r)$. Cases are split into three segments by respect to the radius r.}
\label{fig:cases}
\end{figure}

Appendix A is closely related to Paragraph 2.1. We present a step-by-step procedure that results in a closed form solution of our integral expression (\ref{eq:I}). It describes rotation of the tent-like PDF around the origin (shift $h$ is zero). We focus our attention on the first quadrant. In Fig. \ref{fig:cases} we can see two regions, each corresponding to different expression. We denote the first region of the first quadrant as $R_{1} = \{ (r,\varphi)\in \mathbb{R}_0^+ \times \left[0, \frac{\pi}{2} \right] \quad \vert \quad \sin(\varphi) \leq \frac{L_0}{R_0} \cos(\varphi), r\cos(\varphi)\leq R_0,  r\sin(\varphi)\leq L_0 \}$ and the second region as $R_{2} = \{ (r,\varphi)\in \mathbb{R}_0^+ \times \left[0, \frac{\pi}{2} \right] \quad \vert \quad \sin(\varphi) > \frac{L_0}{R_0} \cos(\varphi), r\cos(\varphi)\leq R_0,  r\sin(\varphi)\leq L_0 \}$. Let us recall that the non-normalized PDF for the first quadrant in polar coordinates is:
\begin{equation}
P(r,\varphi)=
\begin{cases}
\frac{R_0}{R_0+r\cos(\varphi)}, & (r,\phi)\in R_{1} \\
\frac{L_0 - r\sin(\varphi)}{L_0} \frac{R_0^2}{R_0^2 - r^2\cos^2(\varphi)}, & (r,\phi)\in R_{2} \\
0, & elsewhere.
\end{cases}
\label{eq:PDF}
\end{equation}
Our goal is to calculate 

\begin{equation}
I(r) = \int \displaylimits _0^{\frac{\pi}{2}} P(r,\varphi)d\varphi.
\label{eq:I}
\end{equation}
If we assume that $L_0 < R_0$, which is a reasonable assumption due to geometry of a PET scanner and known dimensions of scintillating crystals, we can divide calculation into four segments by respect to the radius $r$: 
\begin{itemize}
  \item $r \in \left[0,L_0\right]$ shown in Fig. (\ref{fig:case1})
  \item $r \in \left<L_0,R_0\right]$ shown in Fig. (\ref{fig:case2})
  \item $r \in \left<R_0,\sqrt{R_0^2+L_0^2}\right]$ shown in Fig. (\ref{fig:case3})
  \item $r \in \left<\sqrt{R_0^2+L_0^2},+\infty \right>$
\end{itemize}
The last part is the simplest: $r>\sqrt{R_0^2+L_0^2}$ is outside of both regions, so $I(r) = 0$.

We define two auxiliary functions:
\begin{equation}
I_1(r;\varphi_x,\varphi_y) = \int \displaylimits _{\varphi_x}^{\varphi_y} \frac{R_0}{R_0-r\cos(\varphi)} d\varphi,
\label{eq:I1}
\end{equation}
and
\begin{equation}
I_2(r;\varphi_x,\varphi_y) = \int \displaylimits _{\varphi_x}^{\varphi_y}\frac{L_0 - r\sin(\varphi)}{L_0} \frac{R_0^2}{R_0^2-r^2\cos^2(\varphi)} d\varphi.
\label{eq:I2}
\end{equation}
Function $I(r)$ can be expressed in terms of $I_1(r;\varphi_x,\varphi_y)$ and $I_2(r;\varphi_x,\varphi_y)$:

\begin{equation}
I(r) = \begin{cases}
I_1(r;0,\varphi_1) + I_2(r;\varphi_1,\frac{\pi}{2}), & r \in \left[0,L_0\right],\\
I_1(r;0,\varphi_1) + I_2(r;\varphi_1,\varphi_2), & r \in \left<L_0,R_0\right],\\
I_1(r;\varphi_3,\varphi_1) + I_2(r;\varphi_1,\varphi_2), & r \in \left<R_0,\sqrt{R_0^2+L_0^2}\right],\\
0, & r \in \left<\sqrt{R_0^2+L_0^2},+\infty \right>,
\end{cases}
\label{eq:IasI1I2}
\end{equation}
where $\varphi_1=\arctan(\frac{L_0}{R_0})$, $\varphi_2=\arcsin(\frac{L_0}{r})$ and $\varphi_3=\arccos(\frac{R_0}{r})$, as shown in Fig. \ref{fig:cases}.

At first, we calculate indefinite integral corresponding to $I_1(r;\varphi_x,\varphi_y)$ and $I_2(r;\varphi_x,\varphi_y)$.
For $I_1$ we have
\begin{equation}
\int \frac{1}{1+\frac{r}{R_0}\cos{\varphi}}d\varphi=\bigg\{t=\tan(\frac{\varphi}{2})\bigg\} = \frac{2R_0}{R_0-r} \int \frac{dt}{\frac{R_0+r}{R_0-r}+t^2}.
\end{equation}
Now, we have to split it into two cases: $r<R_0$ and $r>R_0$.
If $r<R_0$, we get:
\begin{equation}
\frac{2R_0}{R_0-r} \int \frac{dt}{\sqrt{\frac{R_0+r}{R_0-r}}^2+t^2}=\frac{2R_0}{\sqrt{R_0^2-r^2}}\arctan(\sqrt{\frac{R_0-r}{R_0+r}}\tan(\frac{\phi}{2})) + C.
\end{equation}
For $r>R_0$, we have:
\begin{equation}
\frac{2R_0}{r-R_0} \int \frac{dt}{\sqrt{\frac{r+R_0}{r-R_0}}^2-t^2} = \frac{1}{\sqrt{r^2-R_0^2}} \log \left(\frac{\tan(\frac{\varphi}{2}) + \sqrt{\frac{r+R_0}{r-R_0}}}{\tan(\frac{\varphi}{2}) - \sqrt{\frac{r+R_0}{r-R_0}}} \right) + C.
\end{equation}

For completeness, we need to check $r=R_0$ case:
\begin{equation}
\int \frac{1}{1+\cos{\varphi}}d\varphi = \tan(\frac{\varphi}{2}) + C,
\end{equation}
since  $\lim\limits_{r\to R_0^-} \frac{2R_0}{\sqrt{R_0^2-r^2}}\arctan(\sqrt{\frac{R_0-r}{R_0+r}}\tan(\frac{\phi}{2})) = \lim\limits_{r\to R_0^+} \frac{1}{\sqrt{r^2-R_0^2}} \log \left(\frac{\tan(\frac{\varphi}{2}) + \frac{r+R_0}{r-R_0}}{\tan(\frac{\varphi}{2}) - \frac{r+R_0}{r-R_0}} \right) = \tan(\frac{\varphi}{2})$.

The limits were calculated by successive application of the l'Hospital rule. Finally, collected results are:

\begin{equation}
\begin{gathered}
\int \frac{1}{1+\frac{r}{R_0}\cos(\varphi)}d\varphi=
\begin{cases}
\frac{2R_0}{\sqrt{R_0^2-r^2}}\arctan(\sqrt{\frac{R_0-r}{R_0+r}}\tan(\frac{\phi}{2})) + C,& r<R_0,\\
\tan(\frac{\varphi}{2}) + C, & r=R_0,\\
\frac{1}{\sqrt{r^2-R_0^2}} \log \left(\frac{\tan(\frac{\varphi}{2}) + \sqrt{\frac{r+R_0}{r-R_0}}}{\tan(\frac{\varphi}{2}) - \sqrt{\frac{r+R_0}{r-R_0}}} \right)+ C,& r>R_0.
\end{cases}
\end{gathered}
\end{equation}

Indefinite integral corresponding to $I_2(r;\varphi_x,\varphi_y)$ can be calculated as
\begin{equation}
\begin{gathered}
\int \frac{1-\frac{r}{L_0}\sin(\varphi)}{1-\frac{r^2}{R_0^2}\cos^2(\varphi)}d\varphi=\bigg\{t=\tan(\frac{\varphi}{2})\bigg\}=\frac{2R_0^2}{R_0^2-r^2}\int \frac{t^2- \frac{2r}{L_0}t+1}{t^4+2\frac{R_0^2+r^2}{R_0^2-r^2}t^2+1}dt
=\\ \frac{2R_0^2}{R_0^2-r^2}\int \frac{t^2- \frac{2r}{L_0}t+1}{(t^2+\frac{R_0+r}{R_0-r})(t^2+\frac{R_0-r}{R_0+r})}dt = \frac{2R_0^2}{R_0^2-r^2} \bigg(  \frac{1}{2L_0}\frac{R_0^2-r^2}{R_0} \int \frac{tdt}{t^2+\frac{R_0+r}{R_0-r}} + \\ \frac{R_0+r}{2R_0} \int \frac{dt}{t^2+\frac{R_0+r}{R_0-r}} - \frac{1}{2L_0}\frac{R_0^2-r^2}{R_0} \int \frac{tdt}{t^2+\frac{R_0-r}{R_0+r}} + \frac{R_0-r}{2R_0} \int \frac{dt}{t^2+\frac{R_0-r}{R_0+r}}  \bigg)
\end{gathered}
\label{eq:a}
\end{equation}
The result is obtained by decomposition in partial fractions. Again, we split it into two cases.
For $r<R_0$, we get:
\begin{equation}
\begin{gathered}
\frac{R_0}{2L_0}\log\left( \frac{\tan^2(\frac{\varphi}{2}) + \frac{R_0+r}{R_0-r}}{\tan^2(\frac{\varphi}{2}) + \frac{R_0-r}{R_0+r}} \right) + \frac{R_0}{\sqrt{R_0^2-r^2}} \Bigg( \arctan\bigg[\sqrt{\frac{R_0+r}{R_0-r}} \tan(\frac{\varphi}{2})\bigg] +\\ \arctan\bigg[\sqrt{\frac{R_0-r}{R_0+r}} \tan(\frac{\varphi}{2})\bigg]\Bigg) + C
\end{gathered}
\end{equation}

For $r>R_0$, we modify the partial fractions in (\ref{eq:a}) to obtain some simple integrals with known solutions:
\begin{equation}
\begin{gathered}
\frac{2R_0^2}{R_0^2-r^2}\Bigg( \frac{1}{2L_0}\frac{R_0^2-r^2}{R_0} \int \frac{tdt}{t^2-\frac{r+R_0}{r-R_0}} + \frac{R_0+r}{2R_0} \int \frac{dt}{t^2-\frac{r+R_0}{r-R_0}} -  \frac{1}{2L_0} \frac{R_0^2-r^2}{R_0}\times \\ \int \frac{tdt}{t^2-\frac{r-R_0}{r+R_0}} + \frac{R_0-r}{2R_0} \int \frac{dt}{t^2-\frac{r-R_0}{r+R_0}}  \Bigg) =
\frac{R_0}{2L_0}\log\left(\frac{\tan^2(\frac{\varphi}{2}) - \frac{r+R_0}{r-R_0}}{\tan^2(\frac{\varphi}{2}) - \frac{r-R_0}{r+R_0}} \right) \\ + \frac{R_0}{2\sqrt{r^2-R_0^2}} \log\left(\Bigg|\frac{\tan^2(\frac{\varphi}{2}) + \frac{2R_0}{\sqrt{r^2-R_0^2}}\tan(\frac{\varphi}{2}) - 1 }{\tan^2(\frac{\varphi}{2}) - \frac{2R_0}{\sqrt{r^2-R_0^2}}\tan(\frac{\varphi}{2}) - 1}      \Bigg|      \right) +C
\end{gathered}
\end{equation}
For completeness, we check the special case $r=R_0$:
\begin{equation}
\begin{gathered}
\int \frac{1-\frac{R_0}{L_0}\sin(\varphi)}{1-\cos^2(\varphi)} = -\cot(\phi)-\frac{R_0}{L_0}\log(\tan(\frac{\varphi}{2})) + C.
\end{gathered}
\end{equation}
Similarly, by application of the l'Hospital rule, it can be shown that:
\begin{equation}
\begin{gathered}
\lim\limits_{r\to R_0^-} \Big[ \frac{R_0}{2L_0}\log\left( \frac{\tan^2(\frac{\varphi}{2}) + \frac{R_0+r}{R_0-r}}{\tan^2(\frac{\varphi}{2}) + \frac{R_0-r}{R_0+r}} \right) + \frac{R_0}{\sqrt{R_0^2-r^2}} \big( \arctan\bigg[\sqrt{\frac{R_0+r}{R_0-r}} \tan(\frac{\varphi}{2})\bigg] +\\ \arctan\bigg[\sqrt{\frac{R_0-r}{R_0+r}} \tan(\frac{\varphi}{2})\bigg]\big)\Big] = \lim\limits_{r\to R_0^+} \Big[ \frac{R_0}{2L_0}\log\left(\frac{\tan^2(\frac{\varphi}{2}) - \frac{r+R_0}{r-R_0}}{\tan^2(\frac{\varphi}{2}) - \frac{r-R_0}{r+R_0}} \right)\\ + \frac{R_0}{2\sqrt{r^2-R_0^2}} \log\left(\Bigg|\frac{\tan^2(\frac{\varphi}{2}) + \frac{2R_0}{\sqrt{r^2-R_0^2}}\tan(\frac{\varphi}{2}) - 1 }{\tan^2(\frac{\varphi}{2}) - \frac{2R_0}{\sqrt{r^2-R_0^2}}\tan(\frac{\varphi}{2}) - 1}      \Bigg|      \right)\Big]=\\ -\cot(\phi)-\frac{R_0}{L_0}\log(\tan(\frac{\varphi}{2}))
\end{gathered}
\end{equation}
Again, we collect the results:
\begin{equation}
\begin{gathered}
\int \frac{1-\frac{r}{L_0}\sin(\varphi)}{1-\frac{r^2}{R_0^2}\cos^2(\varphi)}d\varphi=\\
\begin{cases}
\frac{R_0}{2L_0}\log\left( \frac{\tan^2(\frac{\varphi}{2}) + \frac{R_0+r}{R_0-r}}{\tan^2(\frac{\varphi}{2}) + \frac{R_0-r}{R_0+r}} \right) + \frac{R_0}{\sqrt{R_0^2-r^2}} \Big( \arctan\bigg[\sqrt{\frac{R_0+r}{R_0-r}} \tan(\frac{\varphi}{2})\bigg] +&\\ \arctan\bigg[\sqrt{\frac{R_0-r}{R_0+r}} \tan(\frac{\varphi}{2})\bigg]\Big) + C,& r<R_0\\
-\cot(\phi)-\frac{R_0}{L_0}\log(\tan(\frac{\varphi}{2})) + C , & r=R_0\\
\frac{R_0}{2L_0}\log\left(\frac{\tan^2(\frac{\varphi}{2}) - \frac{r+R_0}{r-R_0}}{\tan^2(\frac{\varphi}{2}) - \frac{r-R_0}{r+R_0}} \right) + &\\ \frac{R_0}{2\sqrt{r^2-R_0^2}} \log\left(\Bigg|\frac{\tan^2(\frac{\varphi}{2}) + \frac{2R_0}{\sqrt{r^2-R_0^2}}\tan(\frac{\varphi}{2}) - 1 }{\tan^2(\frac{\varphi}{2}) - \frac{2R_0}{\sqrt{r^2-R_0^2}}\tan(\frac{\varphi}{2}) - 1}      \Bigg|      \right) +C,& r>R_0
\end{cases}
\end{gathered}
\end{equation}
\vspace{15pt}
\hspace{10pt}Now, we return to Eq. \ref{eq:IasI1I2}. At first, we highlight some important equations needed for calculation of $I(r)$:
\begin{equation}
\tan[\frac{1}{2}\arctan(\frac{L_0}{R_0})] = \frac{\sqrt{L_0^2+R_0^2}-R_0}{L_0}
\end{equation}
\begin{equation}
\tan^2[\frac{1}{2}\arctan(\frac{L_0}{R_0})] = \frac{L_0^2+2R_0^2-2R_0\sqrt{L_0^2+R_0^2}}{L_0^2}
\end{equation}
\begin{equation}
\tan[\frac{1}{2}\arcsin(\frac{L_0}{r})] = \frac{L_0}{\sqrt{r^2-R_0^2}+r}
\end{equation}
\begin{equation}
\tan^2[\frac{1}{2}\arcsin(\frac{L_0}{r})] = \frac{L_0^2}{2r^2-L_0^2+2r\sqrt{r^2-R_0^2}}
\end{equation}
\begin{equation}
\tan[\frac{1}{2}\arccos(\frac{R_0}{r})] = \sqrt{\frac{r-R_0}{r+R_0}}
\end{equation}
For $r\in[0,L_0]$ we get:
\begin{equation}
\begin{gathered}
I_1(r;0,\arctan(\frac{L_0}{R_0})) + I_2(r;\arctan(\frac{L_0}{R_0}),\frac{\pi}{2}) =\\ \frac{2R_0}{\sqrt{R_0^2-r^2}} \arctan\bigg(\sqrt{\frac{R_0-r}{R_0+r} }\frac{\sqrt{L_0^2+R_0^2}-R_0}{L_0}\bigg) \\+ \frac{R_0}{2L_0} \bigg(\log(\frac{1+\frac{R_0+r}{R_0-r}}{1+\frac{R_0-r}{R_0+r}})-\log(\frac{\frac{L_0^2+2R_0^2-2R_0\sqrt{L_0^2+R_0^2}}{L_0^2}+\frac{R_0+r}{R_0-r}}{\frac{L_0^2+2R_0^2-2R_0\sqrt{L_0^2+R_0^2}}{L_0^2}+\frac{R_0-r}{R_0+r}}) \bigg)+\\
\frac{R_0}{\sqrt{R_0^2-r^2}} \bigg( \arctan\bigg(\sqrt{\frac{R_0+r}{R_0-r}}\bigg) + \arctan\bigg(\sqrt{\frac{R_0-r}{R_0+r}} \bigg) - \\ \arctan\bigg(\sqrt{\frac{R_0+r}{R_0-r}} \frac{\sqrt{L_0^2+R_0^2}-R_0}{L_0}\bigg) - \arctan\bigg(\sqrt{\frac{R_0-r}{R_0+r}}\frac{\sqrt{L_0^2+R_0^2}-R_0}{L_0}\bigg) \bigg) \\  = \frac{2R_0}{\sqrt{R_0^2-r^2}} \arctan(\sqrt{\frac{R_0-r}{R_0+r}} \frac{\sqrt{L_0^2+R_0^2}-R_0}{L_0}) +  \frac{R_0}{2L_0} \times \\ \log(\frac{L_0^2-(R_0+r)(\sqrt{L_0^2+R_0^2}-R_0) }{ L_0^2-(R_0-r)(\sqrt{L_0^2+R_0^2}-R_0)})  + \frac{R_0}{\sqrt{R_0^2-r^2}}(\frac{\pi}{2} - \arctan(\frac{L_0}{\sqrt{R_0^2-r^2}}))
\end{gathered}
\label{eq121}
\end{equation}
Similarly, for $r \in \left<L_0,R_0\right]$:
\begin{equation}
\begin{gathered}
I_1(r;0,\arctan(\frac{L_0}{R_0})) + I_2(r;\arctan(\frac{L_0}{R_0}),\arcsin(\frac{L_0}{r})) = \\ \frac{2R_0}{\sqrt{R_0^2-r^2}} \arctan\bigg(\sqrt{\frac{R_0-r}{R_0+r} }\frac{\sqrt{L_0^2+R_0^2}-R_0}{L_0}\bigg) + \\ \frac{R_0}{2L_0} \bigg(\log\bigg(\frac{\frac{L_0^2}{2r^2-L_0^2+2r\sqrt{r^2-R_0^2}}+\frac{R_0+r}{R_0-r}}{\frac{L_0^2}{2r^2-L_0^2+2r\sqrt{r^2-R_0^2}}+\frac{R_0-r}{R_0+r}}\bigg)-\log\bigg(\frac{\frac{L_0^2+2R_0^2-2R_0\sqrt{L_0^2+R_0^2}}{L_0^2}+\frac{R_0+r}{R_0-r}}{\frac{L_0^2+2R_0^2-2R_0\sqrt{L_0^2+R_0^2}}{L_0^2}+\frac{R_0-r}{R_0+r}}\bigg) \bigg) \\ +
 \frac{R_0}{\sqrt{R_0^2-r^2}}  \bigg( \arctan\bigg(\sqrt{\frac{R_0+r}{R_0-r}} \frac{L_0}{\sqrt{r^2-R_0^2}+r}\bigg) +\\ \arctan\bigg(\sqrt{\frac{R_0-r}{R_0+r}}\frac{L_0}{\sqrt{r^2-R_0^2}+r} \bigg)  - \arctan\bigg(\sqrt{\frac{R_0+r}{R_0-r}} \frac{\sqrt{L_0^2+R_0^2}-R_0}{L_0}\bigg) \\ -\arctan\bigg(\sqrt{\frac{R_0-r}{R_0+r}}\frac{\sqrt{L_0^2+R_0^2}-R_0}{L_0}\bigg) \bigg) =\\  \frac{R_0}{2L_0}\log( \frac{R_0+\sqrt{r^2-L_0^2}}{R_0-\sqrt{r^2-L_0^2}} \cdot \frac{\sqrt{L_0^2+R_0^2}-r}{\sqrt{L_0^2+R_0^2}+r} ) + \frac{2R_0}{\sqrt{R_0^2-r^2}} \times \\ \arctan(\frac{\sqrt{R_0^2-r^2}}{L_0}\frac{L_0^2+(\sqrt{R_0^2+L_0^2}-R_0)(\sqrt{r^2-L_0^2}+r)}{(\sqrt{r^2-L_0^2}+r)(R_0+r)-(\sqrt{R_0^2+L_0^2}-R_0)(R_0-r)})
\end{gathered}
\label{eq122}
\end{equation}
Finally, for $r \in \left<R_0,\sqrt{R_0^2+L_0^2}\right]$:
\begin{equation}
\begin{gathered}
I_1(r;\arccos(\frac{R_0}{r}),\arctan(\frac{L_0}{R_0})) + I_2(r;\arctan(\frac{L_0}{R_0}),\arcsin(\frac{L_0}{r})) = \\  \frac{R_0}{\sqrt{R_0^2-r^2}}\bigg( \log\left( \frac{ \frac{\sqrt{L_0^2+R_0^2}-R_0}{L_0}+\sqrt{\frac{r+R_0}{r-R_0}} }{\frac{\sqrt{L_0^2+R_0^2}-R_0}{L_0}-\sqrt{\frac{r+R_0}{r-R_0}} } \right)  - \log\left( \frac{ \sqrt{\frac{r-R_0}{r+R_0}}+\sqrt{\frac{r+R_0}{r-R_0}} }{\sqrt{\frac{r-R_0}{r+R_0}}-\sqrt{\frac{r+R_0}{r-R_0}} } \right) \bigg) \\ + 
\frac{R_0}{2L_0}\bigg( \log\left(\frac{\frac{L_0^2}{2r^2-L_0^2+2r\sqrt{r^2-L_0^2}}-\frac{r+R_0}{r-R_0}}{\frac{L_0^2}{2r^2-L_0^2+2r\sqrt{r^2-L_0^2}}-\frac{r-R_0}{r+R_0}} \right) - \log\left(\frac{ \frac{L_0^2+2R_0^2-2R_0\sqrt{R_0^2+L_0^2}}{L_0^2} -\frac{r+R_0}{r-R_0}}{ \frac{L_0^2+2R_0^2-2R_0\sqrt{R_0^2+L_0^2}}{L_0^2}-\frac{r-R_0}{r+R_0}} \right)    \bigg) \\ +
\frac{R_0}{2\sqrt{r^2-R_0^2}} \bigg(  \log\left( \frac{\frac{L_0^2}{2r^2-L_0^2+2r\sqrt{r^2-L_0^2}}+\frac{2R_0}{\sqrt{r^2-R_0^2}}\frac{L_0}{\sqrt{r^2-R_0^2}+r} -1  }{ \frac{L_0^2}{2r^2-L_0^2+2r\sqrt{r^2-L_0^2}}-\frac{2R_0}{\sqrt{r^2-R_0^2}}\frac{L_0}{\sqrt{r^2-R_0^2}+r} -1 }  \right) \\- \log\left( \frac{\frac{L_0^2 + 2R_0^2 - 2R_0\sqrt{R_0^2+L_0^2}}{L_0^2}+\frac{2R_0}{\sqrt{r^2-R_0^2}}\frac{\sqrt{R_0^2+L_0^2}-R_0}{L_0} -1   }{ \frac{L_0^2 + 2R_0^2 - 2R_0\sqrt{R_0^2+L_0^2}}{L_0^2}-\frac{2R_0}{\sqrt{r^2-R_0^2}}\frac{\sqrt{R_0^2+L_0^2}-R_0}{L_0} -1  }  \right)   \bigg) \\ = \frac{R_0}{\sqrt{r^2-R_0^2}} \log(\frac{R_0}{r} \frac{ \sqrt{L_0^2+R_0^2}-R_0+L_0 \sqrt{\frac{r+R_0}{r-R_0}} }{ \sqrt{L_0^2+R_0^2}-R_0-L_0 \sqrt{\frac{r+R_0}{r-R_0}} } ) + \\ \frac{R_0}{2L_0} \log( \frac{R_0+\sqrt{r^2-L_0^2}}{R_0-\sqrt{r^2-L_0^2}} \cdot \frac{\sqrt{L_0^2+R_0^2}-r}{\sqrt{L_0^2+R_0^2}+r})  +   \frac{R_0}{2\sqrt{r^2-R_0^2}} g(r)
\end{gathered} 
\label{eq123}
\end{equation}
where 
\begin{equation}
\begin{gathered}
g(r) = \scriptstyle{ \log \bigg| \frac{ \sqrt{r^2-R_0^2} + L_0 }{ \sqrt{r^2-R_0^2} - L_0 }}{\frac{  L_0^3R_0 - 2R_0L_0(r+\sqrt{r^2-L_0^2}) +2r^2(r+\sqrt{r^2-L_0^2})\sqrt{r^2-R_0^2} - L_0^2(2r-\sqrt{r^2-L_0^2})\sqrt{r^2-R_0^2}  }{    -L_0^3R_0 + 2R_0L_0(r+\sqrt{r^2-L_0^2}) +2r^2(r+\sqrt{r^2-L_0^2})\sqrt{r^2-R_0^2} - L_0^2(2r+\sqrt{r^2-L_0^2})\sqrt{r^2-R_0^2} }\bigg|}.
\end{gathered}
\notag
\end{equation}
Expressions in (\ref{eq121}), (\ref{eq122}), and (\ref{eq123}) are in agreement with (\ref{eq:expre_around_center}) which concludes its mathematical derivation.

\section{Triangular crystal-to-crystal response approximation}

Appendix B is closely related to Paragraph 3.2. Here, we calculate an approximation of the response between two crystals. We remind the reader that we defined a helper function $J(r,l)=\frac{1}{\pi}\frac{1}{\sqrt{r^2-l^2}}\mu(r-l)$.

The main challenge is to calculate $I_1(r) = \displaystyle \int \displaylimits_a^b J(r,l)dl$ and $I_2(r) = \displaystyle \int \displaylimits_a^b l \cdot J(r,l)dl$ for $0 \leq a \leq b$. We start with $I_1(r)$.\\
If $r\leq a$, then 
\begin{equation}
\displaystyle \int \displaylimits_a^b J(r,l)dl = \displaystyle \int \displaylimits_a^b \frac{1}{\pi}\frac{1}{\sqrt{r^2-l^2}}\mu(r-l)dl = \bigg( \mu(r-l)=0 \bigg) = 0
\label{eq:rmanje}.
\end{equation}
If $a<r<b$, then 
\begin{equation}
\displaystyle \int \displaylimits_a^b J(r,l)dl = \displaystyle \int \displaylimits_a^b \frac{1}{\pi}\frac{1}{\sqrt{r^2-l^2}}\mu(r-l)dl = \displaystyle \int \displaylimits_a^r \frac{1}{\pi}\frac{1}{\sqrt{r^2-l^2}}dl = \frac{1}{\pi} \bigg(  \frac{\pi}{2} -\arcsin(\frac{a}{r}) \bigg)
\label{eq:rjednako}.
\end{equation}
If $r\geq b$, then 
\begin{equation}
\begin{gathered}
\displaystyle \int \displaylimits_a^b J(r,l)dl = \displaystyle \int \displaylimits_a^b \frac{1}{\pi}\frac{1}{\sqrt{r^2-l^2}}\mu(r-l)dl = \displaystyle \int \displaylimits_a^b \frac{1}{\pi}\frac{1}{\sqrt{r^2-l^2}}dl =\\ \frac{1}{\pi} \bigg( \arcsin(\frac{b}{r}) -\arcsin(\frac{a}{r}) \bigg).
\end{gathered}
\label{eq:rvece}
\end{equation}

Let us show that $I_1(r) = Re\bigg\{\frac{1}{\pi} \bigg( \arcsin(\frac{b}{r}) -\arcsin(\frac{a}{r}) \bigg) \bigg\}$ for $r>0$. Here, we consider the complex inverse sine function. If we restrict the complex function $\arcsin$ to real arguments $x\in \mathbb{R}$, we have:

\begin{equation}
Re\big\{ \arcsin(x) \big\} = 
\begin{cases}\frac{\pi}{2}, & x>1\\ 
\arcsin_{Re}(x), & |x|\leq 1\\ - \frac{\pi}{2}, & x<-1 
\end{cases}.
\end{equation}

If $r\leq a$, then $\frac{b}{r}\geq \frac{a}{r}\geq 1$. Therefore, 
$I_1(r) = Re\bigg\{\frac{1}{\pi} \bigg( \arcsin(\frac{b}{r}) -\arcsin(\frac{a}{r}) \bigg) \bigg\} = 0 $, which corresponds to (\ref{eq:rmanje}).\\
If $a<r<b$, then $0<\frac{a}{r}<1$ and $\frac{b}{r}>1$. Thus, 
$I_1(r) = Re\bigg\{\frac{1}{\pi} \bigg( \arcsin(\frac{b}{r}) -\arcsin(\frac{a}{r}) \bigg) \bigg\} = \frac{1}{\pi} \bigg( \frac{\pi}{2} -\arcsin(\frac{a}{r}) \bigg)$, which corresponds to (\ref{eq:rjednako}).\\
If $r\geq b$, then  $0\leq \frac{a}{r} \leq \frac{b}{r}\leq 1$. Therefore, 
$I_1(r) = Re\bigg\{\frac{1}{\pi} \bigg( \arcsin(\frac{b}{r}) -\arcsin(\frac{a}{r}) \bigg) \bigg\} = \frac{1}{\pi} \bigg( \arcsin(\frac{b}{r}) -\arcsin(\frac{a}{r}) \bigg)$, which corresponds to (\ref{eq:rvece}).

\vspace{.3cm}
Similar technique we apply for calculation of $I_2(r)$.\\
If $r\leq a$, then 
\begin{equation}
\displaystyle \int \displaylimits_a^b l \cdot J(r,l)dl = \displaystyle \int \displaylimits_a^b \frac{1}{\pi}\frac{l}{\sqrt{r^2-l^2}}\mu(r-l)dl = \bigg( \mu(r-l)=0 \bigg) = 0
\label{eq:rmanje2}.
\end{equation}
If $a<r<b$, then 
\begin{equation}
\displaystyle \int \displaylimits_a^b l\cdot J(r,l)dl = \displaystyle \int \displaylimits_a^b \frac{1}{\pi}\frac{l}{\sqrt{r^2-l^2}}\mu(r-l)dl = \displaystyle \int \displaylimits_a^r \frac{1}{\pi}\frac{l}{\sqrt{r^2-l^2}}dl = \frac{1}{\pi} \sqrt{r^2-a^2} 
\label{eq:rjednako2}.
\end{equation}
If $r\geq b$, then 
\vspace{-.5cm}
\begin{equation}
\begin{gathered}
\displaystyle \int \displaylimits_a^b l \cdot J(r,l)dl = \displaystyle \int \displaylimits_a^b \frac{1}{\pi}\frac{l}{\sqrt{r^2-l^2}}\mu(r-l)dl = \displaystyle \int \displaylimits_a^b \frac{1}{\pi}\frac{l}{\sqrt{r^2-l^2}}dl = \\ \frac{1}{\pi} \bigg( \sqrt{r^2-a^2} -\sqrt{r^2-b^2} \bigg)
\end{gathered}
\label{eq:rvece2}.
\end{equation}
We claim that $I_2(r) = Re\bigg\{\frac{1}{\pi} \bigg( \sqrt{r^2-a^2} -\sqrt{r^2-b^2}  \bigg) \bigg\}$ for $r>0$. Here, we consider the complex square root function. If we restrict the complex square to real arguments $x\in \mathbb{R}$, it follows that $Re\big\{ \sqrt{x} \big\} = \begin{cases}0, & x<0\\ \sqrt{x}_{Re}, & x\geq 0 \end{cases}$. \\
If $r\leq a$, then $r^2-b^2\leq r^2-a^2\leq0$. So 
$I_1(r) = Re\bigg\{\frac{1}{\pi} \bigg( \sqrt{r^2-a^2} -\sqrt{r^2-b^2}  \bigg) \bigg\} = 0$, which corresponds to (\ref{eq:rmanje2}).\\
From $a<r<b$, it follows $r^2-b^2<0$ and $r^2-b^2>1$. Thus,  
$I_1(r) = Re\bigg\{\frac{1}{\pi} \bigg( \sqrt{r^2-a^2} -\sqrt{r^2-b^2}  \bigg) \bigg\} = \frac{1}{\pi} \sqrt{r^2-a^2}$, which corresponds to (\ref{eq:rjednako2}).\\
If $r\geq b$, then  $r^2-a^2\geq r^2-b^2 \geq 0$. Therefore,  
$I_1(r) = Re\bigg\{\frac{1}{\pi} \bigg( \sqrt{r^2-a^2} -\sqrt{r^2-b^2}  \bigg) \bigg\} = \frac{1}{\pi} \bigg( \sqrt{r^2-a^2} -\sqrt{r^2-b^2}  \bigg)$, which corresponds to (\ref{eq:rvece2}).

Now, we can easily calculate
$P_a(r;h,R_0,L_0) = \displaystyle \int_{0}^{+\infty} T(l)J(r,l)dl$, where we consider triangular weight function: 
\begin{equation}T(l) = \frac{1}{2L_0R_0}
\begin{cases}
1 - \frac{|l-h|}{L_0} &, |l-h|\leq L_0\\
0 &, elsewhere
\end{cases}.\end{equation}
\vspace{.3cm}
A detailed argument is given for $h\geq L_0$. The same arguments are valid for $h<L_0$.
We have:
\begin{equation}
\begin{gathered}
P_a(r;h,R_0,L_0) = \displaystyle \int_{0}^{+\infty} T(l)J(r,l)dl = \frac{1}{2L_0R_0} \bigg(  \displaystyle \int_{h-L_0}^{h} \frac{L_0+l-h}{L_0}J(r,l)dl +\\  \displaystyle \int_{h}^{h+L_0} \frac{L_0-l+h}{L_0}J(r,l)dl \bigg)=
\frac{1}{2L_0R_0} \bigg(\frac{1}{L_0} \bigg( (L_0-h) \displaystyle \int_{h-L_0}^{h} J(r,l)dl + \\  \displaystyle \int_{h-L_0}^{h} l\cdot J(r,l)dl+   (L_0+h)\displaystyle \int_{h}^{h+L_0} J(r,l)dl - \displaystyle \int_{h}^{h+L_0} l \cdot J(r,l)dl \bigg) \bigg)=\\
\frac{1}{2L_0R_0} Re\bigg\{ \frac{1}{\pi L_0} \bigg( \left( L_0+h \right) \arcsin\left(\frac{L_0+h}{r}\right) -2\cdot h\cdot \arcsin\left(\frac{h}{r}\right)  +  \left( L_0-h \right) \times \\ \arcsin\left(\frac{L_0-h}{r}\right) + 
\sqrt{r^2-(L_0+h)^2} - 2 \sqrt{r^2-h^2} + \sqrt{r^2-(L_0-h)^2}\bigg)\bigg\}.
\end{gathered}
\label{1234}
\end{equation}
By analogy, an expression for square window function $P_{a_2}(r;h,R_0,L_0) = \frac{1}{ 4 L_0 R_0 \pi} Re\bigg\{ \arcsin(\frac{h+L_0}{r}) - \arcsin(\frac{h-L_0}{r})\bigg\}$ can be derived. A careful reader will notice that it is necessary to calculate $I_1(r)$ only. 

Hence, we have proven the correctness of expressions (\ref{eq:proracunt1}) and (\ref{eq:proracunt2}) since they are equal to (\ref{1234}).

\section{Validation of triangular crystal-to-crystal response}
In this appendix, we prove that $P_a(r;0,R_0,L_0)$, according to expression (\ref{eq:proracunt2}), is a valid approximation. We remind the reader that $P_a(r;h=0,R_0,L_0)$ corresponds to the rotation of triangular approximation of the crystal-to-crystal response. In the sequel, we compare $P_a(r;0,R_0,L_0)$ with $P_r(r,R_0,L_0)$, which is an exact integral of the unshifted crystal-to-crystal response ($L_0<<R_0$), and show that they are approximately equal. We focus on $r\in \left[0, R_0 \right]$, since the object captured by the PET scanner is inside a circle of radius $R_0$.  We insert $h=0$ in (\ref{eq:proracunt2}):

\begin{equation}
P_a(r;0,R_0,L_0) = 
\begin{cases}
\frac{1}{2R_0L_0} - \frac{1}{\pi R_0L_0^2}r &, 0 \leq r \leq L_0\\
\frac{2}{\pi R_0L_0}\arctan(\frac{L_0}{\sqrt{r^2-L_0^2}+r}) + &\\ \frac{1}{\pi R_0L_0^2}(\sqrt{r^2-L_0^2}-r) &, L_0<r\leq R_0
\end{cases}
\label{eq:Pa0}.
\end{equation}
To confirm the approximation for the first interval, i.e. for $0 \leq r \leq L_0$, we calculate Taylor expansion of $P_r(r;R_0,L_0)$ at $r=0$, taking into the account $\frac{L_0}{R_0}<<1$:
\begin{equation}
\begin{gathered}
P_{r1}(r;R_0,L_0)=\frac{1}{\pi R_0 L_0}( \frac{2R_0}{\sqrt{R_0^2-r^2}} \arctan(\sqrt{\frac{R_0-r}{R_0+r}} \frac{\sqrt{L_0^2+R_0^2}-R_0}{L_0}) +\\ \frac{R_0}{2L_0}\log(\frac{L_0^2-(R_0+r)(\sqrt{L_0^2+R_0^2}-R_0) }{ L_0^2-(R_0-r)(\sqrt{L_0^2+R_0^2}-R_0)}) +\\ \frac{R_0}{\sqrt{R_0^2-r^2}}(\frac{\pi}{2} - \arctan(\frac{L0}{\sqrt{R_0^2-r^2}}))) = a_0 + a_1 r + \mathcal{O}(r^2),\\
a_0 = P_{r1}(0;R_0,L_0) = \\ \frac{1}{\pi R_0 L_0} (2\arctan(\frac{\sqrt{L_0^2+R_0^2}-R_0}{L_0}) + \frac{\pi}{2}-\arctan{\frac{L_0}{R_0}}) \approx \frac{1}{2R_0L_0},\\
a_1 = \displaystyle \frac{d}{dr} P_{r1}(r;R_0,L_0)\big|_{r=0} = -\frac{1}{\pi R_0 L_0^2}\frac{(R_0^2+L_0^2)(\sqrt{L_0^2 +R_0^2}-R_0)}{R_0(L_0^2+R_0^2-R_0\sqrt{L_0^2+R_0^2})} = \\ -\frac{1}{\pi R_0 L_0^2}\frac{(1+\frac{L_0^2}{R_0^2})(\sqrt{\frac{L_0^2}{R_0^2} +1}-1)}{\frac{L_0^2}{R_0^2}+1-\sqrt{\frac{L_0^2}{R_0^2}+1}} \approx -\frac{1}{\pi R_0 L_0^2}(1+\frac{L_0^2}{R_0^2}) \approx -\frac{1}{\pi R_0 L_0^2}.
\end{gathered}
\end{equation}
Line $l(r) = a_0+a_1\cdot r = \frac{1}{2R_0L_0} -\frac{1}{\pi R_0 L_0^2} r $ is the best fit among all linear functions that approximate $P_{r1}(0;R_0,L_0)$ around $r=0$. Notice that  $l(r)$ is exactly the same as the one in (\ref{eq:Pa0}). Hence, the approximation holds for $0 \leq r \leq L_0$.

For the second interval ($L_0< r \leq R_0$), we split the proof in two parts, each corresponding to one of the terms:

\begin{equation}
\begin{gathered}
\frac{2}{\pi L_0 \sqrt{R_0^2-r^2}} \times \\ \arctan\Big(\frac{\sqrt{R_0^2-r^2}}{L_0} \frac{L_0^2+(\sqrt{R_0^2+L_0^2}-R_0)(\sqrt{r^2-L_0^2}+r)}{(\sqrt{r^2-L_0^2}+r)(R_0+r)-(\sqrt{R_0^2+L_0^2}-R_0)(R_0-r)}\Big)
 \\ \approx \frac{2}{\pi R_0L_0}\arctan(\frac{L_0}{\sqrt{r^2-L_0^2}+r}), 
\label{eq:proof21}
\end{gathered}
\end{equation}
\begin{equation}
\centering
 \frac{1}{2 \pi L_0^2}\log(\frac{R_0+\sqrt{r^2-L_0^2}}{R_0-\sqrt{r^2-L_0^2}} \cdot \frac{\sqrt{L_0^2+R_0^2}-r}{\sqrt{L_0^2+R_0^2}+r}) \approx
 \frac{1}{\pi R_0L_0^2}(\sqrt{r^2-L_0^2}-r).
\label{eq:proof22}
\end{equation}
The following sequence of approximations proof correctness of (\ref{eq:proof21}):
\begin{equation}
\begin{gathered}
\frac{2}{\pi L_0 \sqrt{R_0^2-r^2}} \times\\ \arctan(\frac{\sqrt{R_0^2-r^2}}{L_0} \frac{L_0^2+(\sqrt{R_0^2+L_0^2}-R_0)(\sqrt{r^2-L_0^2}+r)}{(\sqrt{r^2-L_0^2}+r)(R_0+r)-(\sqrt{R_0^2+L_0^2}-R_0)(R_0-r)}) \approx \\
\frac{2}{\pi L_0 \sqrt{R_0^2-r^2}} \arctan(\frac{\sqrt{R_0^2-r^2}}{L_0} \frac{L_0^2+(\sqrt{R_0^2+L_0^2}-R_0)(\sqrt{r^2-L_0^2}+r)}{(\sqrt{r^2-L_0^2}+r)(R_0+r)}) \approx\\
\frac{2}{\pi L_0  R_0 \sqrt{1-\frac{r^2}{R_0^2}}} \arctan(\sqrt{1-\frac{r^2}{R_0^2}} \frac{L_0}{\sqrt{r^2-L_0^2}+r} \frac{R_0 + \frac{1}{2}(\sqrt{r^2-L_0^2}+r)}{R_0+r}) \approx\\
\frac{2}{\pi L_0  R_0 \sqrt{1-\frac{r^2}{R_0^2}}} \arctan(\sqrt{1-\frac{r^2}{R_0^2}} \frac{L_0}{\sqrt{r^2-L_0^2}+r}) \approx\\
\frac{2}{\pi L_0  R_0} \arctan(\frac{L_0}{\sqrt{r^2-L_0^2}+r}).
\end{gathered}
\label{eq:set_of_ap_1}
\end{equation}
The first approximation in (\ref{eq:set_of_ap_1}) is valid since $(\sqrt{r^2-L_0^2}+r)(R_0+r) >> (\sqrt{R_0^2+L_0^2}-R_0)(R_0-r)$. The left hand side expression is increasing, while the right hand side is decreasing by increasing $r$. If we evaluate the previous inequality at $r=L_0$ and show that is holds under the assumption that $\frac{L_0}{R_0}$ is small, then the inequality is true for all $r \in \left[L_0,R_0\right]$.
\begin{equation}
\centering
\begin{gathered}
(\sqrt{r^2-L_0^2}+r)(R_0+r)\big|_{r=L_0} >> (\sqrt{R_0^2+L_0^2}-R_0)(R_0-r)\big|_{r=L_0} \Rightarrow \\  L_0(R_0+L_0) >>\frac{L_0^2}{2R_0}(R_0-L_0) \Rightarrow 2\frac{R_0^2}{L_0^2} + \frac{R_0}{L_0} + 1 >> 0
\end{gathered}
\end{equation}
The second approximation in (\ref{eq:set_of_ap_1}) is valid since $\sqrt{R_0^2+L_0^2}-R_0\approx \frac{L_0^2}{2R_0}$. Next approximation in (\ref{eq:set_of_ap_1}) states that $\frac{R_0 + \frac{1}{2}(\sqrt{r^2-L_0^2}+r)}{R_0+r} \approx 1$. This is correct due to $\sqrt{r^2-L_0^2} \approx r$. 
To prove the last approximation in (\ref{eq:set_of_ap_1}), we are going to split it in two cases - when $r$ is close to $L_0$ ($r\approx L_0$) and when $r$ is much larger then $L_0$ ($r >> L_0$).
If $r$ is close to $L_0$, then $\sqrt{1-\frac{r^2}{R_0^2}} \approx 1$ and the last approximation holds.
If $r$ is much larger then $L_0$, then $\frac{L_0}{\sqrt{r^2-L_0^2}+r} \approx 0$. Therefore, we can use the approximation $\arctan(x) \approx x$ which concludes the first part of the proof.

\begin{equation}
\centering
\begin{gathered}
\arctan(\sqrt{1-\frac{r^2}{R_0^2}} \frac{L_0}{\sqrt{r^2-L_0^2}+r}) \approx \sqrt{1-\frac{r^2}{R_0^2}} \frac{L_0}{\sqrt{r^2-L_0^2}+r} \approx\\ \sqrt{1-\frac{r^2}{R_0^2}} \arctan( \frac{L_0}{\sqrt{r^2-L_0^2}+r} )
\end{gathered}
\end{equation}

Before proving (\ref{eq:proof22}), we simplify the expression. Rewriting (\ref{eq:proof22}) with $ k = \frac{L_0}{R_0} $ and $ x= \frac{r}{R_0} $ we get
\begin{equation}
\begin{gathered}
g(x;k) = \frac{1}{\sqrt{x^2-k^2}-x} \log( f(x;h) ) \approx 2, \\
f(x,k) = \frac{1+\sqrt{x^2-k^2}}{1-\sqrt{x^2-k^2}} \cdot \frac{\sqrt{k^2+1}-x}{\sqrt{k^2+1}+x}.
\end{gathered}
\end{equation}
We prove that $\displaystyle \lim_{k\rightarrow 0} g(x;k) = 2$ which implies (\ref{eq:proof22}).
\begin{equation}
\begin{gathered}
\lim_{k\rightarrow 0} g(x;k) = \lim_{k\rightarrow 0} \frac{1}{\sqrt{x^2-k^2}-x} \log( f(x;k) ) =\left(\frac{0}{0}\right) = (L'Hospital) =\\  \lim_{k\rightarrow 0}  \frac{-\sqrt{x^2-k^2}}{k} \frac{\frac{\partial}{\partial k}f(x;k)}{f(x;k)} = 2.
\end{gathered}
\end{equation}
Therefore, if $k = \frac{L_0}{R_0}$ is sufficiently small, then $g(x;h) \approx 2$, thus finalizing the second part of the proof.

Finally, we conclude that $P_r(r,R_0,L_0)$ is well approximated by $P_a(r,0,R_0,L_0)$ if the ratio $\frac{L_0}{R_0}$ is sufficiently small.

\section{The white image and contribution of weight \textit{w}}
To generate the white image, we use expression 
\begin{equation}
I_{WI}(r) = \frac{1}{N_p\sum_{i,j} w_{ij}}\displaystyle \sum_{i,j} w_{ij}P_a(r;h_{ij},R_{ij},L_{ij}).
\end{equation}
Weight $w_{ij}$ models the contribution of $P_a(r;h_{ij},R_{ij},L_{ij})$ in the entire white image. We focus our attention to calculation of the weights.

We will observe two cases - dependence on the distance between the two crystals ($R_{ij}$) and on the crystal length ($L_{ij}$). A hit length $L_{hit}$ from point source $(x,y)$ is described in Section 2. We know that $\hat{P}(x,y)=  \frac{L_{hit}}{L_{tot}} = \frac{L_{hit}}{4L_0}$ and $\int\int_S \hat{P}(x,y)dxdy = 2L_0R_0$, where $S=[-R_0,R_0]\times[-L_0,L_0]$ is a rectangular support. Therefore, a hit length from point $(x,y)$ is equal to:
\begin{equation}
f(x,y;R_0,L_0)=L_{hit} = 4L_0 
    \begin{cases} 
        \frac{R_0}{R_0+|x|}, & |y| < \frac{L_0}{R_0}|x|, |y| \leq L_0, |x| \leq R_0\\
        \frac{R_0^2}{R_0^2-x^2}\frac{L_0-|y|}{L_0}, & |y| \geq \frac{L_0}{R_0}|x|, |y| \leq L_0, |x| \leq R_0\\
        0, & elsewhere
   \end{cases}
\label{eq:Hit}.
\end{equation}
From all of the above, it follows that $\int\int_S f(x,y)dxdy = 8R_0L_0^2$ is valid.

\begin{figure}[H]
\begin{subfigure}{0.48\textwidth}
  \centering
  \includegraphics[width=1\linewidth]{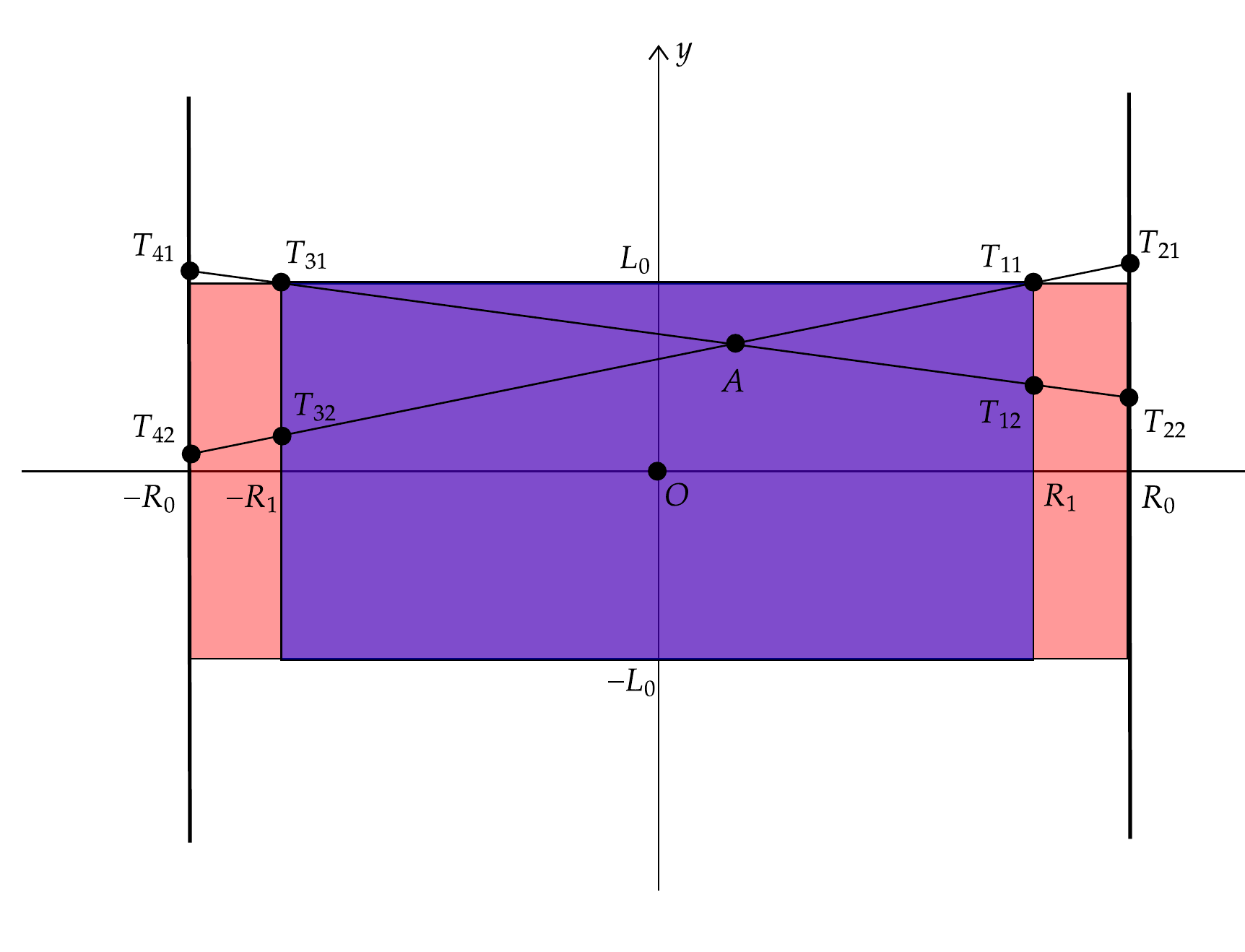}  
  \caption{Shrinking the distance between crystals}
  \label{fig:ShrinkR0}
\end{subfigure}
\begin{subfigure}{0.48\textwidth}
  \centering
  \includegraphics[width=1\linewidth]{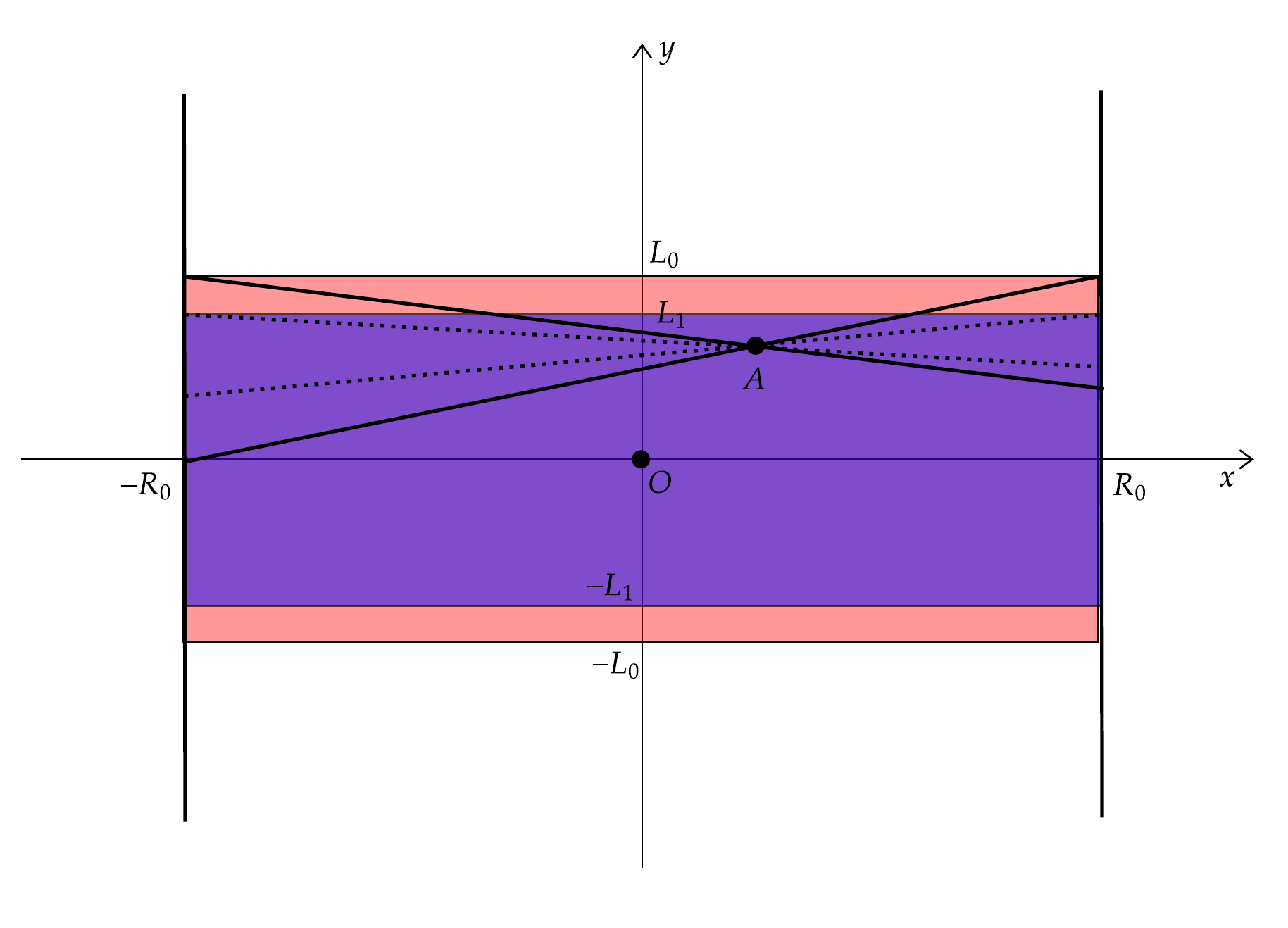}  
  \caption{Shrinking the length of crystals}
  \label{fig:ShrinkL0}
\end{subfigure}
\caption{Dependence on the crystal length and on the distance between the two crystals are examined separately.}
\label{fig:Shrink}
\end{figure}

In Fig. \ref{fig:ShrinkR0}, we observe two pairs of crystals. Both have length $2L_0$, but the larger one has radius $2R_0$ and the smaller one has radius $2R_1$. A hit length for pair of crystals with distance $2R_0$ is $f_{R_0}(x,y)=f(x,y;R_0,L_0)$. When the distance between crystals is $2R_1$, a hit length at the distance $L_1$ is $f(x,y;R_1,L_0)$. If we want to compare two length on same scale, we need to calculate a hit length at $R_0$. As shown in Fig. \ref{fig:ShrinkR0}, a hit length at distance $R_1$ is equal to $L_{hit_1} = |T_{11}T_{12}| + |T_{31}T_{32}|$ and at the distance $L_0$ it is $L_{hit_0} = |T_{21}T_{22}| + |T_{41}T_{42}|$. But, there are relations that connect $|T_{11}T_{12}|$ and $|T_{21}T_{22}|$, and $|T_{31}T_{32}|$ and $|T_{41}T_{42}|$:
\begin{equation}
\begin{gathered}
|T_{11}T_{12}| = \frac{R_1}{R_0}|T_{21}T_{22}|,\;\;\;\; |T_{31}T_{32}| = \frac{R_1}{R_0}|T_{41}T_{42}|.
\end{gathered}
\end{equation}
We insert the last expressions in $L_{hit_0}$ and get $L_{hit_0}=\frac{R_0}{R_1}L_{hit_1}$. Now, we write a hit length at distance $R_0$ from the origin for the pair of crystals that are $2R_1$ apart. A hit length is $f_{R_1}(x,y)=\frac{R_1}{R_0}f(x,y;R_0,L_0)$. We denote $S_0=[-R_0,R_0]\times[-L_0,L_0]$ and $S_1=[-R_1,R_1]\times[-L_0,L_0]$ as supports of functions $f_{R_0}(x,y)$ and $f_{R_1}(x,y)$. A total hit length when two crystals are apart by $2R_0$ is $T_{hit_0} = \int\int_{S_0} f_{R_0}(x,y)dxdy =  \int\int_{S_0} f(x,y;R_0,L_0)dxdy=8R_0L_0^2$. Similarly, a total hit length when two crystals are apart by $2R_1$ is equal to $T_{hit_1} = \int\int_{S_1} f_{R_1}(x,y)\,dxdy =  \int\int_{S_1} \frac{R_0}{R_1} f(x,y;R_1,L_0)\,dxdy= \frac{R_0}{R_1}\, 8R_1L_0^2 = 8R_0L_0^2$. Assuming uniform distribution, the total number of captured events do not depend on the distance between the crystals, since $T_{hit_0}=T_{hit_1}$.

The similar approach is done  for Fig. \ref{fig:ShrinkL0}. First, we denote $S_3=[-R_0,R_0]\times[-L_0,L_0]$ and $S_4=[-R_0,R_0]\times[-L_1,L_1]$ as supports of hit length functions. A hit length from a point source at $(x,y)$ when the crystal length is $2L_0$ is $f_{L_0}(x,y) = f(x,y;R_0,L_0)$. When the length of a crystal is $L_1$ we get $f_{L_1}(x,y) = f(x,y;R_0,L_1)$. A total hit lengths are $T_{hit_3} = \int\int_{S_3} f_{L_0}(x,y)\,dxdy =  \int\int_{S_3} f(x,y;R_0,L_0)dxdy=8R_0L_0^2$ and $T_{hit_4} = \int\int_{S_4} f_{L_1}(x,y)\,dxdy =  \int\int_{S_3} f(x,y;R_0,L_1)dxdy=8R_0L_1^2$. We conclude that $\frac{T_{hit_4}}{T_{hit_3}} = \frac{L_1^2}{L_0^2}$. Assuming uniform distribution, the total number of captured events depend on the length of crystals. 

Hence, the $w_{ij}$ weights are equal to $L_{ij}^2$, as claimed in Section 4. 

\end{document}